\documentclass{elsart}
\usepackage{amsfonts,epsfig,fleqn}
\def\markphotonatomur{\begin{picture}(2,2)(0,0) 
                             \put(2,1){\oval(2,2)[tl]}
                             \put(0,1){\oval(2,2)[br]}
                     \end{picture}
                    }                 
\def\markphotonatomdr{\begin{picture}(2,2)(0,0) 
                             \put(1,0){\oval(2,2)[bl]}
                             \put(1,-2){\oval(2,2)[tr]}
                     \end{picture}
                    }

\def\photonatomright{\begin{picture}(3,1.5)(0,0)
                                \put(0,-0.75){\tencircw \symbol{2}}
                                \put(1.5,-0.75){\tencircw \symbol{1}}
                                \put(1.5,0.75){\tencircw \symbol{3}}
                                \put(3,0.75){\tencircw \symbol{0}}
                      \end{picture}
                     }

\def\photonatomup{\begin{picture}(1.5,3)(0,0)
                             \put(-0.75,3){\tencircw \symbol{3}}
                             \put(-0.75,1.5){\tencircw \symbol{2}}
                             \put(0.75,1.5){\tencircw \symbol{0}}
                             \put(0.75,0){\tencircw \symbol{1}}
                   \end{picture}
                  }

\def\photonright{\begin{picture}(30,1.5)(0,0)
                     \multiput(0,0)(3,0){10}{\photonatomright}
                  \end{picture}
                 }

\def\markphotonurh{\begin{picture}(16,16)(0,0)
                     \multiput(0,0)(2,2){8}{\markphotonatomur}
                  \end{picture}
                 }
\def\markphotondrh{\begin{picture}(16,16)(0,0)
                     \multiput(0,0)(2,-2){8}{\markphotonatomdr}
                  \end{picture}
                 }
\def\photonrighthalf{\begin{picture}(30,1.5)(0,0)
                     \multiput(0,0)(3,0){5}{\photonatomright}
                  \end{picture}
                 }

\def\photonup{\begin{picture}(1.5,30)(0,0)
                  \multiput(0,0)(0,3){10}{\photonatomup}
               \end{picture}
              }
\def\photonuphalf{\begin{picture}(1.5,15)(0,0)
                      \multiput(0,0)(0,3){5}{\photonatomup}
                   \end{picture}
                  }

\def\fermionup{\begin{picture}(1,30)(0,0)
                     \put(0,0){\vector(0,1){15}}
                     \put(0,15){\line(0,1){15}}
               \end{picture}
              }
\def\fermionuphalf{\begin{picture}(1,15)(0,0)
                         \put(0,0){\vector(0,1){7.5}}
                         \put(0,7.5){\line(0,1){7.5}}
                   \end{picture}
                  }

\def\fermionullo{\begin{picture}(24,24)(0,0)
                        \put(0,0){\vector(-1,1){12}}
                        \put(-12,12){\line(-1,1){12}}
                  \end{picture}
                 }

\def\fermionurlo{\begin{picture}(24,24)(0,0)
                        \put(-24,-24){\vector(1,1){12}}
                        \put(-12,-12){\line(1,1){12}}
                  \end{picture}
                 }

\def\fermionull{\begin{picture}(30,15)(0,0)
                        \put(0,0){\vector(-2,1){15}}
                        \put(-15,7.5){\line(-2,1){15}}
                  \end{picture}
                 }
\def\fermionullhalf{\begin{picture}(15,7.5)(0,0)
                        \put(0,0){\vector(-2,1){7.5}}
                        \put(-7.5,3.75){\line(-2,1){7.5}}
                  \end{picture}
                 }
\def\fermionurr{\begin{picture}(30,15)(0,0)
                        \put(-30,-15){\vector(2,1){15}}
                        \put(-15,-7.5){\line(2,1){15}}
                  \end{picture}
                 }
\def\fermionurrhalf{\begin{picture}(15,7.5)(0,0)
                        \put(-15,-7.5){\vector(2,1){7.5}}
                        \put(-7.5,-3.75){\line(2,1){7.5}}
                  \end{picture}
                 }

\newenvironment{Feynman}[3]{\begin{center}
                            \setlength{\unitlength}{#3 mm}
                            \begin{picture}(#1)(#2)
                            \thicklines
                           }{\end{picture} \end{center}}
\input{psfig}
\begin{document}
\hyphenation{brems-strah-lung}
\renewcommand{\thesection}{\arabic{section}}
\renewcommand{\theequation}{\thesection.\arabic{equation}}
\renewcommand{\medskip}{\vspace{.4cm} \\}
\renewcommand{\bigskip}{\vspace{.5cm} \\}
\newcommand{\ezero}{\setcounter{equation}{0}}
\newcommand{\ds}{\displaystyle}
\newcommand{\nl}{\nonumber \\}
\newcommand{\mycos}{\cos\!}
\newcommand{\mysin}{\sin\!}
\newcommand{\myln}{\ln\!}
\newcommand{\mysp}{{\rm Li_2\!}}
\newcommand{\mytri}{{\rm Li_3\!}}
\newcommand{\ba}{\begin{eqnarray}}
\newcommand{\ea}{\end{eqnarray}}
\newcommand{\beq}{\begin{equation}}
\newcommand{\eeq}{\end{equation}}
\newcommand{\xsec}{cross-sec\-tion}
\newcommand{\xsecs}{cross-sec\-tions}
\newcommand{\miniskip}{\vspace{.15cm} \\}
\newcommand{\smskip}{\vspace{.3cm} \\}
\newcommand{\hugeskip}{\vspace{.8cm} \\}
\newcommand{\numreal}{I\!\!R}
\newcommand{\numcomp}{{\rm C} \put(-5.5,.3){\line(0,1){7.4}}
                              \put(-5.2,.2){\line(0,1){7.6}} }
\newcommand{\ieps}{{\rm i}\varepsilon}
\newcommand{\iospi}{\frac{{\rm i}}{16\,\pi^2}}
\newcommand{\lz}{l_0}
\newcommand{\ri}{{\rm i}}
\newcommand{\cc}{{\mathcal C}}
\newcommand{\cd}{{\mathcal D}}
\newcommand{\cf}{{\mathcal F}}
\newcommand{\cg}{{\mathcal G}}
\newcommand{\cl}{{\mathcal L}}
\newcommand{\cm}{{\mathcal M}}
\newcommand{\co}{{\mathcal O}}
\newcommand{\cp}{{\mathcal P}}
\newcommand{\cs}{{\mathcal S}}
\newcommand{\ct}{{\mathcal T}}
\newcommand{\cu}{{\mathcal U}}
\newcommand{\oal}{${\mathcal O}(\alpha)$}
\newcommand{\zz}{$Z$}
\newcommand{\wpl}{$W^+$}
\newcommand{\wmi}{$W^-$}
\newcommand{\epl}{$e^+$}
\newcommand{\emi}{$e^-$}
\newcommand{\ee}{\epl\emi}
\newcommand{\fone}{$f_1$}
\newcommand{\bftwo}{$\bar f_2$}
\newcommand{\fthree}{$f_3$}
\newcommand{\bffour}{$\bar f_4$}
\newcommand{\fpone}{$\fone\bfone$}
\newcommand{\fptwo}{$\ftwo\bftwo$}
\newcommand{\ccthree}{{\tt CC3}}
\newcommand{\nctwo}{{\tt NC2}}
\newcommand{\nceight}{{\tt NC8}}
\newcommand{\re}{\mathrm{Re}}
\newcommand{\im}{\mathrm{Im}}
\newcommand{\sone}{$s_1$}
\newcommand{\stwo}{$s_2$}
\newcommand{\slam}{$\sqrt{\lambda}$}
\newcommand{\ZZ}{Z^0}
\newcommand{\WPL}{W^+}
\newcommand{\WMI}{W^-}
\newcommand{\EPL}{e^+}
\newcommand{\EMI}{e^-}
\newcommand{\EE}{\EPL\EMI}
\newcommand{\SONE}{s_{1}}
\newcommand{\STWO}{s_{2}}
\newcommand{\TONE}{t_{12}}
\newcommand{\TTWO}{t_{34}}
\newcommand{\DONE}{d_{1}}
\newcommand{\DTWO}{d_{2}}
\newcommand{\ME}{m_e}
\newcommand{\MES}{m_e^2}
\newcommand{\FONE}{f_1}
\newcommand{\FTWO}{f_2}
\newcommand{\SIW}{\sin{\!\theta_w}}
\newcommand{\COW}{\cos{\!\theta_w}}
\newcommand{\SWS}{\sin^2{\!\theta_w}}
\newcommand{\CWS}{\cos^2{\!\theta_w}}
\newcommand{\SWSQ}{s^2_w}
\newcommand{\CWSQ}{c^2_w}
\newcommand{\SLAM}{\sqrt{\lambda}}
\newcommand{\SLAMP}{\sqrt{\lambda^{'}}}
\newcommand{\LAMP}{\lambda^{'}}
\newcommand{\SLAMB}{\sqrt{\bar{\lambda}}}
\newcommand{\LAMB}{\bar{\lambda}}
\newcommand{\sprm}{s'_{\!-}}
\newcommand{\sprp}{s'_{\!+}}
\newcommand{\SPR}{s'}
\newcommand{\dagg}[1]{#1 \hspace{-.19cm} / \hspace{.06cm}}
%
%
%
\begin{frontmatter}

\vspace{-1.5cm}
\begin{flushleft}
  DESY 96--028 \\
  February 1996
\end{flushleft}
\vspace{.5cm}

\title{Complete Initial State QED Corrections to Off-Shell Gauge Boson
  Pair Production in \ee~Annihilation}
\author{Dima Bardin$^1$,  Dietrich Lehner$^2$, Tord Riemann} 
\address{Deutsches Elektronen-Synchrotron DESY,
 Institut f\"ur Hochenergiephysik IfH~Zeuthen,
 Platanenallee 6, D-15738 Zeuthen, Germany}
\thanks{On leave of absence from Bogoliubov Theor.
   Lab., JINR, ul. Joliot-Curie 6, RU-141980 Dubna, Moscow
   Region, Russia.}
\thanks{Now at Fakult\"at f\"ur Physik, Albert-Ludwigs-Universit\"at, 
        Hermann-Herder-Str.~3, D-79104 Freiburg, Germany}
\begin{abstract}
We study Standard Model four-fermion production in $e^+ e^-$
annihilation at LEP2 energies and above using a semi-analytical
approach. We derive the complete QED initial state corrections (ISR)
to the reactions $\EE \rightarrow (\ZZ\ZZ) \rightarrow
f_1\bar{f_1}f_2\bar{f_2}$~and $\EE \rightarrow (\WPL\WMI) \rightarrow
{\bar f}_1^u f_1^d f_2^u {\bar f}_2^d$ with $f_1\neq
f_2$~and $f_i\neq e^\pm, \stackrel{_{(-)}}{\nu_e}$. As compared to
the well-known universal {\it s}-channel ISR, additional complexity
arises due to non-universal, process-dependent ISR contributions from
{\it t}- and {\it u}-channel fermion exchanges. The full set of
formulae needed to perform numerical calculations is given
together with samples of numerical results. 
\end{abstract}
\end{frontmatter}
%
%
\section{Introduction}
\label{intro}
\ezero
%
The LEP2 \ee~accelerator will finally operate at energies between 176
and 205 GeV~\cite{lep2rep} and thus pass the production thresholds for
$W^{\pm}$~and \zz~pairs. 
At LEP2, a typical process will be four-fermion production which is
much more complex than fermion pair production as known from LEP1. 
This complexity shows up at tree level already, because of the
many Feynman diagrams involved in four-fermion production. 
As tree-level amplitudes are not sufficient to describe experimental
data, radiative corrections to four-fermion production are needed.
It would be desirable to derive complete \oal~electroweak and QCD 
corrections to four-fermion production, but this has not yet been
achieved, although many partial results were reported
in~\cite{lep2ww,lep2wgen,lep2sm,lep2exgen} and in references quoted
therein.  
In view of the anticipated experimental precision of LEP2 and
future \ee colliders, it would be desirable to have theoretical predictions
for \xsecs\ and distributions accurate at the level of half a percent.

In recent years, three major approaches to four-fermion production in
\ee~annihilation have been developed, namely Monte Carlo
approaches~\cite{lep2wgen}, the semi-analytical
approach~\cite{muta,WWnuni,teuw94,nc8a,nc8b,ZZnuni,dldiss,cc11}, and 
the ``deterministic approach''~\cite{passarino94}.
Monte Carlo and deterministic techniques use numerical integration for
all phase space variables.  
Typically, the semi-analytical method performs analytical integrations
over the five (seven, if ISR is included) angular degrees of freedom
and uses high precision numerical integration for the remaining two
(three, if ISR is included) squared invariant masses.  
It represents an approach to the high-dimensional, highly
singular phase space integration problem inherent in this type of
physical problem, which is elegant, fast, and numerically stable.  
Thus, it may serve as an ideal source of benchmarks for the other
two approaches.
From LEP1, we know that semi-analytical calculations are also relevant
to experimentalists.

\begin{figure}[t]
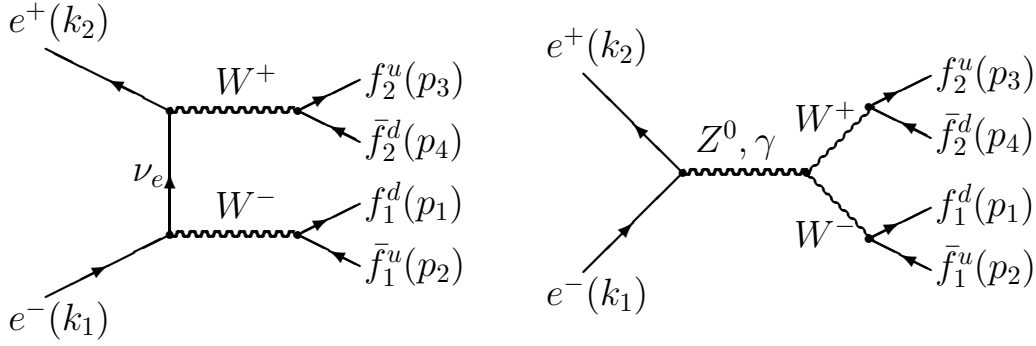

\vspace*{.6cm}
\begin{Feynman}{150,60}{-25,5}{0.55}
%
\large
\thicklines
%
\put(-35,15){\fermionup}
\put(-35,15){\fermionurr}
\put(-35,45){\fermionull}
\put(-35,45){\photonright}
\put(-35,15){\photonright}
\put(11,37.5){\fermionullhalf}
\put(11,52.5){\fermionurrhalf}
\put(11,7.5){\fermionullhalf}
\put(11,22.5){\fermionurrhalf}
\put(-35,15){\circle*{2}}
\put(-35,45){\circle*{2}}
\put(-4,15){\circle*{2}}
\put(-4,45){\circle*{2}}
\put(-24,49){$W^+$}
\put(-24,19){$W^-$}
\put(-74,64){$e^+(k_2)$}
\put(-74,-8){$e^-(k_1)$}
\put(-44,28){$\nu_e$}
\put(12,50){$f_2^u (p_3)$}
\put(12,35){${\bar f}_2^d(p_4)$}
\put(12,20){$f_1^d(p_1)$}
\put(12,5){${\bar f}_1^u(p_2)$}
%
\put(89,30){\fermionurlo}
\put(89,30){\fermionullo}
\put(89,30){\photonright}
\put(119,30){\markphotonurh}
\put(119,30){\markphotondrh}
\put(149,38.5){\fermionullhalf}
\put(149,53.5){\fermionurrhalf}
\put(149,6.5){\fermionullhalf}
\put(149,21.5){\fermionurrhalf}
\put(89,30){\circle*{2}}
\put(119,30){\circle*{2}}
\put(134,14){\circle*{2}}
\put(134,46){\circle*{2}}
\put(56,58){$e^+(k_2)$}
\put(56,-2){$e^-(k_1)$}
\put(92,35){$Z^0,\gamma$}
\put(116,40){$W^+$}
\put(116,12){$W^-$}
\put(150,51.5){$f_2^u (p_3)$}
\put(150,36.5){${\bar f}_2^d(p_4)$}
\put(150,19.5){$f_1^d(p_1)$}
\put(150,4.5){${\bar f}_1^u(p_2)$}
\end{Feynman}
\vspace{1.0cm}
\caption[Tree level $W$~pair production Feynman diagrams]
{\it The tree level Feynman diagrams for off-shell $W$~pair
  production (the \ccthree\ process). Left: t-channel diagram. Right:
  s-channel diagram. The particle momenta are given by the $k_i$~and
  $p_j$.}
\label{cc3diag}
\vspace{1.2cm}
\end{figure}

\begin{figure}[t]
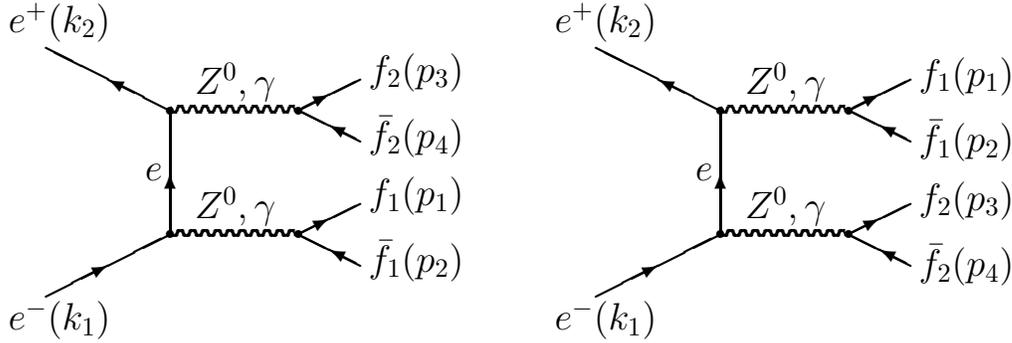

\vspace*{.6cm}
\begin{Feynman}{75,60}{50,0}{0.55}
%
\large
\thicklines
%
%
\put(5,15){\fermionurr}
\put(5,45){\fermionull}
\put(5,15){\fermionup}
\put(5,45){\photonright}
\put(5,15){\photonright}
\put(5,45){\circle*{2}}
\put(5,15){\circle*{2}}
\put(36,15){\circle*{2}}
\put(36,45){\circle*{2}}
\put(51,7.5){\fermionullhalf}
\put(51,22.5){\fermionurrhalf}
\put(51,37.5){\fermionullhalf}
\put(51,52.5){\fermionurrhalf}
\put(-1,28){$e$}
\put(-34,64){$e^+(k_2)$}
\put(-34,-8){$e^-(k_1)$}
\put(11,49){$\ZZ,\gamma$}
\put(11,19){$\ZZ,\gamma$}
\put(53,06){${\bar f}_1(p_2)$}
\put(53,22){$f_1(p_1)$}
\put(53,36){${\bar f}_2(p_4)$}
\put(53,52){$f_2(p_3)$}
%
%
\put(138,15){\fermionurr}
\put(138,45){\fermionull}
\put(138,15){\fermionup}
\put(138,45){\photonright}
\put(138,15){\photonright}
\put(138,45){\circle*{2}}
\put(138,15){\circle*{2}}
\put(169,15){\circle*{2}}
\put(169,45){\circle*{2}}
\put(184,7.5){\fermionullhalf}
\put(184,22.5){\fermionurrhalf}
\put(184,37.5){\fermionullhalf}
\put(184,52.5){\fermionurrhalf}
\put(132,28){$e$}
\put(98,64){$e^+(k_2)$}
\put(98,-8){$e^-(k_1)$}
\put(144,49){$\ZZ,\gamma$}
\put(144,19){$\ZZ,\gamma$}
\put(186,5){${\bar f}_2(p_4)$}
\put(186,21){$f_2(p_3)$}
\put(186,35){${\bar f}_1(p_2)$}
\put(186,51){$f_1(p_1)$}
\end{Feynman}
\vspace{1.0cm}
\caption[Tree level neutral gauge boson Feynman diagrams]
{\it The tree level Feynman diagrams for off-shell neutral gauge boson
  pair production (\nceight\ process).
  Left: t-channel. Right: u-channel. The
  \nctwo\ process is obtained by neglecting diagrams with exchange
  photons.}
\label{nc8diag}
\vspace{1.5cm}
\end{figure}

In this article we will present semi-analytical results for
the gauge boson pair production reactions 
\\
$\begin{array}{lllcllr}
  ~{\tt CC3}:~~ &
    \EE & \! \rightarrow \! & (\WPL\WMI) & \! \rightarrow \! &
    {\bar f}_1^u f_1^d f_2^u {\bar f}_2^d (\gamma)~,
    \hspace{.45cm} & f_i\!\neq\!e^\pm,\stackrel{_{(-)}}{\nu_e} \\
  ~{\tt NC2}:~~ &
    \EE & \! \rightarrow \! & (\ZZ\ZZ) & \! \rightarrow \! &
    f_1\bar{f_1}f_2\bar{f_2}(\gamma)~,
    \hspace{.45cm} &
    f_1\!\neq\!f_2,~f_i\!\neq\!e^\pm,\stackrel{_{(-)}}{\nu_e} \\
  ~{\tt NC8}:~~ & \EE & \! \rightarrow \! &
    (\ZZ\ZZ,\ZZ\gamma,\gamma\gamma) & \! \rightarrow \! &
    f_1\bar{f_1}f_2\bar{f_2}(\gamma)~,
    \hspace{.45cm} & f_1\!\neq\!f_2,~f_i\!\neq\!e^\pm,
    \stackrel{_{(-)}}{\nu_e}
\end{array}$
\vspace{-1.05cm}
\beq
  \label{4fproc}
\eeq
with complete initial state QED corrections.
The three Feynman diagrams for the charged current \ccthree\ process
at tree level are given in figure~\ref{cc3diag}.  
The two or eight diagrams for the neutral current \nctwo\ and
\nceight\ processes are depicted in figure~\ref{nc8diag}\footnote{
Strictly speaking, the \nctwo\ process is well-defined
(i.e. observable) only as on-shell reaction.}. 
Classifications of four-fermion processes may be found in
references~\cite{teuw94,berends94}.
%
%
%
%
%
%

Initial state QED corrections (ISR) represent a dominant correction
in \ee\ annihilation.
Complete ISR to four-fermion production separates into a universal,
factorizing, process-independent contribution and a non-universal,
non-factorizing, process-dependent part.
With the index $J$ labelling the \ccthree, \nctwo, and \nceight\
processes, the ISR corrected \xsecs\ can be generically written as
\vspace{.7cm}
\ba
  \frac{\d\sigma_{J,\mathrm{QED}}(s)}{\d s_1 \d s_2} & = &
    \int 
    \frac{\d s'}{s} 
    \Bigl[  G(s'/s) \, \sigma_{J,0}(s',s_1,s_2)  + 
    \sigma_{J,\mathrm{QED}}^{\mathrm{non-univ}}(s,s',s_1,s_2) \Bigr]
    \nl
  \label{BBnuni}
\ea
\\
with invariant boson masses $s_1$~and $s_2$, reduced center of mass
energy squared $s'$, and tree level four-fermion production \xsec\
$\sigma_{J,0}$. The factor $G(s'/s)$ contains all mass singularities
$\ln(s/\MES)$ and incorporates the process-indepen\-dent ISR radiators
as known from {\it s}-channel \ee~annihilation~\cite{QEDhigher}. 
In addition, there is a non-universal contribution
$\sigma_{J,\mathrm{QED}}^{\mathrm{non-univ}}$.
It appears together with {\it t}-channel and {\it u}-channel
amplitudes and is mass singularity free.
For pure {\it s}-channel contributions it is absent.  

Complete initial state QED corrections were shortly communicated for
the \ccthree\ process in~\cite{WWnuni} and for the \nctwo\ and 
\nceight\ processes in~\cite{ZZnuni}.
While a definition of initial state radiation is straightforward in
the neutral current process, there is an arbitrariness for $W$~pair
production in its definition.
In~\cite{WWnuni}, we restored the U(1)-invariance of the initial state
photon emission by adding an auxiliary current.
This arbitrariness is characteristic of charged current processes.  
Of course, it is the sum of the corrections what will be finally
observable. 

In this paper, the complete analytical formulae for the non-universal
corrections, supplemented by a study of their numerical
importance, will be presented for the first time. 

For on-shell production,
\vspace{.7cm}
\ba
  \EE & \rightarrow & \WPL\WMI(\gamma), 
  \label{onww} \nl
  \EE & \rightarrow & \ZZ\ZZ(\gamma),
  \label{onzz}
\ea
\\
the generic cross-section may be obtained
as follows:  
\vspace{.7cm}
\ba
{\bar \sigma}_{J,\mathrm{QED}}(s) & \, = \, &
  \int\limits_{4M_V^2}^s
  \frac{\d s'}{s} \;
  \Bigl[ \: G(s'/s) \, {\bar \sigma}_{J,0}(s',M_V^2,M_V^2)
  \nl & & \hspace{3.5cm} + \;
  {\bar \sigma}_{J,\mathrm{QED}}^{\mathrm{non-univ}}(s,s',M_V^2,M_V^2) \Bigr].
  \label{BBnunon}
\ea
\\
Here, $M_V$ represents the $W$ or $Z$ boson mass. The relation
between ${\bar \sigma}_{J,0}$\ and $\sigma_{J,0}$ on one hand and
and ${\bar \sigma}_{J,\mathrm{QED}}^{\mathrm{non-univ}}$ and
$\sigma_{J,\mathrm{QED}}^{\mathrm{non-univ}}$ on the other hand will
be discussed in section~\ref{xsISR1}.

For $Z$~pair production, there are no additional QED corrections for
the on-shell case, 
while for $W$~pair production there are final state corrections and
initial-final interferences. 
The non-universal initial state QED corrections 
were not known before as explicit analytical expressions.
However, they have been determined as a part of the complete
electroweak corrections to the processes~(\ref{onzz}) with numerical
integrations in~\cite{on1loop}.

The outline of this report is as follows.
Section~\ref{xsISR1} presents general features of our approach to the
complete initial state QED corrections. 
In section~\ref{xsISR2} we give a detailed presentation of the
non-universal \xsec\ contributions.
Section~\ref{numres} contains numerical results and section~\ref{sum}
concluding remarks.
In a series of appendices, we give technical details of the performed
computations.
Some notations are introduced in appendix~\ref{appborncg}.
Our phase space parametrization and all relevant relations between
particle four-momenta and phase space variables are presented in
appendix~\ref{ps2to5}.
In appendix~\ref{appmatel}, the tree level, the real ISR, and the
virtual ISR matrix elements are given.
These matrix elements represent the starting point for the
calculations in this paper.
The analytical integrals needed for the integration of the angular
phase space variables and, in the case of virtual corrections, loop
momenta are collected in appendix~\ref{appints}.  
%
%
\section
{General Structure of Initial State QED Corrections
\label{xsISR1}
}
\ezero
%
To include initial state QED corrections (ISR) to~(\ref{4fproc}), the
five-particle phase space is required to take into account four final
state fermions and a bremsstrahlung photon with momentum $p$.
We make use of the following para\-me\-tri\-zation: 
\vspace{.5cm}
\ba
  \d\Gamma_5 & = & \frac{1}{(2\pi)^{14}} \:
                  \frac{\sqrt{\lambda(s,s',0)}}{8s} \:
                  \frac{\sqrt{\lambda(s',\SONE,\STWO)}}{8s'} \:
                  \frac{\sqrt{\lambda(\SONE,m_1^2,m_2^2)}}{8\SONE} \:
                  \frac{\sqrt{\lambda(\STWO,m_3^2,m_4^2)}}{8\STWO}
                  \nl
             &   & \hspace{3.8cm}\times\; \d s' \; \d \SONE \; \d \STWO \;
                  \d \! \mycos\theta \,
                  \d \Omega_R \, \d \Omega_1 \, \d \Omega_2.
  \label{ph25el}
\ea
\\
In equation~(\ref{ph25el}), the azimuth angle around the beam has already
been integrated.
We have adopted the usual definition of the $\lambda$
function,
\vspace{.5cm}
\ba
  \lambda(a,b,c) &=& a^2+b^2+c^2-2ab-2ac-2bc, \nl
  \lambda &\equiv& \lambda(s,s_1,s_2).
  \label{lambda}
\ea
\\
We use $k_1$~and $k_2$ as the initial electron and positron
four-momenta, while $p_1$, $p_2$, $p_3$, and $p_4$ label the final state
fermion momenta as indicated in figures~\ref{cc3diag}
and~\ref{nc8diag}.
The relevant invariant masses are\footnote[4]{In our metric space-like
  four-vectors $k$ have positive $k^2$. Thus $k^2=-m^2$ for on-shell
  particles of mass $m$.}
\vspace{.5cm}
\ba
  s  & = & - (k_1+k_2)^2 \; = \; - (p_1+p_2+p_3+p_4+p)^2, 
\nl
  s' & = & - (p_1+p_2+p_3+p_4)^2,
\nl
  s_1 & = & - (p_1+p_2)^2,
\nl
  s_2 & = & - (p_3+p_4)^2.
\ea
\\
The photon scattering angle $\theta$\ is defined as the angle
between $\pol{p}$\ and $\pol{k}_2$ in the center of mass system. 
The solid angle
\beq
  \d \Omega_R \; = \; \d \! \mycos\theta_R \d\phi_R
\eeq
represents the production solid angle of the boson three vector
$\pol{v}_1 \!=\! \pol{p}_1 \!+\! \pol{p}_2$\ in the four-fermion rest
frame $\pol{p}_1 \!+\! \pol{p}_2 \!+\! \pol{p}_3 \!+\! \pol{p}_4
\!=\!0$. 
$\Omega_1$\ [$\Omega_2$] is the solid angle of $\pol{p}_1$\
[$\pol{p}_3$] in the two-fermion rest frame
$\pol{p}_1\!+\!\pol{p}_2\!=\!0$\ [$\pol{p}_3\!+\!\pol{p}_4\!=\!0$].
In this frame, the three-vectors $\pol{p}_1$\ and $\pol{p}_2$\
[$\pol{p}_3$\ and $\pol{p}_4$] are back to back.
The polar and azimuthal decay angles in the above two-particle rest
frames are defined via
\beq
  \d \Omega_i = \d \!\mycos\theta_i \: \d \phi_i, \hspace{1cm}
  i=1,2.
\eeq

Further details of the kinematics and the five-particle phase space
may be found in appendix~\ref{ps2to5}.

\begin{figure}[t]
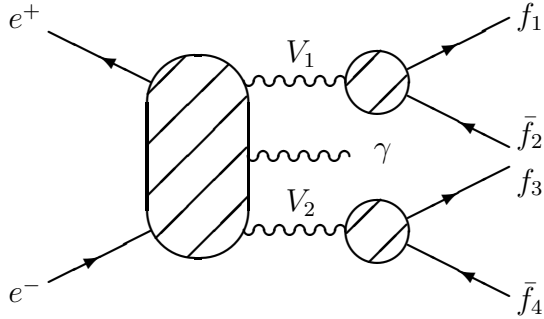

  \vspace*{.2cm}
  \begin{Feynman}{75,60}{-14.5,0}{0.9}
%
    \put(5,19){\fermionurrhalf}
    \put(5,41){\fermionullhalf}
    \put(10.5,15.2){\line(1,1){8.9}}
    \put(10.5,22.7){\line(1,1){8.7}}
    \put(10.5,22.7){\line(-1,-1){4.7}}
    \put(10.5,30.2){\line(1,1){8.7}}
    \put(10.5,30.2){\line(-1,-1){5.85}}
    \put(10.5,37.7){\line(1,1){6}}
    \put(10.5,37.7){\line(-1,-1){5.85}}
    \put(10.1,44.8){\line(-1,-1){5.5}}
    \put(12,30){\oval(15,30)}
    \put(18.65,19){\photonrighthalf}
    \put(19.25,30){\photonrighthalf}
    \put(18.65,41){\photonrighthalf}
    \put(34,40.7){\line(1,1){5}}
    \put(37.3,36.5){\line(1,1){5.5}}
    \put(34,18.7){\line(1,1){5}}
    \put(37.3,14.5){\line(1,1){5.5}}
    \put(38.5,19){\circle{10}}
    \put(38.5,41){\circle{10}}
    \put(58,50.5){\fermionurrhalf}
    \put(58,28.5){\fermionurrhalf}
    \put(58,31.5){\fermionullhalf}
    \put(58,9.5){\fermionullhalf}
    \put(-16,8.5){\emi}
    \put(-16,49){\epl}
    \put(25,44){$V_1$}
    \put(38,30){$\gamma$}
    \put(25,22){$V_2$}
    \put(59,7.5){${\bar f}_4$}
    \put(59,25){$ f_3$}
    \put(59,32){${\bar f}_2$}
    \put(59,49.5){$ f_1$}
  \end{Feynman}
  \vspace{-.5cm}
  \caption[Generic two-boson plus photon production Feynman diagram
  for 
  \ee~annihilation]
    {\it Generic two-boson production and decay Feynman diagram
      with real photon initial state radiation.}
  \label{twores}
  \vspace{1cm}
\end{figure}

The processes~(\ref{4fproc}) have the generic structure shown in
figure~\ref{twores}.
%
%
%
%
Their ISR-corrected \xsecs\ get two types of contributions, namely
universal and non-universal ones.
The first type is universal in the sense that it arises from photonic
insertions to any of the basic diagrams and is independent of the
details of the subsequent interactions.
These universal corrections are, of course, related to the collinear
divergences of the (radiating) initial state electrons and positrons
and appear to
be known, including higher order terms, from, e.g., the study of ISR
corrections to the (single) $Z$ line shape.    
They are exactly those of the $s$-channel ISR contributions.
For diagrams with $t$- and $u$-channel exchanges and interferences of
these diagrams among themselves and with $s$-channel diagrams, there
are additional corrections which are not logarithmically enhanced and
which depend on the details of the interfering amplitudes.
These additional corrections are thus non-universal.   

\begin{figure}[tb]
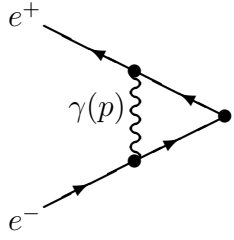

\begin{Feynman}{150,45}{12,28}{0.8}
%
%
\put(69,60){$e^+$}
\put(69,26){$e^-$}
\put(79,45){$\gamma(p)$}
\put(105,45){\fermionullhalf}
\put(90,52.5){\fermionullhalf}
\put(105,45){\fermionurrhalf}
\put(90,37.5){\fermionurrhalf}
\put(90,37.5){\photonuphalf}
\put(105,45){\circle*{2}}
\put(90,37.5){\circle*{2}}
\put(90,52.5){\circle*{2}}
\end{Feynman}
\caption{\em The amputated s-channel virtual initial state QED Feynman
  diagram.}
\label{svirt}
\vspace{.5cm}
\end{figure}

\begin{figure}[tb]
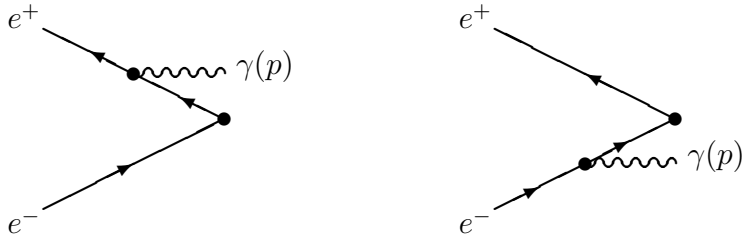

\begin{Feynman}{150,46}{-100,28}{0.8}
%
\put(-81,60){$e^+$}
\put(-81,26){$e^-$}
\put(-43,52.5){$\gamma(p)$}
\put(-45,45){\fermionurr}
\put(-45,45){\fermionullhalf}
\put(-60,52.5){\photonrighthalf}
\put(-60,52.5){\fermionullhalf}
\put(-45,45){\circle*{2}}
\put(-60,52.5){\circle*{2}}
%
\put(-6,60){$e^+$}
\put(-6,26){$e^-$}
\put(32,37.5){$\gamma(p)$}
\put(30,45){\fermionurrhalf}
\put(30,45){\fermionull}
\put(15,37.5){\photonrighthalf}
\put(15,37.5){\fermionurrhalf}
\put(30,45){\circle*{2}}
\put(15,37.5){\circle*{2}}
\end{Feynman}
\caption{\it The amputated s-channel initial state bremsstrahlung Feynman
  diagrams.}
\label{annISR}
\vspace{.7cm}
\end{figure}

ISR-corrected cross-sections for the processes~(\ref{4fproc}) with
soft photon exponentiation may be described by the ansatz
\vspace{.5cm}
\ba
  \sigma^{\rm ISR}_J(s) & = &  \int \d \SONE \int \d \STWO \;\;
  \int
\frac{\d s'}{s} \;\;
    \sum_k \frac{\d^3\Sigma_J^{(k)}
      (s,s';\SONE,\STWO)}{\d \SONE \d \STWO\d s'}
  \label{ISRxstot}
\ea
\\
with the threefold differential \xsec\ 
\beq
  \frac{\d^3\Sigma_J^{(k)}(s,s';s_1,s_2)}{\d s_1 \d s_2 \d s'}
  \; = \;\;  {\mathcal C}_J^{(k)}(s',\SONE,\STWO)
  \left[ \beta_e v^{\beta_e - 1} {\mathcal S}_J^{(k)}+
         {\mathcal H}_J^{(k)} \right].
  \label{ISRxsdif}
\eeq
where
\vspace{.5cm}
\ba
  \beta_e & = & \frac{2\,\alpha}{\pi} \!
    \left[ \ln \left(\frac{s}{m_e^2}\right) - 1 \right],
  \nl
  v & = & 1 - \frac{s'}{s}.
\ea
\\
In equation~(\ref{ISRxsdif}) we have used several additional notations
which will now be explained.
The subscript $J$ labels different processes, $J \, \in \left\{ {\tt
  CC3,NC2,NC8} \right\}$ and the superscript index $k$ stands for
\xsec\ contributions which stem from squared amplitudes or
interferences with distinct Feynman topologies and coupling
structures.
Using
\beq
  \sigma_J^{(k,0)}(s;s_1,s_2) \equiv
  \frac{\sqrt{\lambda}}{\pi s^2} \, \cg^{(k)}_J(s;s_1,s_2),
\eeq
the soft+virtual contributions ${\mathcal S}_J^{(k)}$\ and the hard
contributions ${\mathcal H}_J^{(k)}$\ take the form
\vspace{.5cm}
\ba
  {\mathcal S}_J^{(k)} (s,s';s_1,s_2) & = &
    \left[1 + {\bar S}(s) \right] \sigma_J^{(k,0)}(s';s_1,s_2)
    + \sigma_{{\hat S},J}^{(k)}(s';s_1,s_2), 
\nl \nl
  {\mathcal H}_J^{(k)}(s,s';s_1,s_2) & = &
  \underbrace{{\bar H}(s,s') \:
    \sigma_J^{(k,0)}(s';s_1,s_2)\;\;\;\;}_{Universal~Part}
  + \underbrace{\sigma_{{\hat H},J}^{(k)}
    (s,s';s_1,s_2)}_{Non-universal~Part}
\label{defnonu}
\ea
\\
with the well-known soft+virtual radiator ${\bar S}$\ and the hard
radiator ${\bar H}$, 
\vspace{.5cm}
\ba
  {\bar S}(s) & = & \frac{\alpha}{\pi}
                    \left(  \frac{\pi^2}{3} - \frac{1}{2} \right)
                    + \frac{3}{4} \, \beta_e + {\mathcal O}(\alpha^2),
                    \nl \nl
  {\bar H}(s,s') & = & - \frac{1}{2}
                    \left(1+\frac{s'}{s}\right)\beta_e
                    + {\mathcal O}(\alpha^2).
\ea
\\
Higher order terms~\cite{QEDhigher} may be implemented exactly as
described in~\cite{lep2ww,lep2wgen,cc11}. 
We draw the attention to the definition of the (differential) effective
tree level cross-section
\vspace{.5cm}
\ba
  \frac{\d \sigma_{J,0}(s';\SONE,\STWO)}{\d \SONE \d \STWO} & \, = \, &
\frac{\sqrt{\lambda(s',s_1,s_2)}}{\pi s'^2}
\sum_k 
    \cc_J^{(k)}(s',s_1,s_2) \cg_J^{(k)}(s';s_1,s_2)
  \label{totsig}
\ea
\\
which is inherent in the universal ISR correction part.
Coupling constants and boson propagators are collected in
$\cc_J^{(k)}$, while the $\cg_J^{(k)}$ represent kinematical functions
obtained from the analytical angular integration.
Explicit expressions for $\cc_J^{(k)}$\ and $\cg_J^{(k)}$\ may be found
in appendix~\ref{appborncg}.
If the index $k$ is associated with $s$-channel \ee\ annihilation
only, non-universal ISR contributions are absent.
The $s$-channel ISR diagrams are generically shown in
figures~\ref{svirt} and~\ref{annISR}. 
They contribute to $W$ pair production.
The non-universal \xsec\ contributions originate from the angular
dependence of initial state {\em t}- and {\em u}-channel propagators
and contribute both to $W$- and $Z$-pair production. 
 
To close the section, we comment on the on-shell 
cross-section~(\ref{BBnunon}). 
With the aid of equation~(\ref{rholim}) it is obtained by the
following replacements: 
\vspace{.5cm}
\ba
  \lefteqn{
  {\bar \sigma}_{J,0}(s,M_V^2,M_V^2) \; = \;
    \lim_{\Gamma_V \to 0} \int \! \d s_1 \int \! \d s_2 \;
    \sigma_{J,0}(s,s_1,s_2)
} 
\nl
  & \hspace*{.5cm} = \, & \int \! \d s_1 \int \!\d s_2 \;
    \lim_{\Gamma_V \to 0} \left[ 
    \frac{\sqrt{\lambda}}{\pi s^2}
    \sum_k \cc_J^{(k)}(s,s_1,s_2) \,\cg_J^{(k)}(s,s_1,s_2) \right]
  \label{xsosborn}  
\ea
\\
and
\vspace{.5cm}
\ba
  \lefteqn{
{\bar \sigma}_{J,\mathrm{QED}}^{\mathrm{non-univ}}(s,s',M_V^2,M_V^2)
 \; = \;
 \lim_{\Gamma_V \to 0} \int \! \d s_1 \int \! \d s_2 \;
  \sigma_{J,\mathrm{QED}}^{\mathrm{non-univ}}(s,s',s_1,s_2)  
} 
\nl
  & \hspace*{.1cm} \equiv\, & 
\int \! \d s_1 \int \!\d s_2 \;
   \sum_k \lim_{\Gamma_V \to 0}
 {\mathcal C}_J^{(k)}(s',\SONE,\STWO)
  \Biggl[ \beta_e v^{\beta_e - 1} 
\sigma_{{\hat S},J}^{(k)}(s';s_1,s_2)
\nl
  & & \hspace{6.5cm} + \;
\sigma_{{\hat H},J}^{(k)}(s,s';s_1,s_2)
\biggr].
\label{offon}
\ea
%
\section{Non-universal Initial State Corrections
\label{xsISR2}
}
\ezero
%
In this section, we present the final analytical results for the
ISR corrected threefold differential \xsecs for the
processes~(\ref{4fproc}).
The starting point of our calculations are matrix elements
of \ccthree\, \nctwo\, and \nceight\ processes, which are given in
Appendix~\ref{appmatel}. 
With the help of the computer algebra packages {\tt SCHOONSCHIP},
{\tt FORM}, and {\tt Mathematica}~\cite{mathematica}, matrix elements
were squared, spin summation was performed, the scalar products were
expressed in terms of the phase space variables, and algebraic
manipulations of the resulting expressions were carried out.
All analytical integrations were obtained from hand-made tables of
canonical integrals.
The kinematical relations needed for the treatment of real
bremsstrahlung may be found in appendix~\ref{ps2to5}.
For the virtual corrections, tree level kinematics may be used as it
was explained in appendix~B of~\cite{cc11}. 
In the course of performing the various steps of analytical
integrations we proceeded as follows. 

The {\em virtual photonic corrections} have been treated as a net sum
of all contributing diagrams. 
The infrared singularity was isolated and, as a part of the
universal corrections, subtracted from the net correction.
Thus, the remaining, non-universal virtual corrections are by
construction free of infrared problems\footnote[5]
{The same was done with the
real photonic corrections. Thus, non-universal virtual and
real corrections may be treated completely separately.
The interested reader may find details in~\cite{dldiss}.
}.
After tensor integration over the final state angular
variables (see appendix~\ref{tenint}), the loop momentum integrations
were performed with the aid of the integrals of appendix~\ref{loopint1}
and the final integration over the vector boson production angle
$\vartheta$ in the center of mass system with the aid of
appendix~\ref{loopint2}. 

Also the {\em real photonic corrections} were first integrated over the
final state angular variables (see again appendix~\ref{tenint}).
Then we integrated over the production angles $\phi_R$ and $\theta_R$
of the vector boson in the two-boson rest frame (for the canonical
integrals see appendix~\ref{bremint1}) and finally over the photon
production angle $\theta$ in the center of mass system (see the table
of canonical integrals in appendix~\ref{bremint2}).   

We have used the ultrarelativistic approximation for final state
fermions and initial state electrons, i.e. the masses of these
particles are neglected wherever possible. 
The various tables of canonical integrals were checked by
{\tt Fortran} programs with a precision of typically $10^{-10}$.
At the end of our calculations, algebraic manipulations were carried out
by hand to yield compactification of our final results\footnote[6]{The latter
  was cross-checked against {\tt FORM} outputs with the aid 
of auxiliary {\tt FORM} codes.} given in
equations~(\ref{Vnunit}),~(\ref{Vnunitu}),~(\ref{brnunit}),
(\ref{brnunitu}),~(\ref{Vnunist}),~(\ref{brnunist}).

In the following two subsections, the non-universal corrections
introduced in equation~(\ref{defnonu}) will be explicitly given and
commented.
%
\subsection{The Neutral-Current Case
\label{nccase}
}
\begin{figure}[tb]
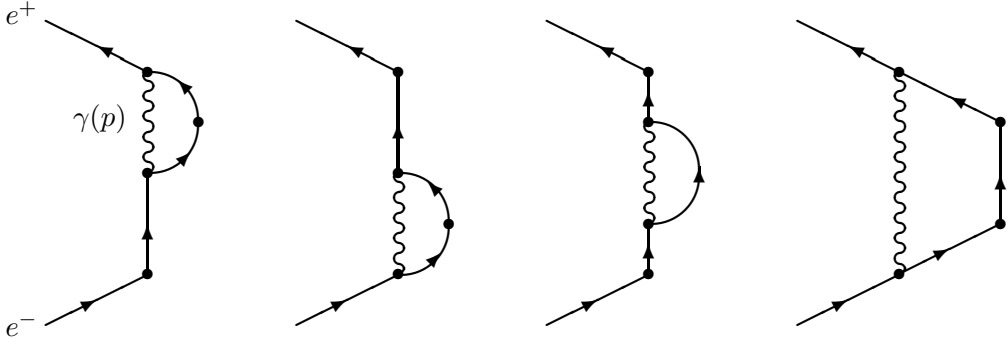

\begin{center}
\begin{Feynman}{150,55}{-43,8}{0.9}
%
\small
\put(-42,52){$e^+$}  
\put(-42,6){$e^-$}
\put(-21,15){\fermionurrhalf}
\put(-21,45){\fermionullhalf}
\put(-21,15){\fermionuphalf}
\put(-21,30){\photonuphalf}
\put(-32,37){$\gamma(p)$}
\put(-21,37.5){\oval(15,15)[r]}
\put(-15.6,42.8){\vector(-1,1){1}}  
\put(-15.55,32.2){\vector(1,1){1}}  
\put(-21,45){\circle*{1.5}}
\put(-21,15){\circle*{1.5}}
\put(-21,30){\circle*{1.5}}
\put(-13.5,37.5){\circle*{1.5}}
%
\put(16,15){\fermionurrhalf}
\put(16,45){\fermionullhalf}
\put(16,30){\fermionuphalf}
\put(16,15){\photonuphalf}
\put(16,22.5){\oval(15,15)[r]}
\put(21.35,27.8){\vector(-1,1){1}}  
\put(21.35,17.2){\vector(1,1){1}}  
\put(16,45){\circle*{1.5}}
\put(16,15){\circle*{1.5}}
\put(16,30){\circle*{1.5}}
\put(23.50,22.5){\circle*{1.5}}
%
\put(53,15){\fermionurrhalf}
\put(53,45){\fermionullhalf}
\put(53,15){\line(0,1){7.5}}
\put(53,18.75){\vector(0,1){1}}  
\put(53,41.25){\vector(0,1){1}}  
\put(53,37.5){\line(0,1){7.5}}
\put(53,22.5){\photonuphalf}
\put(53,30){\oval(15,15)[r]}
\put(60.48,30){\vector(0,1){1}}  
\put(53,15){\circle*{1.5}}
\put(53,22.5){\circle*{1.5}}
\put(53,37.5){\circle*{1.5}}
\put(53,45){\circle*{1.5}}
%
\put(90,15){\fermionurrhalf}
\put(105,22.5){\fermionurrhalf}
\put(90,45){\fermionullhalf}
\put(105,37.5){\fermionullhalf}
\put(105,22.5){\fermionuphalf}
\put(90,15){\photonup}
\put(90,45){\circle*{1.5}}
\put(90,15){\circle*{1.5}}
\put(105,22.5){\circle*{1.5}}
\put(105,37.5){\circle*{1.5}}
\end{Feynman}
\end{center}
\caption{\em The amputated t- and u-channel virtual initial state QED
  Feynman diagrams.}
\label{tvirt}
\vspace{.8cm}
\end{figure}

\begin{figure}[t]
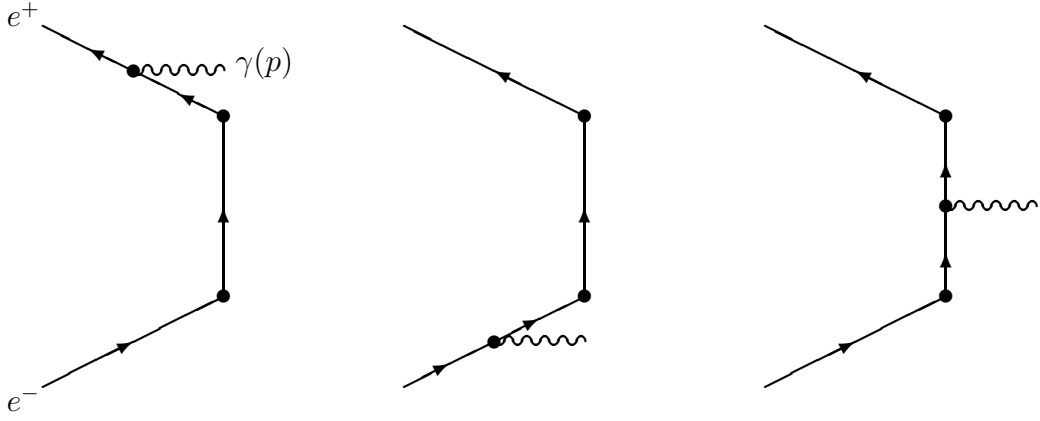

\begin{Feynman}{150,65}{-70,-3}{0.8}
%
\put(-81,60){$e^+$}
\put(-81,-4){$e^-$}
\put(-43,52.5){$\gamma(p)$}
\put(-45,15){\fermionup}
\put(-45,15){\fermionurr}
\put(-45,45){\fermionullhalf}
\put(-60,52.5){\photonrighthalf}
\put(-60,52.5){\fermionullhalf}
\put(-45,15){\circle*{2}}
\put(-45,45){\circle*{2}}
\put(-60,52.5){\circle*{2}}
%
%
\put(15,15){\fermionup}
\put(15,15){\fermionurrhalf}
\put(15,45){\fermionull}
\put(0,7.5){\photonrighthalf}
\put(0,7.5){\fermionurrhalf}
\put(15,15){\circle*{2}}
\put(15,45){\circle*{2}}
\put(0,7.5){\circle*{2}}
%
%
\put(75,15){\fermionuphalf}
\put(75,30){\fermionuphalf}
\put(75,30){\photonrighthalf}
\put(75,15){\fermionurr}
\put(75,45){\fermionull}
\put(75,15){\circle*{2}}
\put(75,30){\circle*{2}}
\put(75,45){\circle*{2}}
%
\end{Feynman}
\caption{\it The amputated t- and u-channel initial state bremsstrahlung
  Feynman diagrams.}
\label{convISR}
\vspace{.8cm}
\end{figure}

%
In the neutral current case, initial state QED corrections are
represented by the Feynman diagrams of figures~\ref{tvirt}
and~\ref{convISR}.
 
The non-universal \nceight\ corrections are
\\
\ba
  \sigma_{{\hat S},\nceight}(s;\SONE,\STWO) & = & \frac{\alpha}{\pi}
    \;\frac{\SONE\STWO}{8 \pi {s}^2} \;
    \sigma_{V,\nceight}^{\mathrm{non-univ}}(s;\SONE,\STWO),
  \nl \nl
  \sigma_{{\hat H},\nceight}(s,s';\SONE,\STWO) & = & \frac{\alpha}{\pi}
    \;\frac{\SONE\STWO}{\pi s} \;
    \sigma_{R,\nceight}^{\mathrm{non-univ}}(s,s';\SONE,\STWO),
  \label{nunixsNC}
\ea
\\
where the subindex $V$ stands for virtual and the subindex $R$ for
real photonic corrections. 

The virtual contribution is a sum of contributions due to the different
interferences of loop diagrams with tree level graphs:
\beq
  \sigma_{V,\nceight}^{\mathrm{non-univ}}(s;\SONE,\STWO) \; = \;
    \sigma_{V,t}^{\mathrm{non-univ}} 
\; + \;
    2 \sigma_{V,tu}^{\mathrm{non-univ}} 
\; + \;
    \sigma_{V,u}^{\mathrm{non-univ}}. \nl
\eeq
The squared $t$- and $u$-channel contributions are equal:
\\
\ba
  \lefteqn{ \sigma_{V,t}^{\mathrm{non-univ}}(s;\SONE,\STWO) \;\; = \;\;
    \sigma_{V,u}^{\mathrm{non-univ}}(s;\SONE,\STWO)} 
\nl\nl
    & = & \; \frac{3\,\cl_2}{2}
          \left[ \rule[0cm]{0cm}{.35cm}
          s\,l_+ + \delta\,l_- - s \left( s-\sigma \right)\,{\rm I_{12q}}
          \right] 
        \nl
    & & \;+\; \cl_1  \left[ \rule[0cm]{0cm}{.35cm}
          (s-2\sigma)(s-\sigma)\,{\rm I_{12q}} +
          \frac{3\,\sigma\delta}{2\,s}\,l_- -
          \frac{s-4\sigma}{2}\;l_+ + s - \sigma \right]
        \nl
    & & \;+\; \cl_0  \left[ \rule[0cm]{0cm}{.35cm}
          - \left( \lambda + \sigma^2 + \frac{s^2}{2} \right)\,{\rm I_{12q}}
          - \frac{5\,\delta}{2}\,l_- - 4\sigma\,l_+ + 8\sigma - 9s
          \right] 
        \nl \nl
    & & \;+\;
          2\,\SLAM \left[4\left( \rule[0cm]{0cm}{.35cm} 2-l_+\right) \; + \;
          \left( \rule[0cm]{0cm}{.35cm} 3s-\sigma \right)\,
          {\rm I_{12q}}\right].
  \label{Vnunit} 
\ea
\\ 
Here, the following notations have been used:
\\
\ba
  \sigma \; & = & \SONE + \STWO, \nl
  \delta \; & = & \SONE - \STWO, \nl \nl
  l_+ \; & = & \myln \frac{\SONE}{s} \; + \; \myln \frac{\STWO}{s}\;,
  \nl \nl
  l_- \; & = & \myln \frac{\SONE}{s} \; - \; \myln \frac{\STWO}{s}
    \;\; = \;\; \myln \frac{\SONE}{\STWO}\;, \nl \nl
  \cl_0 & = & \ln \frac{s-\sigma+\sqrt{\lambda}}
                       {s-\sigma-\sqrt{\lambda}}\;,  \nl \nl
  \cl_1 & = & \frac{\,s\,(s-\sigma)\,}{\lambda}
    \left[ \rule[0cm]{0cm}{.6cm}
           \cl_0 \; - \; \frac{2\SLAM}{s-\sigma} \right], \nl \nl
  \cl_2 & = & \frac{\,s\,(s-\sigma)^3}{\lambda^2} \left[
    \rule[0cm]{0cm}{.6cm} \cl_0 \; - \; \frac{2\SLAM}{s-\sigma} \;-\;
    \frac{2}{3} \left( \frac{\SLAM}{s-\sigma} \right)^{\!3}\, \right],
    \nl\nl
  {\rm I_{12q}} & = & \frac{1}{\SLAM} \left(\cd_+ \; - \; \frac{1}{2}
                      l_{-} \cl_{-} \right),
               \nl\nl
  \cl_- & = & \cl_{12} - \cl_{34}\;, \nl \nl
  \cl_{12} & = & \ln \frac{s+\delta+\sqrt{\lambda}}
                          {s+\delta-\sqrt{\lambda}}\;,  \nl \nl
  \cl_{34} & = & \ln \frac{s-\delta+\sqrt{\lambda}}
                          {s-\delta-\sqrt{\lambda}}\;,  \nl \nl
  \cd_+ & = & \mysp \left( -\,\frac{t_{max}}{\SONE} \right) \:-\:
              \mysp \left( -\,\frac{t_{min}}{\SONE} \right) \:+\:
              \mysp \left( -\,\frac{t_{max}}{\STWO} \right) \:-\:
              \mysp \left( -\,\frac{t_{min}}{\STWO} \right),
              \hspace{-.2cm} \nl\nl
  t_{min}  & = & \frac{1}{2} \left(s-\sigma-\SLAM\right), \nl \nl
  t_{max}  & = & \frac{1}{2} \left(s-\sigma+\SLAM\right).
  \label{nunivnota1}
\ea
\vspace{.2cm} \\

The virtual corrections due to the $tu$-interferences are: 
\\
\ba
  \lefteqn{ \sigma_{V,tu}^{\mathrm{non-univ}}(s;\SONE,\STWO) 
\; = \;   \frac{3\,\cl_2}{2}
          \left[ \rule[0cm]{0cm}{.35cm}
          s\,l_+ + \delta\,l_- - s (s-\sigma)\,{\rm I_{12q}} \right]}
          \nl
    & & \hspace{1cm}+\; \cl_1 \left[ \rule[0cm]{0cm}{.6cm}
          (s-\sigma)\,\left( \rule[0cm]{0cm}{.35cm}
            1 + 5\,s\,{\rm I_{12q}} + \frac{\delta}{2s}\,l_- \right) -
          \frac{9}{2} \left( \rule[0cm]{0cm}{.35cm} s\,l_+ +
            \delta\,l_- \right) \right] 
          \nl
    & & \hspace{1cm}+\; \cl_0 \left\{ \rule[0cm]{0cm}{.7cm}
            \frac{s^2}{s-\sigma} \left[ \rule[0cm]{0cm}{.6cm} \!
            -\frac{7}{3}\,\cl_0^2 - 16\,\cd_d + 4\,\cd_{d+} -
            4 \left( \rule[0cm]{0cm}{.35cm}
              l_\sigma - 2\,l_+ +\frac{3}{2} \right) l_\sigma
            \right. \right.
            \nl & & \hspace{4.5cm}+\;
            \left. \rule[0cm]{0cm}{.6cm} 4\,l_-\,l_{d-} -
            4 \left( \rule[0cm]{0cm}{.35cm} 1 - \frac{3\pi^2}{2} \right)
            - 3 \left( \rule[0cm]{0cm}{.35cm} l_+ - 3 \right)l_+
          \right] \nl & & \hspace{2.19cm} + \;
            2\,s^2 \left[ \rule[0cm]{0cm}{.6cm} \!
            \left( \rule[0cm]{0cm}{.35cm} 3\,h_{1+} + h_{2+} \right)
            \!
            \left( \rule[0cm]{0cm}{.35cm} l_\sigma - \frac{3}{4}l_+\right)
            + \frac{3}{4}
             \left( \rule[0cm]{0cm}{.35cm}\,3\,h_{1-} + h_{2-} \right)
             \right. 
             l_-\,
             \nl & &
            \hspace{10.3cm} \left. \rule[0cm]{0cm}{.6cm}
            + 2\,h_{1+} \right] \nl
          & & \hspace{2.63cm} + \;
            s \left[ \rule[0cm]{0cm}{.6cm} 2\,l_+ - 6\,l_\sigma -
            \frac{15}{2}\,s\,{\rm I_{12q}} - \frac{1}{d_1} - \frac{1}{d_2} 
            - 3\; \right] \nl
          & & \left. \rule[0cm]{0cm}{.7cm} \hspace{2.5cm} + \;
            \sigma \left( \rule[0cm]{0cm}{.35cm} l_+ - 2\,l_\sigma
          \right)
            + \frac{\delta}{2}\,l_-  \right\}
          \nl
    & & \hspace{1cm}+\; \frac{8\,s^2}{s-\sigma}
           \left[ \rule[0cm]{0cm}{.7cm} \cf_{d-} - \cf_{t-} -
            \cf_\sigma + \left( \rule[0cm]{0cm}{.35cm} \frac{3}{4} 
                       - \frac{1}{2}l_+
          \right)
            \cd_\sigma + \frac{1}{2}l_+\cd_{d-} \right]
          \nl \nl
    & & \hspace{1cm}+\; s^2 \left( \rule[0cm]{0cm}{.35cm}
          3\,h_{1+} + h_{2+} \right) \; \left( \rule[0cm]{0cm}{.35cm}
            \SLAM\;{\rm I_{12q}} - 2\,\cd_\sigma \right)
          \nl \nl
    & & \hspace{1cm}+\; s \;
          \SLAM\; \left(\rule[0cm]{0cm}{.35cm}2\,{\rm I_{12q}}
          + h_{1-} l_-\right) + 2\,\left(3s + \sigma  \right)\;
          \cd_\sigma~. 
  \label{Vnunitu}
\ea
\vspace{.1cm} \\
In equation~(\ref{Vnunitu}), the following additional 
notations have been used:
\\
\ba
  h_{1\pm} & = & \frac{1}{s-\SONE} \; \pm \; \frac{1}{s-\STWO}\;, \nl \nl
  h_{2\pm} & = & \frac{\STWO}{\,\left(s-\SONE\right)^{2}\,} \; \pm \;
                 \frac{\SONE}{\,\left(s-\STWO\right)^{2}\,}\;, \nl \nl
  \DONE & = & \frac{s-\STWO}{\SONE}\;, \nl \nl
  \DTWO & = & \frac{s-\SONE}{\STWO}\;, \nl \nl
  l_\sigma \; & = & \myln \frac{s-\sigma}{s}\;, \nl \nl
  l_{d-} & = & \ln \DONE \; - \; \ln \DTWO \nonumber\;,
\nl\nl
  \cd_\sigma & = &  \mysp \left( \frac{t_{max}}{\,s-\sigma\,} \right)
            \; - \; \mysp \left( \frac{t_{min}}{\,s-\sigma\,} \right),
            \nl\nl
  \cd_d & = & {\mathrm Re} \left[ \rule[0cm]{0cm}{.35cm}
                \mysp \left(\DONE\right) \; + \;
                \mysp \left(\DTWO\right) \right], \nl\nl
  \cd_{d+}\!\! & = & \mysp \left( \frac{t_{max}}{\,s-\STWO} \right)
               \:+\: \mysp \left( \frac{t_{min}}{\,s-\STWO} \right)
               \:+\: \mysp \left( \frac{t_{max}}{\,s-\SONE} \right)
               \:+\: \mysp \left( \frac{t_{min}}{\,s-\SONE} \right),
               \nl\nl
  \cd_{d-}\!\! & = & \mysp \left( \frac{t_{max}}{\,s-\STWO} \right)
               \:-\: \mysp \left( \frac{t_{min}}{\,s-\STWO} \right)
               \:-\: \mysp \left( \frac{t_{max}}{\,s-\SONE} \right)
               \:+\: \mysp \left( \frac{t_{min}}{\,s-\SONE} \right),
               \nl\nl
  \cf_\sigma & = & \mytri \left( \frac{t_{max}}{\,s-\sigma\,} \right)
           \; - \; \mytri \left( \frac{t_{min}}{\,s-\sigma\,} \right),
                   \nl\nl
  \cf_{d-} & = & \mytri \left( \frac{t_{max}}{\,s-\STWO} \right) \:-\:
                 \mytri \left( \frac{t_{min}}{\,s-\STWO} \right) \:+\:
                 \mytri \left( \frac{t_{max}}{\,s-\SONE} \right) \:-\:
                 \mytri \left( \frac{t_{min}}{\,s-\SONE} \right),
                 \nl\nl
  \cf_{t-} & = & \mytri \left( -\,\frac{t_{max}}{\DONE\,t_{min}} \right)
           \:-\: \mytri \left( -\,\frac{t_{min}}{\DONE\,t_{max}} \right)
                 \nl \nl & & \hspace{1cm}
           \:+\: \mytri \left( -\,\frac{t_{max}}{\DTWO\,t_{min}} \right)
           \:-\: \mytri \left( -\,\frac{t_{min}}{\DTWO\,t_{max}} \right).
  \label{nunivnota2}
\ea
\vspace{.2cm} \\

The non-universal real bremsstrahlung contribution to the \nceight\
process is given by:
\\
\ba
  \sigma_{R,\nceight}^{\mathrm{non-univ}}(s,s';\SONE,\STWO) \; = \;
    \sigma^{\mathrm{non-univ}}_{R,t} \; + \;
    \sigma^{\mathrm{non-univ}}_{R,tu} \; + \;
    \sigma^{\mathrm{non-univ}}_{R,u}.
\ea
\\
Again, the squared $t$- and $u$-channel contributions are equal:
\\
\ba
  \lefteqn{\hspace{-.5cm}
    \sigma^{\mathrm{non-univ}}_{R,t}(s,s';\SONE,\STWO) \; = \;
    \sigma^{\mathrm{non-univ}}_{R,u}(s,s';\SONE,\STWO) \; = \;
    \SLAMP \left( \frac{1}{\sprm} - \frac{\sigma}{\,\LAMB\,} \right)}
    \nl \nl & &+\;
    \frac{\delta}{4\sprm}\;
    \left[ \rule[0cm]{0cm}{.85cm} 
          \; \frac{s'_-}{s}\left(1-\frac{ss'_+}{s'^2}\right)
          + \frac{\,\sigma\left(3s-\sprm-\sigma\right)\,}{\LAMB}
    \right. \nl & &
    \hspace{4.8cm}+ \left. \rule[0cm]{0cm}{.85cm}
      \frac{s}{\,\sprm\,} \left( \rule[0cm]{0cm}{.65cm}
      1+\frac{\sigma^2}{\,\LAMB\,} \right) +
      12 \, \frac{\,s\,\SONE\,\STWO\,\sigma\,}{\LAMB^2}
      \right] \left( L_{c1} -  L_{c2} \right)
    \nl & &+\;
      \frac{1}{\,4\,}
        \left[ \rule[0cm]{0cm}{.75cm} 1 +
               \frac{\sigma s'_+}{ss'} - \frac{\,s-\sigma\,}{\sprm} -
               \frac{\,2s\sigma\,}{{\sprm}^2} \, + \,
               \frac{\,2\sigma\,}{\LAMB}\left(\sprp-\sigma\right)
               \right.
    \nl & &
    \hspace{7.65cm}+\; \left. \rule[0cm]{0cm}{.75cm}
    24\,\frac{\,s\,\SONE\,\STWO\,\sigma\,}{\LAMB^2} \; \right]
     \left( L_{c1} +  L_{c2} \right)
    \nl & &
    + \; \SLAMP \left( \rule[0cm]{0cm}{.65cm}
         \frac{s'_+}{\,4 {s'}^2} +
         \frac{\sigma}{\,2 \LAMB\,} +
         \frac{3\,\sigma\SONE\STWO}{\LAMB^2} \right)
         \left( L_{c3} + L_{c4} \right)
    \;+\;
    \frac{\sigma s'_+s'_-}{2ss'^2} L_{c5}
    \nl \nl & &-\;
    \frac{1}{4\SLAMB} \left[ \rule[0cm]{0cm}{.85cm} \;
      2\sigma + s + \sprm + \frac{\sigma}{\LAMB}
      \left( \rule[0cm]{0cm}{.65cm} s\,\left(\sprm + \sigma\right)
             - 4s_1s_2 \right) \right.
    \nl & &\hspace{6.0cm}+\;
      \left. \rule[0cm]{0cm}{.85cm}
      12 \, \frac{\,s\,\SONE\,\STWO\,\sigma
      \left( \sprm+\sigma\right)\,}{\LAMB^2}
        \; \right]D^t_{12}
    \nl & &+\;
    \frac{\left(\sigma - s'\right)s'_+}{\,4s'^2}
    \left( \rule[0cm]{0cm}{.65cm} D^t_1 \!+\! D^t_2 \!\right)
    \;+\;
    \frac{\sigma s'_+ - s's'_-}{\,4 s'^2}
    \left( \rule[0cm]{0cm}{.65cm} D_{z1t2} \!+\! D_{z2t1} \right).
    \label{brnunit} 
\ea
\\
In the above equation we have introduced the symbols:
\\
\ba
  s'_{\pm} & \equiv &  s \pm s', \qquad
  \lambda'  \equiv  \; \lambda(s',\SONE,\STWO),
\nl  \nl
  {\bar \lambda}  & \equiv & \lambda(\sprm,-\SONE,-\STWO) \; =
  \; {\sprm}^2 + 2\sprm\sigma + \delta^2.
\ea
\\
The additional notations 
are introduced in appendix~\ref{bremint2}.
For easy reference, the number of each integral
corresponding to the enumeration in appendix~\ref{bremint2} is listed
in table~\ref{notafind1}.

The most cumbersome contribution is due to the $tu$-interference: 
\\
\ba
  \lefteqn{ \hspace{-.5cm}
    \sigma^{\mathrm{non-univ}}_{R,tu}(s,s';\SONE,\STWO) \; = \;
    \frac{2\,\sprm \sigma}{s' \left(s'-\sigma\right)} \; L_{c5}
      \;\; + \;\; \frac{\sprp}{s'}
      \left( \rule[0cm]{0cm}{.35cm} D^t_1 + D^t_2 \right)}
  \nl \nl & & - \;
      \frac{s^2+{s'}^2}{\,2 \, \sprm\left(s'\!-\!\sigma\right)\,}
        \left( \rule[0cm]{0cm}{.35cm} D_{z1t1} \!+\! D_{z2t2} \right)
      \; - \;
      \left( \rule[0cm]{0cm}{.7cm} \frac{s}{\,s'} \, + \,
             \frac{s^2+{s'}^2}{\,2\,\sprm\left(s'\!-\!\sigma\right)\,}
           \right)
      \left( \rule[0cm]{0cm}{.35cm} D_{z2t1} \!+\! D_{z1t2} \right) \!
  \nl \nl & & - \;
      \frac{\,s^2+{s'}^2-4\sigma\left(s'-\sigma\right)\,}
           {2\,\left(s'-\sigma\right) \left(\sprp-2\sigma\right)}
        \; \left[ \rule[0cm]{0cm}{.45cm}
                     \left( \rule[0cm]{0cm}{.35cm} L_{c6} + L_{c7}
                     \right)L_{c8} \, + \, D^{tu}_{a1} + D^{tu}_{a2}
              \right]
  \nl \nl & & - \;
      \frac{\,s-2\SONE\,}{\sqrt{s\SONE}} \, D_{t1u2} \;\; - \;\;
      \frac{\,s-2\STWO\,}{\sqrt{s\STWO}} \, D_{t2u1} \;\; + \;\;
      \frac{s^2+{s'}^2}{\,4\,\sprm \left(s'-\sigma \right)}
        \left( \rule[0cm]{0cm}{.35cm} D^z_{t1u2} + D^z_{t2u1} \right)
  \nl \nl & & + \;
      \frac{\,\SLAMP \left[ \rule[0cm]{0cm}{.35cm}
                            s^2 + {s'}^2 - 4\sigma (s'-\sigma) \right]
            \left[ \rule[0cm]{0cm}{.35cm}
                   s (s'-\sigma) - \SONE (\sprp - 2\sigma) \right]}
           {2\,(\sprp - 2\sigma) (s'-\sigma)} \; D^a_{t1u2}
 \nl \nl & & + \;
      \frac{\,\SLAMP \left[ \rule[0cm]{0cm}{.35cm}
                            s^2 + {s'}^2 - 4\sigma (s'-\sigma) \right]
            \left[ \rule[0cm]{0cm}{.35cm}
                   s (s'-\sigma) - \STWO (\sprp - 2\sigma) \right]}
           {2\,(\sprp - 2\sigma) (s'-\sigma)} \; D^a_{t2u1}.
  \label{brnunitu}
\ea

Again, additional notations are introduced and the number of each
integral corresponding to the enumeration in appendix~\ref{bremint2}
is listed in table~\ref{notafind2}.

\begin{table}[tb]
  \vspace{.3cm}
  \caption{\it 
Notations for $\sigma^{\mathrm{non-univ}}_{R,t}(s,s';\SONE,\STWO)$.
The numbers refer to appendix~\ref{bremint2}
}
  \vspace{.3cm}
  \begin{center}
  \begin{tabular}{|c||c|c|c|c|c|c|c|c|c|c|} \hline
    Notation & $L_{c1}$ & $L_{c2}$ & $L_{c3}$ & $L_{c4}$ & $L_{c5}$
    & $D^t_1$ &$D^t_2$  & $D^t_{12}$ & $D_{z1t2}$ &$D_{z2t1}$
\\ \hline
    Integral No  
& 7) & 8) & 7) & 8) & 7) & 14)&15) & 31)& 11) & 12)
    \\ \hline 
  \end{tabular}
  \end{center}
  \label{notafind1}
\vspace{.8cm}
\end{table}
%
\begin{table}[tb]
  \vspace{.3cm}
  \caption{\it 
Notations for $\sigma^{\mathrm{non-univ}}_{R,tu}(s,s';\SONE,\STWO)$.
The numbers refer to appendix~\ref{bremint2}
}
  \vspace{.3cm}
  \begin{center}
  \begin{tabular}{|c||c|c|c|c|c|c|c|} \hline
Notation  &$L_{c6}$ &$L_{c7}$ &$L_{c8}$ &$D_{z1t1}$ &$D_{z2t2}$ &
    $D^{tu}_{a1}$ & $D^{tu}_{a2}$ 
\\ \hline 
Integral No & 18) & 19) & 18) & 10) & 13) & 18) & 19)
\\ \hline \hline
Notation &$D_{t1u2}$
& $D_{t2u1}$ &$D^z_{t1u2}$  &$D^z_{t2u1}$ &$D^a_{t1u2}$
    &$D^a_{t2u1}$  &  
\\ \hline
Integral No
& 36) & 37) & 38) & 39) &  40) &  41) &
\\ \hline
  \end{tabular}
  \end{center}
  \label{notafind2}
  \vspace{1.2cm}
\end{table}

%
\subsection{The Charged-Current Case
\label{cccase}
}
%
It is in order to mention that, for the \ccthree\ process, the
separation of initial and final state radiation is not unique.
Since, in the {\em t}-channel contribution, there is electric charge
flow from the initial state to the final state, electric current
conservation is violated and one faces the problem to find a
gauge-invariant definition of initial state radiation. 
As a solution we proposed in~\cite{WWnuni} what we call the
{\em current splitting technique} ({\tt CST}).
In brief, the
{\tt CST} splits the electrically neutral $t$-channel neutrino flow
into two oppositely flowing electric charges +1 and --1. Charge --1
is assigned to the initial state, charge +1 to the final state.
This enables a gauge invariant definition of ISR with photon emission
and absorption from the $t$-channel exchange particle.
An auxiliary current is added to the naive charged current 
for this purpose in appendix~\ref{appmatel}. 
Thus, the \ccthree\ $t$-channel receives the same ISR corrections as
the neutral current $t$-channel.
Of course, when performing a complete calculation, the final state
corrections have to take into account opposite auxiliary terms so that
the net auxiliary effect will vanish. 

The non-universal corrections are due to the interferences of the
$s$- and $t$-channel diagrams of figures~\ref{svirt} and~\ref{tvirt}
with the corresponding Born diagrams  and of the diagrams of
figures~\ref{annISR} and~\ref{convISR} among themselves.
These interferences show less symmetry than the interferences of the
$t$- and $u$-channel diagrams of the neutral current case.    
However, as mentioned earlier, the pure $s$-channel corrections have
no non-universal parts:
\beq
  \sigma^s_{{\hat S},\ccthree} \; = \; \sigma^s_{{\hat H},\ccthree}
    \; = \; 0.
\eeq
Further,
for the pure $t$-channel non-universal contributions to the \ccthree\
process one finds, of course, expressions equal to those for the
pure $t$-channel non-universal contributions to the \nceight\ process,
namely
\\
\ba
  \sigma^t_{{\hat S},\ccthree}(s;\SONE,\STWO) & = &
    \frac{\alpha}{\pi} \; \frac{\SONE\STWO}{\,8\,\pi \,{s}^2\,}
    \; \sigma_{V,t}^{\mathrm{non-univ}}(s;\SONE,\STWO),
  \nl \nl
  \sigma^t_{{\hat H},\ccthree}(s,s';\SONE,\STWO) & = &
    \frac{\alpha}{\pi} \; \frac{\SONE\STWO}{\pi s} \;
    \sigma^{\mathrm{non-univ}}_{R,t}(s,s';\SONE,\STWO)
  \label{nunixst}
\ea
\\
with $\sigma_{V,t}^{\mathrm{non-univ}}$\ from equation~(\ref{Vnunit})
and $\sigma^{\mathrm{non-univ}}_{R,t}$\ from equation~(\ref{brnunit}).

What remains to be calculated in addition are the virtual and real
{\em st}-inter\-fe\-rence contributions.

The virtual correction is given by
\beq
  \sigma^{st}_{{\hat S},\ccthree}(s;\SONE,\STWO) \; = \;
    \frac{\alpha}{\pi} \; \frac{\SONE\STWO}{\,8\,\pi \,{s}^2\,}
    \; \sigma_{V,st}^{\mathrm{non-univ}}(s;\SONE,\STWO)
  \label{nunixssts}
\eeq
with
\\
\ba
  \lefteqn{ \sigma_{V,st}^{\mathrm{non-univ}}(s;\SONE,\STWO) \; = \;
    -\;\SLAM \left[ 7 s + 3 \sigma
    +\, \delta\;l_- \; +  \;
    \left( 4 s \sigma + 2 \lambda - \delta^2 \right){\rm I_{12q}}\;\right]}
 \nl \nl & &+\;
    \cl_0 \left[ \rule[0cm]{0cm}{.7cm}
      5 s \sigma + 3 \lambda -
      \frac{3 \sigma^2}{2} - \frac{7 \delta^2}{2}
      + \left( 2 s \sigma + \lambda - \frac{\delta^2}{2} \right)l_+
      + \delta \left( s + \frac{\sigma}{2} \right) l_- \,
      \right]. \nl
  \label{Vnunist}
\ea
\\
The real bremsstrahlung contribution reads
\beq
  \sigma^{st}_{{\hat H},\ccthree}(s,s';\SONE,\STWO) \; = \;
    \frac{\alpha}{\pi} \; \frac{\,\SONE\STWO\,}{8\,\pi}
    \; \sigma_{R,st}^{\mathrm{non-univ}}(s,s';\SONE,\STWO),
  \label{nunixssth}
\eeq
with
\\
\ba
  \lefteqn{ \sigma_{R,st}^{\mathrm{non-univ}}(s,s';\SONE,\STWO)
    \; = \; - \;
    \frac{\delta \, \sprm \, \sprp}{s^2\,s'}
    \left( \rule[0cm]{0cm}{.7cm} \:
      \frac{\sigma\,\sprp}{2\,s\,s'} + 1
    \right) \; \left( L_{c1} - L_{c2} \right)}
  \nl & & + \;
    \frac{\sprp}{s} \,
    \left[ \rule[0cm]{0cm}{.8cm} \:
      \frac{1}{2} \left( \frac{\sigma}{s} - 1 \right)
                  \left( \frac{\sigma}{s} + 3 \right)
      + \frac{\SONE\,\STWO}{s}
        \left( \frac{2}{s'} - \frac{1}{s} \right)
    \right] \; \left( L_{c1} + L_{c2} \right)
  \nl & & + \;
    \frac{\SLAMP}{s'} \,
    \left[ \rule[0cm]{0cm}{.8cm} \:
      \frac{\sprm}{s} \left( \frac{\sigma\,\sprp}{s\,s'} - 8 \right)
      \; + \;
      \frac{\sprp}{2\,s} \left( \frac{\sigma}{s'} + 3 \right) \;
        \left( L_{c3} + L_{c4} \right) \:
    \right]
  \nl & & + \;
    \frac{\sprm}{s} \,
    \left[ \rule[0cm]{0cm}{.8cm} \:
      \frac{\sigma^2 \, {\sprp}^2}{s^2\,{s'}^2} +
      \frac{2\sprm}{s'}
        \left( \frac{\SONE\,\STWO\,\sprp}{s^2\,s'} - \frac{\sigma}{s}
        \right)
      + 4 \:
    \right] \; L_{c5}
  \nl & & - \;
    \frac{2\,\delta}{s}
    \left( \rule[0cm]{0cm}{.35cm} D^t_1 - D^t_2 \right) \; + \;
    2\:\left( \frac{\SONE\,\STWO\,\sprp}{s\,{s'}^2} +
              \frac{\sigma}{s'} - \frac{\sprp}{s}
        \right)
    \left( \rule[0cm]{0cm}{.35cm} D^t_1 + D^t_2 \right)
  \nl & & - \;
    2\:\left( \frac{\SONE\,\STWO\,\sprp}{s\,{s'}^2} +
              \frac{\sigma}{s'} - 1
        \right)
    \left( \rule[0cm]{0cm}{.35cm} D_{z1t2} + D_{z2t1} \right).
  \label{brnunist}
\ea
\\
%
\subsection{Discussion
\label{nonunivdiss}
}
%
To close the section we will now discuss some formal features
of our results.
Non-universal \xsec\ contributions are analytically rather
involved and contain many di- and trilogarithms.
Although expected, it is noteworthy that interferences
are more involved than matrix element squared contributions, which is
especially true for the $tu$-interference.
Looking at the integrals collected in Appendix D, one sees that
some of them are very complicated (in appendix D.5 e.g. the integral
31) for the real $t$-channel correction and integrals 36), 38), 40)
for the real $tu$-interference).
However, the final answers for $t$-, $u$-, and $st$-contributions, both
virtual and real, are remarkably compact. 
The virtual contributions contain only one
true dilogarithm, which arises from the three-point scalar integral
${\rm I_{12q}}$ (see equation~(\ref{nunivnota1}) and integral 15) in
appendix~\ref{loopint1}).   
Both these virtual and real contributions do not contain true
trilogarithms. 
In contrast, the $tu$-interferences are rather cumbersome.
This may be traced back to the angular integrations involving products
of different space-like fermionic propagators. 
Virtual $tu$-interferences contain several true
trilogarithms and real bremsstrahlung $tu$-interference contributions
exhibit complicated complex-valued dilogarithms. 
The latter, however, could indicate that some simplifications were
overlooked. 

We mention that the four non-universal cross-section contributions to
the \ccthree\ process became much more compact after inclusion of the
auxiliary terms than without them.
 
The attentive reader will have noticed that the virtual+soft
non-universal contributions are evaluated at $s'$ rather than at $s$.
This is to avoid an unphysical, $\delta$-distribution-like
concentration of non-universal virtual+soft ISR corrections at zero
radiative energy loss.

In equations~(\ref{nunixsNC}), (\ref{nunixst}), (\ref{nunixssts}),
and (\ref{nunixssth}), both reasonable threshold and high-energy
properties may be observed.
For $\sqrt{\lambda}$ or $\sqrt{\lambda'}$ approaching zero, non-universal
corrections vanish, which may be verified by inspection of the explicit
expressions.
Thus also the ISR corrected \xsec\ vanishes for the kinematical zero of
the tree level or the universally corrected \xsec.
At high energies, the {\em screening property} may be
verified to hold.
By screening we mean the assurance that cross-sections fall sufficiently
fast with rising $s$.
For the \ccthree\ case, this is ensured in Born approximation due to
an interplay between the $s$- and $t$-channel matrix elements and
relies a certain relation between the gauge couplings (see
e.g.~\cite{buras}).
In the \nctwo\ and \nceight\ cases, the same is achieved by an
interplay between the $t$- and $u$-channel amplitudes leading to a
\xsec\ proportional to the {\em screening factor} $s_1s_2/s^2$ (the
couplings of the two channels are equal)~\cite{teuw94}.
From the universal QED corrections, no additional problems arise,
because the interplay between the various contributing tree level
matrix elements is not disturbed. 
This is completely different for the non-universal corrections. 
Here, the various interferences get different corrections and the
interplay between them is destroyed.
Fortunately, one may see from the explicit expressions that the
screening factor observed in the net \nctwo\ and \nceight\ tree level
cross-sections is rediscovered not only for full \xsecs, but even for
individual non-universal contributions.
So, unitarity is not violated.
We were not able to find the screening property for
non-universal \ccthree\ corrections without including the auxiliary
corrections. 
Also in this respect, auxiliary corrections seem to be a quite natural,
if not necessary ingredient of the calculation.
%
%
\section{Numerical results}
\label{numres}
\begin{figure}[t]
\vspace*{-1.35cm}
  \begin{center}
    \mbox{
      \hspace*{-1.8cm}
      \epsfysize=14cm
      \epsffile{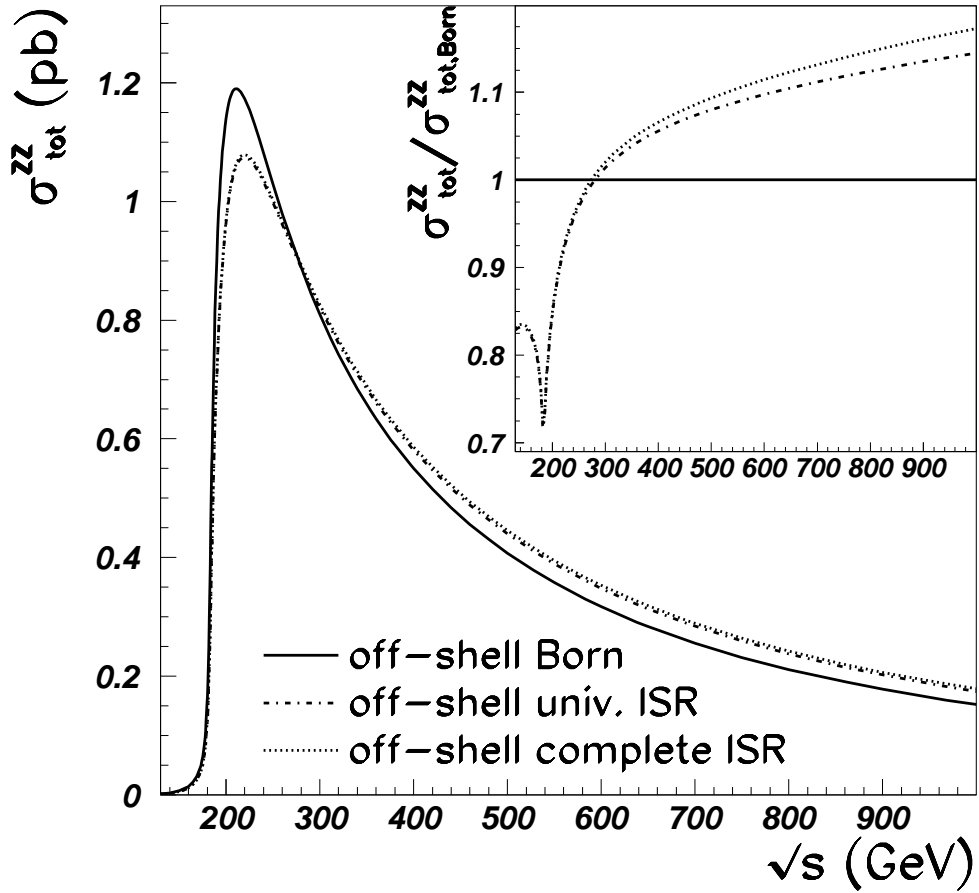} \hspace*{-1cm} }
  \end{center}
  \vspace{-.8cm}
  \caption[The \zz~pair \xsec\ with initial state radiative corrections]
    {\it The inclusive total off-shell \zz~pair production \xsec\
      $\sigma^{ZZ}_{tot}(s)$.
      In the inset, the relative deviations of the universally and
      completely ISR corrected \xsecs\ from the tree level \xsec\
      are given.}
  \label{xsZZ}
  \vspace{.6cm}
\end{figure}

\begin{figure}[t]
  \vspace*{2cm}
  \begin{center}
    \mbox{
      \hspace*{-2.5cm}
      \epsfysize=14cm
      \epsffile{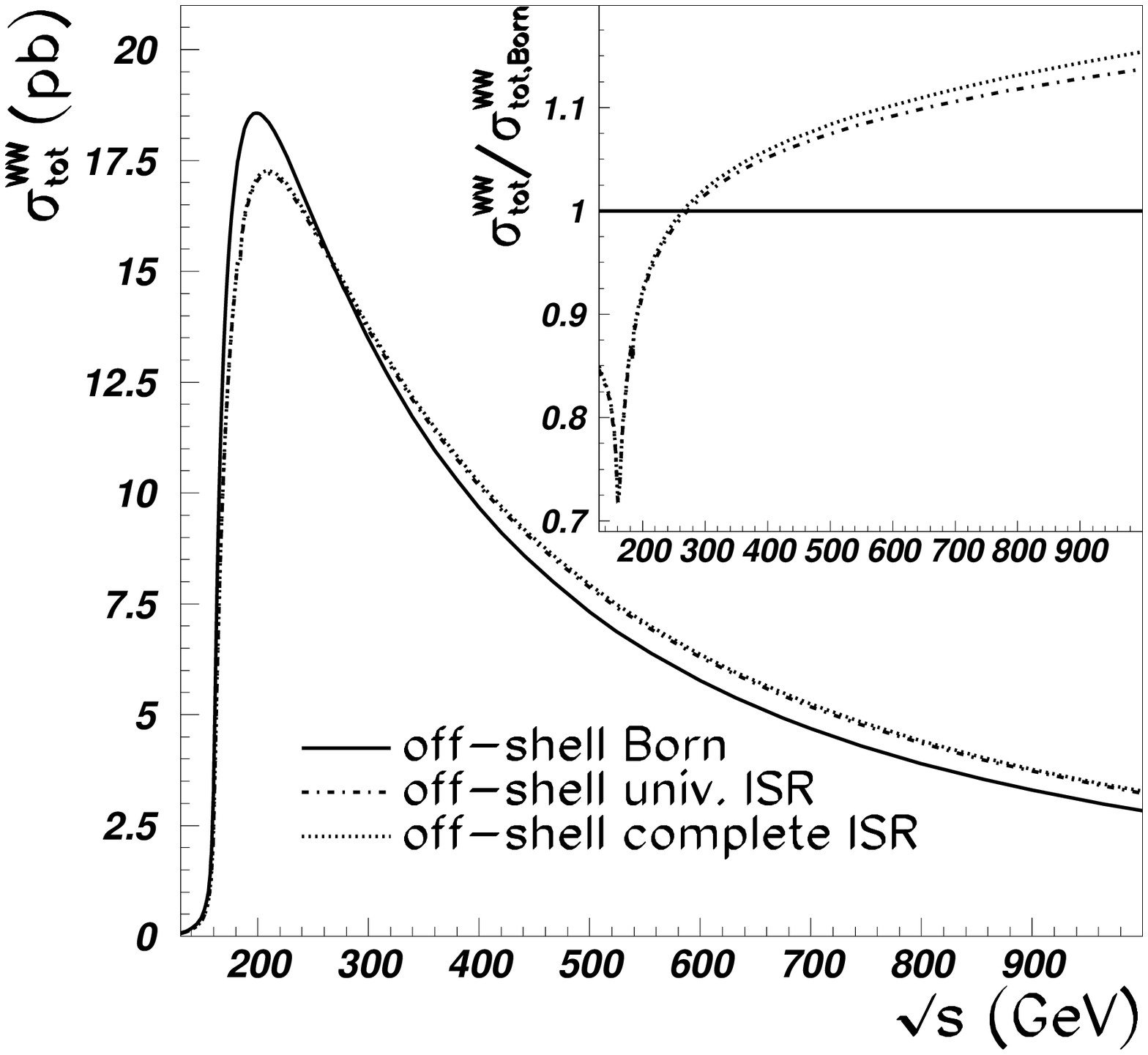} \hspace*{-1cm} }
  \end{center}
  \vspace{-4.4cm}
  \caption[The $W$~pair \xsec\ with initial state radiative corrections]
    {\it The inclusive total off-shell $W$\ pair production \xsec\
      $\sigma^{WW}_{tot}(s)$.
      In the inset, the relative deviations of the universally and
      completely ISR corrected \xsecs\ from the tree level \xsec\
      are given.}
  \label{xsWW}
  \vspace{.6cm}
\end{figure}

\ezero
%
To illustrate the analytical results of sections~\ref{xsISR1}
and~\ref{xsISR2}, we present numerical results~\cite{monster}.
The effect of ISR on the boson pair production processes \nctwo\ and
\ccthree\ is seen from figures~\ref{xsZZ} and~\ref{xsWW} where
inclusive \xsecs\ are presented.
%
\noindent
From both figures one realizes that, as discussed in
section~\ref{xsISR1}, the dominant part of the initial state QED
corrections originates from universal ISR.
Universal ISR is typically of $\co(10\%)$.
Non-universal ISR on the contrary is suppressed and rises from a few
parts per thousand in the LEP2 energy region to $\co(1\%)$ at 1~TeV.
For the \nctwo\ process at 1~TeV the non-universal contribution to the
\xsec\ is 2.5\%, whereas for \ccthree\ at 1~TeV it is 1.4\%.
For reference we have given \xsec\ values for both \zz\ and $W$~pair
production in table~\ref{xsecvals}.
The numerical precision is better than $10^{-4}$.
As numerical input for table~\ref{xsecvals} we have used the standard
LEP2 input (table~5 in~\cite{lep2wgen}) which is reproduced in
table~\ref{lep2input}. 
The weak mixing angle is determined by
\beq
  \sin^2\theta_W = \frac{\pi\,\alpha(2M_W)}{\sqrt{2} G_F M_W^2}
\eeq
and we use the relation $G_{\mu}/\sqrt{2} = g^2/(8M_W^2)$.
\begin{table}[t]
  \caption{\em Inclusive total off-shell \xsec\ values for the
    processes \nctwo\ and \ccthree\ with and without initial state QED
    corrections}
  \vspace{.3cm}
  \begin{center}
    \begin{tabular}{|c||c|c|c||c|c|c||} \hline
      $\sqrt{s}$ & \multicolumn{3}{c||}{$\sigma_\nctwo$\ [pb]}
                 & \multicolumn{3}{c||}{$\sigma_\ccthree$\ [pb]}
      \\
      & $\sigma_\nctwo^{\mathrm{tree}}$
      & $\sigma_\nctwo^{\mathrm{univ}}$
      & $\sigma_\nctwo^{\mathrm{ISR}}$
      & $\sigma_\ccthree^{\mathrm{tree}}$
      & $\sigma_\ccthree^{\mathrm{univ}}$
      & $\sigma_\ccthree^{\mathrm{ISR}}$
      \\\hline \hline
      161 & 0.0154 & 0.0127 & 0.0127 & 4.8087  & 3.4521  & 3.4648
      \\\hline
      175 & 0.0648 & 0.0506 & 0.0508 & 15.9168 & 13.2903 & 13.3408
      \\\hline
      183 & 0.3811 & 0.2733 & 0.2743 & 17.6819 & 15.4036 & 15.4637
      \\\hline
      192 & 0.9783 & 0.7788 & 0.7817 & 18.4479 & 16.5958 & 16.6625
      \\\hline
      205 & 1.1833 & 1.0274 & 1.0313 & 18.5078 & 17.1879 & 17.2600
      \\\hline
      500 & 0.4097 & 0.4424 & 0.4479 & 7.3731  & 7.9168  & 7.9810
      \\\hline
      800 & 0.2124 & 0.2388 & 0.2435 & 3.9971  & 4.4501  & 4.5026
      \\\hline
    \end{tabular}
  \end{center}
  \label{xsecvals}
  \vspace{.8cm}
\end{table}
\begin{table}[t]
  \caption{\em Input parameters for table~\ref{xsecvals}}
  \vspace{.3cm}
  \def\arraystretch{1.2}
  \begin{center}
    \begin{tabular}{|c|c||c|c|}\hline
      Quantity   & Value         & Quantity       & Value
      \\\hline\hline
      $M_Z$      & $91.1888$ GeV & $\alpha(0)$    & $1/137.0359895$
      \\\hline 
      $\Gamma_Z$ & $2.4974$ GeV  & $\alpha(2M_W)$ & $1/128.07$
      \\\hline 
      $M_W$      & $80.23$ GeV   & $G_{\mu}$          &
        $1.16639\cdot10^{-5}$ GeV$^{-2}$ \\\hline
      $\Gamma_W$     & $3G_{\mu} M_W^3/(\sqrt{8}\pi)$ &$V_{CKM}$      &
      $\mathbf{1}$        \\\hline
    \end{tabular}
  \end{center}
  \label{lep2input}
  \vspace{.8cm}
\end{table}

\begin{figure}[t]
  \vspace*{2cm}
  \begin{center}
    \mbox{
      \hspace*{-2.5cm}
      \epsfysize=14cm
      \epsffile{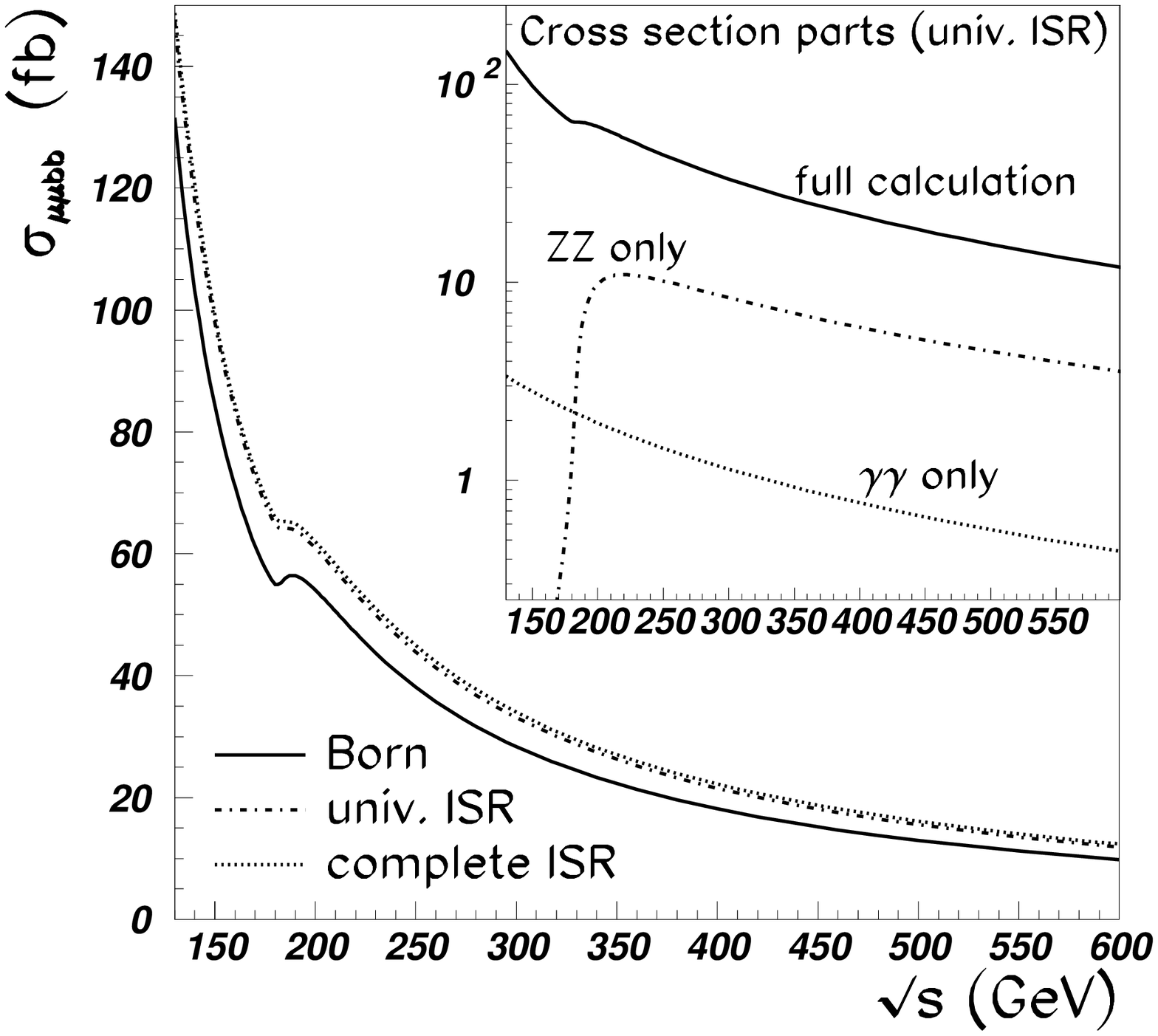} \hspace*{-1cm} }
  \end{center}
  \vspace{-4.7cm}
  \caption[The $W$~pair \xsec\ with initial state radiative corrections]
    {\it The total off-shell \xsec\ for process~\ref{eeNC4f}.
      The inset shows how \zz and photon pair production contribute to
      the \nceight\ \xsec.
      For the particle masses we used numerical values from~\cite{pdb94}.}
  \label{xsNC}
  \vspace{.6cm}
\end{figure}

In figure~\ref{xsNC}, we present, as an example for the \nceight\
process, the \xsec\ for
\beq
  \EE \rightarrow (\ZZ\ZZ,\ZZ\gamma,\gamma\gamma) \rightarrow
  \mu^+\!\mu^-  b \bar{b} \, (\gamma) .
  \label{eeNC4f}
\eeq
%
Due to the negative slope of the \xsec\ curve ISR corrections 
are always positive.
Universal ISR corrections vary between approximately 12\% at
$\sqrt{s}$=130 GeV and 21\% at 600 GeV.
Non-universal corrections rise from 0.9\% at 130 GeV to 4.2\% at 600 GeV.
Thus, compared to the \nctwo\ case, the \nceight\ non-universal
corrections are considerably enhanced.

Finally it is worth mentioning that non-universal ISR tends to be
harder than universal ISR, a fact which emerges from a numerical
analysis of \xsecs\ with lower cuts on $s'$~\cite{dldiss}.
This may be understood as follows: \oal~non-universal corrections are
infrared finite, whereas the infrared divergent universal corrections
are resummed by means of the soft photon exponentiation. 
Thus the universal corrections contain important soft resummed parts,
and the non-universal corrections do not.
%
%
\section{Concluding Remarks}
\label{sum}
\ezero
%
In this paper we have presented the analytical results of the first
complete gauge-invariant calculations of initial state QED
corrections to off-shell vector boson pair production in $e^+ e^-$
annihilation.
We have decomposed the corrections into a dominant universal
contribution and a numerically smaller non-universal contribution.  
Non-universal corrections were found to be smoothly behaving around
the two boson production threshold and to be explicitly screened from
unitarity violations at high energies.
As a by-product, the angular distribution of radiated photons is
available from intermediate steps of the calculation but has not been
presented here.

By construction, we have not taken into account genuine weak
corrections~\cite{ew} or QED corrections related to the final
state~\cite{on1loop}.
Compared to ISR, both are known to be smaller but may reach several
percent nevertheless.
Therefore they are comparable in size to the non-universal corrections
calculated here.
See also references~\cite{lep2ww,teubeen}.
Special suppressions of the order ${\mathcal
  O}(\Gamma_V/M_V)$ are estimated for interference
effects between initial and final state radiation in resonant boson
pair production and inter-bosonic final state interferences~\cite{yakmel}.
The complete photonic corrections to four-fermion
production are not available although important steps towards their
calculation have been achieved~\cite{vanO}.

The presented calculation should be considered in connection with the
program of semi-analytical treatments of complete processes at tree
level~\cite{teuw94}. 
It would be quite interesting to compute complete ISR not only for the
off-shell production of vector boson pairs, but also for true
four-fermion processes.  
Simplest are the so-called {\tt NC24}~\cite{nc8a} and
{\tt CC11}~\cite{cc11} processes, but also the combination of {\tt NC24}
with Higgs production~\cite{nc8b} should be straightforward. 
Because in these reactions all background Feynman diagrams, i.e. those
diagrams that must be added to \ccthree\ [\nceight] to get the sets
for {\tt CC11} [{\tt NC24}], are {\em s}-channel diagrams, there is
nothing but universal ISR for the {\em pure} background.
So the only missing pieces are the non-universal ISR contributions to
the interferences between the boson pair signal and the background.
The results may be of numerical relevance in the {\tt NC24} case below
the $ZZ$ production threshold. 
For {\tt CC11} and for the inclusion of Higgs production they will be 
of mere theoretical interest.
%
\section*{Acknowledgments}
We would like to thank F.~Jegerlehner, A.~Leike, A.~Olchevski, and
U.~M\"uller for helpful discussions.
The contribution of M.~Bilenky to the calculation of the virtual
corrections to the {\tt CC3} process in 1992 is gratefully
acknowledged.
We are indebted to J. Bl\"umlein for carefully reading the manuscript.
%
%
\appendix
\def\theequation{\Alph{section}.\arabic{equation}}
%
%
\section{Couplings and Factorizing Kinematical Functions
\label{appborncg}
}
\ezero
%
\subsection{Couplings
\label{subc}
}
The electroweak couplings and neutral boson propagators enter the Born
and QED corrected cross-sections in certain combinations.

The \nceight\ process is described by one coupling
function~\cite{nc8a}:
\\
\ba
  \lefteqn{\cc_\nceight \;\; = \;\;
    \frac{2}{\,(6\pi^2)^2\,} \;
      \sum_{V_i,V_j,V_k,V_l=\gamma,Z}
      \re \: \rule[-.2cm]{0cm}{1cm}
        \frac{1}{D_{V_i}(\SONE) D_{V_j}(\STWO) D_{V_k}^*(\SONE)
          D_{V_l}^*(\STWO)}}
      \nl
    & & \hspace{2.75cm} \times \left[ \rule[0cm]{0cm}{.4cm}
      L(e,V_i) \, L(e,V_k) \, L(e,V_j) \, L(e,V_l) \right. \nl
    & & \hspace{3.8cm} \left. \rule[0cm]{0cm}{.4cm} + \;
      R(e,V_i) \, R(e,V_k) \, R(e,V_j) \, R(e,V_l) \, \right] \nl
    & & \hspace{2.75cm} \times \left[ \rule[0cm]{0cm}{.4cm}
      L(\FONE,V_i) \, L(\FONE,V_k) \right. \nl
    & & \hspace{3.8cm} \left. \rule[0cm]{0cm}{.4cm} + \;
      R(\FONE,V_i) \, R(\FONE,V_k) \, \right]
      \; N_c(f_1) \, N_p(V_i,V_k,m_1,\SONE) \nl
    & &
      \hspace{2.75cm} \times
      \left[ \rule[0cm]{0cm}{.4cm}
      L(\FTWO,V_j) \, L(\FTWO,V_l) \right. \nl
    & & \hspace{3.8cm} \left. \rule[0cm]{0cm}{.4cm} + \;
      R(\FTWO,V_j) \, R(\FTWO,V_l) \, \right]
      \; N_c(f_2) \, N_p(V_j,V_l,m_2,\STWO). \nl
    \label{nc8c} \\ \nonumber
\ea
We have used the left- and right-handed fermion-boson couplings
\\
\ba
  L(f,\gamma) \; = \; \frac{\,e\,Q_{\!f}\,}{2},
  \hspace{1.83cm} & &
  L(f,Z) \; = \;   \frac{g}{\,2\, c_W} \,
    \left(I_3^f - s_W^2 \, Q_{\!f}\right),
  \nl \nl
  R(f,\gamma) \; = \; \frac{\,e\,Q_{\!f}\,}{2},
  \hspace{1.8cm} & &
  R(f,Z) \; = \; - \frac{g}{\,2\, c_W}\, s_W^2 \,Q_{\!f}
  \label{LRcoup}
\ea
\\
where $Q_{\!f}$\ is the fermion charge in units of the positron charge
$e$, $I_3^f$\ is the fermion's weak isospin third component, and
$s_W$\ denotes the sine of the weak mixing angle.
$Q_{\!e}\!=\! -1$\, and the fine structure constant is given by
$e \! = \! g s_W \! = \! \sqrt{4\pi\alpha\,}$.
In addition, equation~(\ref{nc8c}) contains boson propagators
\beq
  D_V(s) = s - M_V^2 + \ri s \Gamma_V / M_V \;.
  \label{Vprop}
\eeq
The photon has zero mass and width, $M_\gamma \! = \! \Gamma_\gamma
\! = \! 0$. 
The color factor $N_c(f)$ is unity for leptons and three for quarks.
The phase space factors
\beq
  N_p(V_m,V_n,m,s) \, = \, \left\{ \, \rule[-.3cm]{0cm}{1.2cm}
  \!\!\!
    \begin{array}{ccl}
      1 & \;\; {\rm for} \;\; & V_m\neq\gamma \;\; {\rm or} \;\;
        V_n\neq\gamma \vspace{.2cm} \\
      \sqrt{1 \! - \! \frac{4m^2}{s}} \; (1 \! + \! \frac{2m^2}{s})
      & \;\; {\rm for} \;\; & V_m = V_n = \gamma
    \end{array} \right.
  \label{psfact}
\eeq
in equation~(\ref{nc8c}) correctly take into account that fermion
masses may be neglected if the corresponding fermion pair couples to a
\zz~resonance but may not if it only couples to a photon. 

The \xsec\ for the \nctwo\ process is obtained by omitting exchange
photons in equation~(\ref{nc8c}).

The \ccthree\ process is described by three coupling functions
$\cc_{\tt CC3}^{(k)}$,
one for the {\it t}-channel, one for the {\it s}-channel, and one
for the {\it st}-interference contribution~\cite{muta,cc11}:
\\
\ba
   \cc_{\tt CC3}^t & = & \frac{2}{(6\pi^2)^2} \;
     \frac{1}{\left|D_W(s_1)\right|^2 \left|D_W(s_2)\right|^2}
     \nl & & \times \rule[0cm]{0cm}{.4cm} L^4(e,W) \;
       L^2(f_1,W) \, N_c(f_1) \; L^2(f_2,W) \, N_c(f_2),
     \label{cc3ct} 
\\ \nl
   \cc_{\tt CC3}^s & = &
     \frac{2}{\left(6\pi^2\right)^2} \; \sum_{V_i,V_j=\gamma,Z} \;
     \re \:
     \frac{1}{\left|D_W(s_1)\right|^2\left|D_W(s_2)\right|^2
              D_{V_i}(s) D_{V_j}^*(s)} \nl
     & & \times \left[ \rule[-.1cm]{0cm}{.5cm} L(e,V_i)L(e,V_j)
               + R(e,V_i)R(e,V_j)\right] \nl
     & & \times \rule[0cm]{0cm}{.4cm} \; g_3(V_i) g_3(V_j) \;
         L^2(f_1,W) N_c(f_1) \; L^2(f_2,W) \, N_c(f_2),
     \label{cc3cs} 
\\ \nl
   \cc_{\tt CC3}^{st}  & = &
     \frac{2}{\left(6\pi^2\right)^2} \;  \sum_{V_i=\gamma,Z} \;\,
     \re \:
     \frac{1}{\left|D_W(s_1)\right|^2\left|D_W(s_2)\right|^2
              D_{V_i}(s)} \; \! L^2(e,W) \nl
     & & \times \rule[0cm]{0cm}{.4cm} \; L(e,V_i) \, g_3(V_i) \;
         L^2(f_1,W) \, N_c(f_1) \; L^2(f_2,W) \, N_c(f_2) .
     \label{cc3cst} \\ \nonumber
\ea
The left-handed couplings $L(f,W)$\ of a weak isodoublet
with the fermion $f$\ to a $W$\ boson are given by
\beq
  L(f,W) \; = \; \frac{g}{\,2 \sqrt{2}\,}.
  \label{Wcoup}
\eeq
We neglect Cabibbo-Kobayashi-Maskawa mixing effects.
The right-handed couplings $R(f,W)$\ vanish.
The couplings $g_3(V_i)$\ originate from the three-boson vertices and
are
\\
\ba
  g_3(\gamma) & = & g \, s_W , 
 \nl
  g_3(Z) & = & g \, c_W.
  \label{tricoup}
\ea
\\
Using $G_{\mu}/\sqrt{2} = g^2/(8M_W^2)$ in the $\mathcal C$ functions,
one easily rewrites the \xsecs\ for heavy boson pair production, namely
the processes \ccthree\ and \nctwo, in terms of Breit-Wigner densities
\beq
  \rho_V(s_i) \; = \; \frac{1}{\pi}
    \frac {s_i\Gamma_V/M_V }
          {|s_i - M_V^2 + \ri s_i \Gamma_V /M_V |^2} \times
    {\mathrm{BR}}(i), \hspace{.7cm} V=Z,W^{\pm}
  \label{rho}
\eeq
with branching fractions ${\mathrm{BR}}(i)$ into the appropriate
fermion pairs. 

For the process \nctwo\ one gets
\beq
  \cc_\nctwo \; = \;
    \frac{\,\rho_Z(\SONE) \, \rho_Z(\STWO)\,}{\SONE\STWO} \;
    \frac{\left( G_{\mu} M_Z^2 \right)^2}{4} \;
    \left[ \left( 1 - 2 s_W^2 \right)^4 + \left( 2 s_W^2 \right)^4
    \right]. 
  \label{ZZtoNC}
\eeq
For $W^\pm$\ pair production, the situation is slightly more
involved. 
Defining
\beq
  \cc_\ccthree \; = \; \frac{\left(G_{\mu} M_W^2 \right)^2}{s_1s_2}
    \rho_W(s_1) \rho_W(s_2),
  \label{cc3crho}
\eeq
one obtains by inspection of equations~(\ref{cc3ct})--(\ref{cc3cst})
\\
\ba
  \cc_\ccthree^t & \; = \; & \cc_\ccthree \; c_{\nu\nu},
  \label{cc3tcrho} 
\nl \nl
  \cc_\ccthree^s 
  &\; = \; & \frac{4}{s^2} \cc_\ccthree \;
    \left( c_{\gamma\gamma} + c_{\gamma Z} + c_{ZZ} \right),
  \label{cc3scrho} 
\nl \nl
  \cc_\ccthree^{st} & \; = \; &
  \frac{1}{s} \cc_\ccthree \; \left( c_{\nu\gamma} + c_{\nu Z} \right)
\label{cc3stcrho}
\ea
\\
with factors $c_{ab}$ as defined in~\cite{teuw94} (see also
\cite{muta}). 

For on-shell heavy boson pair production, $\Gamma_V
\!\to\! 0$, Breit-Wigner densities are replaced by
$\delta$-distributions,
\beq
  \rho_V(s) \stackrel{\Gamma_V \rightarrow 0}{\longrightarrow}
  \delta(s - M_V^2) \times {\mathrm {BR}}(i).
  \label{rholim}
\eeq
\subsection{Kinematical Functions for the Factorizing QED Corrections
\label{subg}
}
The kinematical function for the {\tt NC} case is~\cite{cnc2}:
\\
\ba
\cg_\nctwo(s;\SONE,\STWO) &=& \cg_\nceight(s;\SONE,\STWO)
  = \SONE\STWO \left[ \rule[0cm]{0cm}{.4cm}
    \frac{\,s^2+(\SONE+\STWO)^2\,}{s-\SONE-\STWO}\,\cl - 2 \right].
\label{nc8g} 
\ea
\\
For the {\tt CC} case, one finds~\cite{muta}:
\\
\ba
  \cg_{\tt CC3}^t (s;s_1,s_2) & = & \frac{1}{48}
    \left[ \rule[-.1cm]{0cm}{.5cm} \: \lambda +
           12 \, s \, \left( s_1 + s_2 \right) - 48 s_1 s_2 \right.
    \nl & & \left. \hspace{2.62cm} \rule[-.1cm]{0cm}{.5cm}
           + 24 \left(s - s_1 - s_2\right) s_1 s_2
           \; \cl \right],
    \label{cc3gt} \\ \nl
  \cg_{\tt CC3}^s (s,s_1,s_2) & = & \frac{\lambda}{192}
    \left[ \rule[-.1cm]{0cm}{.5cm} \lambda + 12\left(s s_1 + s s_2 +
           s_1 s_2 \right)\right],
    \label{cc3gs} \\ \nl
  \cg_{\tt CC3}^{st} (s;s_1,s_2) & = & \frac{1}{48}
    \left\{ \rule[-.2cm]{0cm}{.7cm} (s-s_1-s_2)
            \left[ \rule[-.1cm]{0cm}{.5cm}
                   \lambda + 12(s s_1 + s s_2 + s_1s_2) \right] \right.
    \nl & &  \left. \rule[-.2cm]{0cm}{.7cm} \hspace{2.8cm}
                   - 24 \left( s s_1 + s s_2 + s_1 s_2 \right) s_1 s_2
                   \; \cl \right\}
    \label{cc3gst} \\ \nonumber
\ea
for the kinematical functions. 
The logarithm $\cl$ contained in integrated $t$- or $u$-channel
contributions is defined by 
\beq
  \cl \; = \; \cl(s;\SONE,\STWO) \; = \;
    \frac{1}{\sqrt{\lambda}} \,
    \ln \frac{s-\SONE-\STWO+\sqrt{\lambda}}
             {s-\SONE-\STWO-\sqrt{\lambda}} \; .
\eeq
%
%
\section{The 2$\,\to\,$5 Particle Phase Space
\label{ps2to5}
}
\ezero
When considering initial state photon bremsstrahlung to a 2$\,\to\,$4
process, a photon with momentum $p$ appears as fifth particle in the
final state. 
The intermediate vector bosons (two-fermion systems) have momenta 
$v_1 = p_1 + p_2$, $v_2 = p_3 + p_4$.
The five-particle phase space has eleven kinematical degrees of
freedom.
It is convenient to parametrize the five-particle Lorentz
invariant phase space by a decomposition into subsequently decaying
particles:
\\
\ba
  \d\Gamma_5 & = & \frac{1}{(2\pi)^{14}} \: 
                  \frac{\sqrt{\lambda(s,s',0)}}{8s} \:
                  \frac{\sqrt{\lambda(s',\SONE,\STWO)}}{8s'} \:
                  \frac{\sqrt{\lambda(\SONE,m_1^2,m_2^2)}}{8\SONE} \:
                  \frac{\sqrt{\lambda(\STWO,m_3^2,m_4^2)}}{8\STWO}
                  \nl
            &   & \times\; \d s' \; \d\SONE \; \d\STWO \;
                  \d\!\mycos\theta \; \d\phi_{R} \,
                  \d\!\mycos\theta_{\!R} \;
                  \d\phi_1 \, \d\!\mycos\theta_1 \;
                  \d\phi_2 \, \d\!\mycos\theta_2,
\label{gamma5}
\ea
\\
where $m_i$\ is the mass of the final state fermion with momentum
$p_i$\ and \\
\vspace{.3cm}
\begin{tabular}{rcl}
  $\!\phi$        & : & Azimuth angle around the beam direction. \\
  $\!\theta$      & : & Polar (scattering) angle of the photon with
                        respect to the \epl
                        \vspace{-.2cm} \\
                  &   & direction $\pol{k}_2$ in the center of mass
                        frame.  \\
  $\!\phi_R$      & : & Azimuth angle around the photon direction
                        $\pol{p}$.\\
  $\!\theta_R$    & : & Polar angle of $\pol{v}_1$ in the
                        $(\pol{v}_1\!+\!\pol{v}_2)$ rest frame with
                        z axis along $\pol{p}$. \\
  $\!\phi_1$   & : & Azimuth angle around the $\pol{v}_1$
                        direction.\\
  $\!\theta_1$ & : & Decay polar angle of $\pol{p}_1$ in the
                        $\pol{v}_1$ rest frame with axis along
                        $\pol{v}_1$.\\
  $\!\phi_2$   & : & Azimuth angle around the $\pol{v}_2$
                        direction.\\
  $\!\theta_2$ & : & Decay polar angle of $\pol{p}_3$ in the
                        $\pol{v}_2$ rest frame with axis along
                        $\pol{v}_2$.\\
  $\!\SONE$       & : & Invariant mass of the final state fermion pair
                        \fone\bftwo~:~~$\SONE=-v_1^2$. \\
  $\!\STWO$       & : & Invariant mass of the final state fermion pair
                        \fthree\bffour~:~~$\STWO=-v_2^2$. \\
  $\!s'$          & : & Reduced center of mass energy squared. This is
                        the invariant
                        \vspace{-.2cm} \\
                  &   & mass of the final state four-fermion
                        system:
                        \vspace{-.2cm} \\
                  &   & $s'= -(v_1+v_2)^2 = -(p_1+p_2+p_3+p_4)^2$.
\end{tabular}
\vspace{.3cm}

The ranges of the kinematical variables are \\
$\begin{array}{cccccc}
  \hspace{.75cm}
  & \left(m_1 + m_2 + m_3 + m_4\right)^2 & \leq & s' & \leq & s \\
  & \left(m_1 +m_2\right)^2 & \leq & \SONE & \leq & \left( \sqrt{s'} 
- m_3- m_4 \right)^2 \\
  & \left(m_3 +m_4\right)^2 & \leq & \STWO & \leq & \left( \sqrt{s'} - \sqrt{\SONE}
                                     \right)^2 \\
  & -1 & \leq & \mycos\theta & \leq & +1       \\
  &~~0 & \leq & \phi_{\{R,1,2\}} & \leq & 2 \pi \\
  & -1 & \leq & \mycos\theta_{\{R,1,2\}} & \leq & +1
\end{array}$
\vspace{-1cm}
\beq
  \label{par25a}
\eeq
or, alternatively,
\medskip \\
$\begin{array}{cccccc}
  \hspace{.75cm}
  & \left(m_1 + m_2\right)^2 & \leq & \SONE & \leq & \left( \sqrt{s} -
  m_3- m_4 \right)^2 
\\
  & \left(m_3 + m_4\right)^2 & \leq & \STWO & \leq & \left( \sqrt{s} - \sqrt{\SONE}
                                     \right)^2 \\
  & \left( \sqrt{\SONE} + \sqrt{\STWO} \right)^2 & \leq & s' & \leq &
  s \\
  & -1 & \leq & \mycos\theta & \leq & +1       \\
  &~~0 & \leq & \phi_{\{R,1,2\}} & \leq & 2 \pi \\
  & -1 & \leq & \mycos\theta_{\{R,1,2\}} & \leq & \hspace{1.4cm} +1
  \hspace{1.3cm}.
\end{array}$
\vspace{-1cm}
\beq
  \label{par25b}
\eeq

\begin{figure}[t]
\begin{center}
\vspace{2cm}
\mbox{
\epsfysize=7.87cm
\epsffile{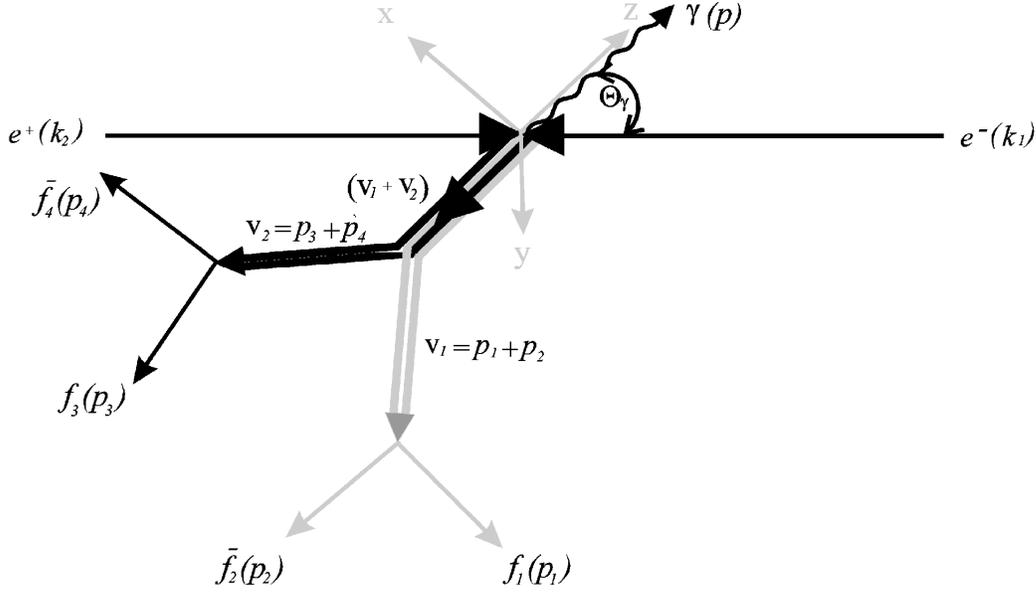}
}
\vspace{-.2cm}
\end{center}
\caption[The 2$\to$5 body phase space]
{\it Graphical representation of a 2$\,\to\,\!$5 particle reaction. In
  general, the vectors
  $\pol{v}_1,\pol{v}_2,\pol{p}_1,\pol{p}_2,\pol{p}_3,$~and
  $\pol{p}_4$ do not lie in the plane of the picture. In the figure,
  the angle $\theta$~is called $\theta_\gamma$ to emphasize that it
  belongs to the direction of the radiated photon.
  \label{ps25}}
\vspace{.5cm}
\end{figure}

An illustration of the 2$\,\to\,$5 particle phase space is given in
figure~\ref{ps25}.
%
%

For the processes under consideration, the integrations
over the final state fermion decay angles $\phi_1,~\theta_1$\ and
$\phi_2,~\theta_2$\ are separated from the other integrations and are
carried out analytically with the help of invariant tensor
integration (see appendix~\ref{tenint}).
The remaining integrations are nontrivial.
We will now explain, how the scalar products appearing in the squared
matrix elements can be expressed in the center of mass frame with a
Cartesian coordinate system as drawn in figure~\ref{ps25} using the
phase space variables introduced in equation~(\ref{gamma5}). 

The initial state vectors $k_1$~and $k_2$
are given by
\\
\ba
  k_1 & = & \left( k^0,~k\,\mysin\theta,~0,~-k\,\mycos\theta
            \right),
  \nl
  k_2 & = & \left( k^0,~-k\,\mysin\theta,~0,~k\,\mycos\theta
            \right)
  \label{kvectrad}
\ea
\\
with 
\\
\ba
  k^0  =  \frac{\sqrt{s}}{2},
  \hspace{1.5cm}
  k    =  \frac{\sqrt{s}}{2} \, \beta,
  \hspace{1.5cm}
  \beta & \equiv & \sqrt{1-\frac{4\MES}{s}}
  \label{kpars}
\ea
\\
where $m_e$\ is the electron mass. 

For the momentum of the photon radiated along the $z$ axis one finds in
the  center of mass  system:
\beq
  p \; =\; (p^0,~0,~0,~p^0) \hspace{1.3cm} {\rm with} \hspace{1.3cm}
  p^0 \; =\; \frac{s-s'}{2\sqrt{s}}\;.
  \label{photmom}
\eeq

Thus, the invariant products $k_1k_2$\ and $k_ip$\ may be expressed
easily. 

The virtual boson pair recoils from the photon in the center of mass
system:
\beq
(v_1+v_2) \; =\;
  \left( \frac{s+s'}{2\sqrt{s}},~0,~0,~-\frac{s-s'}{2\sqrt{s}} \right).  
\label{v1v2}
\eeq

The vectors $v_1$ and $v_2$ in the two-boson rest frame (the $R$-system
with $\pol{v}_{1,R} \!+\! \pol{v}_{2,R} \!=\! 0$) are easily derived: 
\\
\ba
  v_{1,R} &\equiv& \left( p_{12}^{R,0}, 0, 0, \cp_R 
         \right)
 =  \left( \rule[0cm]{0cm}{.95cm}
              \frac{s'+\SONE-\STWO}{2\sqrt{s'}},~0,~0,~~
              \frac{\sqrt{\lambda(s',\SONE,\STWO)}}{2\sqrt{s'}}
            \right), 
\nl
  v_{2,R} &\equiv& \left( p_{34}^{R,0}, 0, 0, -\cp_R 
         \right)
 = \left( \rule[0cm]{0cm}{.95cm}
               \frac{s'-\SONE+\STWO}{2\sqrt{s'}},~0,~0,~
              -\frac{\sqrt{\lambda(s',\SONE,\STWO)}}{2\sqrt{s'}}
            \right).
  \label{vvect}
\ea
\\
Both $v_{1,R}$\ and $v_{2,R}$\ do not depend on angular variables.

We need $v_i$ in the  center of mass system.
The boost  between the center of mass frame and
the two-boson rest frame $R$ with zero component $\gamma^{(0)}$ and
modulus of the spatial components $\gamma$ is   
\\
\ba
  \gamma^{(0)} \; = \; \frac{s+s'}{2\sqrt{ss'}}\;, \hspace{2cm}
  \gamma   \; = \; \frac{s-s'}{2\sqrt{ss'}}\;.
  \label{kingamma}
\ea
\\
Taking into account that the production angles of the virtual vector
bosons are not fixed, equation~(\ref{vvect}) yields the following center
of mass system four-vectors:
\\
\ba
  v_1 & = & \left( \gamma^{(0)}\,p_{12}^{R,0} - \gamma \cp_R
  \mycos\theta_R,~
                   \cp_R \mysin\theta_R \mycos\phi_R,~
                  -\cp_R \mysin\theta_R \mysin\phi_R, \right. \nl
    & & \hspace{7.5cm} \left.
                   \gamma^{(0)} \cp_R \mycos\theta_R - \gamma
                   \,p_{12}^{R,0}
            \right), \nl
  v_2 & = & \left( \gamma^{(0)}\,p_{34}^{R,0} + \gamma \cp_R
  \mycos\theta_R,~
                  -\cp_R \mysin\theta_R \mycos\phi_R,~
                   \cp_R \mysin\theta_R \mysin\phi_R, \right. \nl
    & & \hspace{7.5cm} \left.
                  -\gamma^{(0)} \cp_R \mycos\theta_R - \gamma
                  \,p_{34}^{R,0}
            \right). \nl
  \label{v5vect}
\ea
\\
The sum $(v_{1,R}+v_{2,R})$ in the $R$-system is simply $(\sqrt{s'},0,0,0)$.
The boost into the center of mass system yields $(v_1+v_2)= (
\gamma^{(0)} 
\sqrt{s'},0,0,-\gamma \sqrt{s'} )$, which is exactly~(\ref{v1v2}).     

The above formulae allow to express the products $k_iv_j$\ and $pv_j$. 
The product $p_1p_2$ [and analogously $p_3p_4$] is most easily
calculated in the  center of mass system $R_1$ [$R_2$] defined by
$\pol{p}_1\!+\!\pol{p}_2 = 0$ [$\pol{p}_3\!+\!\pol{p}_4 = 0$]. 
This is explained in equations~(B.5)--(B.11) of~\cite{cc11} for the
2$\to$4 process.
The discussion of reference~\cite{cc11} remains valid also with an
additional photon in the initial state.  
For the remaining scalar products $p p_i, k_jp_i, v_j p_i$, and $p_i
p_j$ we used explicit expressions for the vectors $p_i$ in the center
of mass system.
The vectors $p_i$ are obtained from the vectors $p_{i,R}$ in the
$R$-system by rotating and applying the Lorentz
boost~(\ref{kingamma}):
\\
\ba
  p_i & = & \left(\!\!
  \begin{array}{c}
    \gamma^{(0)}\,p_{i,R}^0(s') - \gamma \left[ \rule[0cm]{0cm}{.35cm}
      p_{i,R}^z(s')\mycos\theta_R - p_{i,R}^x(s')\mysin\theta_R
     \right]
 \\ 
    \left[ \rule[0cm]{0cm}{.35cm}
      p_{i,R}^z(s')\mysin\theta_R + p_{i,R}^x(s')\mycos\theta_R
    \right] \mycos\phi_R  + p_{i,R}^y(s')\mysin\phi_R 
\\
    - \left[ \rule[0cm]{0cm}{.35cm}
      p_{i,R}^z(s')\mysin\theta_R + p_{i,R}^x(s')\mycos\theta_R
    \right]
      \mysin\phi_R  + p_{i,R}^y(s')\mycos\phi_R 
\\
    \gamma^{(0)} \left[ \rule[0cm]{0cm}{.35cm}
      p_{i,R}^z(s')\mycos\theta_R - p_{i,R}^x(s')\mysin\theta_R \right]
      - \gamma\,p_{i,R}^0(s')
   \end{array}
  \!\!\right)\;.
  \label{5pvect}
\ea
\\
The four-momenta $p_{i,R}(s')$ are exactly the final state momenta
$p_i$ shown in equation~(B.21) of~\cite{cc11}, when these are evaluated
at $s'$ instead of $s$.
The frame $R$ there has to be understood as frame $R_1$ or $R_2$
introduced here. 

Now, the derivation of all momenta in the center of mass system in
terms of the integration variables is completed for the 2$\to$5
process and all scalar products appearing in the initial state
bremsstrahlung calculation can be expressed in terms of phase space
variables.

For the calculation of virtual and soft photon corrections, the
2$\to$4 phase space is needed together with the corresponding momentum
representations.
It may be taken over completely from appendix~B of~\cite{cc11}.  
%
%
\section{Matrix Elements}
\label{appmatel}
\ezero
%
In this appendix, we will present the matrix elements for the
\ccthree\ and \nctwo\ processes. 
The matrix elements for the \nceight\ process are easily obtained from
the \nctwo\ matrix elements by adding amplitudes where $Z$ couplings
(propagators) have been replaced by photon couplings (propagators).
%
%
\subsection{Tree Level Amplitudes}
\label{bornamp}
%
For the calculation of soft and virtual corrections,
the tree level matrix elements for the \ccthree\ and \nctwo\ processes are
given in terms of $s$-channel, $t$-channel, and $u$-channel amplitudes
corresponding to the Feynman diagrams of figures~\ref{cc3diag}
and~\ref{nc8diag}:
\\
\ba
  \cm^B_{\ccthree} & = & \cm^B_{s,Z} + \cm^B_{s,\gamma} +
                         \cm^B_{t,W}, 
\\
  \cm^B_{\nctwo} & = & \cm^B_{t,Z} + \cm^B_{u,Z}
\ea
\\
with
\\
\ba
  \cm^B_{s,Z} & = &
    \frac{g_{\gamma \gamma'}}{\,D_Z(s)\,}\;
    \frac{g_{\beta \beta'}}{\,D_W(\SONE)\,}\:
    \frac{g_{\alpha \alpha'}}{\,D_W(\STWO)\,} \;
    B^{\gamma'}_{s,Z}\,M_{12,W}^{\beta'}\,M_{34,W}^{\alpha'} \;
    g_3(Z)\,T^{\alpha \beta \gamma},
  \label{sZamp}
  \\ \nl
  \cm^B_{s,\gamma} & = &
    \frac{g_{\gamma \gamma'}}{\,D_\gamma(s)\,}\;
    \frac{g_{\beta \beta'}}{\,D_W(\SONE)\,}\:
    \frac{g_{\alpha \alpha'}}{\,D_W(\STWO)\,} \;
    B^{\gamma'}_{s,\gamma}\,M_{12,W}^{\beta'}\,M_{34,W}^{\alpha'}
    \; g_3(\gamma)\,T^{\alpha \beta \gamma},
  \label{sgamp}
  \\ \nl
  \cm^B_{t,W} & = &
    \frac{g_{\beta \beta'}}{\,D_W(\SONE)\,}\:
    \frac{g_{\alpha \alpha'}}{\,D_W(\STWO)\,}\;
    \frac{1}{\,q_t^2\,}\;
    B^{\alpha \beta}_{t,W}\,M_{12,W}^{\beta'}\,M_{34,W}^{\alpha'},
  \label{tWamp}
  \\ \nl
  \cm^B_{t,Z} & = &
    \frac{g_{\beta \beta'}}{\,D_Z(\SONE)\,}\:
    \frac{g_{\alpha \alpha'}}{\,D_Z(\STWO)\,}\;
    \frac{1}{\,q_t^2\,}\;
    B^{\alpha \beta}_{t,Z}\,M_{12,Z}^{\beta'}\,M_{34,Z}^{\alpha'},
  \label{tZamp}
  \\ \nl
  \cm^B_{u,Z} & = &
    \frac{g_{\alpha \alpha'}}{\,D_Z(\SONE)\,}\:
    \frac{g_{\beta \beta'}}{\,D_Z(\STWO)\,}\;
    \frac{1}{\,q_u^2\,} \;
    B^{\alpha \beta}_{u,Z}\,M_{12,Z}^{\alpha'}\,M_{34,Z}^{\beta'}.
  \label{uZamp}
\ea
\\
We have used the vector boson propagator denominators $D_V$\ defined
in equation~(\ref{Vprop}), the trilinear coupling $g_3(V^0)$\ given in 
equation~(\ref{tricoup}), and
\beq
  T^{\alpha \beta \gamma} \; = \; (v_1-v_2)^\gamma \, g^{\alpha \beta}
    - 2\,v_1^\alpha \, g^{\beta \gamma} + 2\,v_2^\beta \, g^{\alpha
      \gamma}.
\eeq
In the massless approximation, the annihilation matrix elements
$B_{s,V^0}^{\gamma'}$, $B_{t,V}^{\alpha\beta}$, $B_{u,V}^{\alpha\beta}$,
and the decay matrix elements $M_{ij,V}^{\mu}$ are given by: 
\\
\ba
  B_{s,V^0}^{\gamma'}  & = & {\bar u}(-k_2)  \gamma^\mu
    \left[  \rule[0cm]{0cm}{.35cm}
            L(e,V^0)  \left( 1  + \gamma_5 \right)
          + R(e,V^0)  \left( 1  - \gamma_5 \right)
    \right]  u(k_1),
  \label{scurrent}
  \\ \nl
  B^{\alpha \beta}_{t,V} & = & 2\,{\bar u}(-k_2)  \gamma^\alpha
    \left[  \rule[0cm]{0cm}{.35cm}
            L^2(e,V)  \left( 1+\gamma_5 \right)
          + R^2(e,V)  \left( 1-\gamma_5 \right)
    \right]
\dagg{q}_{\!t} \, \gamma^\beta u(k_1), 
\nl
  \label{tcurrent}
  \\ \nl
  B^{\alpha \beta}_{u,Z} & = & 2\,{\bar u}(-k_2)  \gamma^\alpha
    \left[  \rule[0cm]{0cm}{.35cm}
            L^2(e,Z)  \left( 1+\gamma_5 \right)
          + R^2(e,Z)  \left( 1-\gamma_5 \right)
    \right]
\dagg{q}_{\!\!u} \, \gamma^\beta u(k_1), 
\nl
  \label{ucurrent}
  \\ \nl
  M_{ij,V}^\mu & = & {\bar u}(p_i) \gamma^\mu
    \left[  \rule[0cm]{0cm}{.35cm}
            L(f_i,V)  \left( 1+\gamma_5 \right)
          + R(f_i,V)  \left( 1-\gamma_5 \right)
    \right]  u(-p_j)
  \label{fcurrent}
\ea
\\
with left- and right-handed couplings as defined in
equations~(\ref{LRcoup}) and~(\ref{Wcoup}).
We have further used 
\\
\ba
  q_t & = & k_1 - v_1 \; = \; -k_2 + v_2  \; = \;
            \frac{1}{2} \left( k_1 - k_2 - v_1 + v_2\right), 
\\ \nl
  q_u & = & k_1 - v_2  \; = \; -k_2 + v_1  \; = \;
    \frac{1}{2} \left( k_1 - k_2 + v_1 - v_2 \right).
\ea
\\
%
%
\subsection{Virtual Initial State Corrections}
\label{virtamp}
%
The amplitudes for virtual initial state corrections are easily
obtained from the matrix elements in equations~(\ref{sZamp})
to~(\ref{uZamp}) by substituting the initial state currents.

%
The $t$-channel \nctwo\ virtual QED matrix element $\cm^V_{t,Z}$\ is
represented by the Feynman diagrams of figure~\ref{tvirt} and given by 
\beq
  \cm^V_{t,Z} \; = \;\frac{g_{\beta \beta'}}{\,D_Z(\SONE)\,}\:
                     \frac{g_{\alpha \alpha'}}{\,D_Z(\STWO)\,}
                     \; V^{\alpha \beta}_{t,Z}
                     M_{12,Z}^{\beta'}\,M_{34,Z}^{\alpha'}
\eeq
with 
\\
\ba
  V^{\alpha\beta}_{t,Z} & = & 
     e^2  \left\{
    \frac{C^{\alpha\beta}_{t,Z}}{16\pi^2} \right.
\nl
& - & \ri \left. \left. \mu^{(4-n)}  \int  \frac{\d^n p}{(2\pi)^n}
    \left( V^{\alpha\beta}_{t,{\rm vert1}} +
           V^{\alpha\beta}_{t,{\rm vert2}} +
           V^{\alpha\beta}_{t,{\rm self}}  +
           V^{\alpha\beta}_{t,{\rm box}} \right)
     \right|_{\mu=m_e}  \right\}
  \label{tloop}
\ea
\\
where dimensional regularization is applied with $p$ being the loop
photon momentum, and where
\\
\ba
  V^{\alpha\beta}_{t,{\rm vert1}} & = &
    {\bar u}(-k_2) \;
    \frac{ \left( -2 k_2^\mu - \gamma^\mu \dagg{p} \right)
           \gamma^\alpha \left(\dagg{q}_{\!t} - \dagg{p} \right)
           \gamma_\mu \, \dagg{q}_{\!t} \, \gamma^\beta }
         { \left[ \rule[0cm]{0cm}{.35cm} p^2 - \ieps \right] \,
           \left[ \rule[0cm]{0cm}{.35cm} (p - q_t)^2 -\ieps
           \right] \,
           \left[ \rule[0cm]{0cm}{.35cm} p^2 + 2 k_2 p- \ieps \right] \,
           q_t^2}
    \nl & &         
 \hspace{2.2cm}\times
    2 \left[  L^2(e,Z) \;\left( 1+\gamma_5 \right)
            + R^2(e,Z) \;\left( 1-\gamma_5 \right)
      \right] \; u(k_1)~,
  \nl \nl
  V^{\alpha\beta}_{t,{\rm vert2}} & = &
    {\bar u}(-k_2) \;
    \frac{ \gamma^\alpha \, \dagg{q}_{\!t} \, \gamma^\mu
           \left(\dagg{q}_{\!t} - \dagg{p} \right) \gamma^\beta
           \left( 2 k_{1,\mu} - \dagg{p} \gamma_\mu \right) }
         { \left[ \rule[0cm]{0cm}{.35cm} p^2 - \ieps \right] \,
           \left[ \rule[0cm]{0cm}{.35cm} (p - q_t)^2 - \ieps
           \right] \,
           \left[ \rule[0cm]{0cm}{.35cm} p^2 - 2 k_1 p- \ieps \right] \,
           q_t^2}
    \nl & &         
 \hspace{2.2cm}
\times    2 \left[  L^2(e,Z) \;\left( 1+\gamma_5 \right)
            + R^2(e,Z) \;\left( 1-\gamma_5 \right)
      \right] \; u(k_1)~,
  \nl \nl
  V^{\alpha\beta}_{t,{\rm self}} & = &
    {\bar u}(-k_2) \;
    \frac{ \gamma^\alpha \, \dagg{q}_{\!t} \, \gamma^\mu
           \left(\dagg{q}_{\!t} - \dagg{p} \right) \gamma_\mu \,
           \dagg{q}_{\!t} \, \gamma^\beta }
         { \left[ \rule[0cm]{0cm}{.35cm} p^2 - \ieps \right] \,
           \left[ \rule[0cm]{0cm}{.35cm} (p - q_t)^2 - \ieps
           \right]
           \left[ \rule[0cm]{0cm}{.35cm} q_t^2 \right]^2 }
\nl & &     
\hspace{2.2cm}
 \times    2 \left[  L^2(e,Z) \;\left( 1+\gamma_5 \right)
            + R^2(e,Z) \;\left( 1-\gamma_5 \right)
      \right] \; u(k_1)~,
  \nl \nl
  V^{\alpha\beta}_{t,{\rm box}} & = &
    {\bar u}(-k_2)
  \nl & & \hspace{.4cm} \times
    \frac{ \left( -2 k_2^\mu - \gamma^\mu \dagg{p} \right)
           \gamma^\alpha \left(\dagg{q}_{\!t} - \dagg{p} \right)
           \gamma^\beta
           \left( 2 k_{1,\mu} - \dagg{p} \gamma_\mu \right) }
         { \left[ \rule[0cm]{0cm}{.35cm} p^2 - \ieps \right] \,
           \left[ \rule[0cm]{0cm}{.35cm} p^2 - 2 k_1 p - \ieps \right] \,
           \left[ \rule[0cm]{0cm}{.35cm} p^2 + 2 k_2 p - \ieps \right] \,
           \left[ \rule[0cm]{0cm}{.35cm} (p - q_t)^2 - \ieps \right] }
  \nl & & \hspace{.4cm} \times 
    2 \left[  L^2(e,Z) \;\left( 1+\gamma_5 \right)
            + R^2(e,Z) \;\left( 1-\gamma_5 \right)
      \right] \; u(k_1)~.
  \label{tloopparts}
\ea
\\
In the on-shell renormalization scheme, the $t$-channel counterterm
part in equation~(\ref{tloop}) originates from the
counterterms of two vertex and one electron self energy loops.
It reads 
\beq
  C^{\alpha\beta}_{t,Z} \; = \;
    \frac{B^{\alpha \beta}_{t,Z}}{q_t^2}
    \left( \rule[0cm]{0cm}{.35cm} 2 {\rm P} + 4 {\rm P^{IR}} - 4
    \right)
  \label{tloopct}
\eeq
with $B^{\alpha \beta}_{t,V}$ from equation~(\ref{tcurrent}) and the
ultraviolet and infrared poles
\\
\ba
  {\rm P} & \equiv & \frac{1}{n-4} + \frac{1}{2} \, \gamma_E +
               \myln \frac{\ME}{\mu} - \myln \left(2\sqrt{\pi}\right)
               \; \rule[-.45cm]{.02cm}{1.15cm}_{\;\mu=\ME}
  \hspace{1cm} n \; = \; 4 - \varepsilon,
  \label{poleUV} \\ \nl
  {\rm P^{IR}} & \equiv & \frac{1}{n-4} + \frac{1}{2}\, \gamma_E +
               \myln \frac{\ME}{\mu} - \myln \left(2\sqrt{\pi}\right)
               \; \rule[-.45cm]{.02cm}{1.15cm}_{\;\mu=\ME}
  \hspace{1cm} n \; = \; 4 + \varepsilon .
  \label{poleIR}
\ea
\\
Here, $\gamma_E$ is Euler's constant.

%
The $u$-channel virtual QED matrix element $\cm^V_{u,Z}$\ is obtained
from $\cm^V_{t,Z}$\ by the interchanges
\beq
  v_1 \leftrightarrow v_2, \hspace{2cm}
  \cm_{12,Z}^\mu \leftrightarrow \cm_{34,Z}^\mu .
  \label{bornmatsym}
\eeq
The symmetry~(\ref{bornmatsym}) implies that, after the angular
integrations, $\cm^V_{u,Z}$\ is not explicitly needed any more,
because the terms corresponding to the interference of $\cm^V_{u,Z}$\
with the tree level matrix element is obtained from the interference
of $\cm^V_{t,Z}$\ with the tree level matrix element by interchanging
\sone\ and \stwo.

The virtual initial state QED matrix element for the \ccthree\
$t$-channel is:
\beq
  \cm^V_{t,W} \; = \; \frac{g_{\beta \beta'}}{\,D_W(\SONE)\,}\,
                    \frac{g_{\alpha \alpha'}}{\,D_W(\STWO)\,}
                    \; V^{\alpha \beta}_{t,W}
                    M_{12,W}^{\beta'}\,M_{34,W}^{\alpha'}
\eeq
with
\\
\ba
  V^{\alpha\beta}_{t,W} & = & e^2 
  \left\{ 
     V^{\alpha\beta}_{t,{\rm aux}} 
     - \ri \left. \, \mu^{(4-n)} \! \int \! \frac{\d^n p}{(2\pi)^n}
           V^{\alpha\beta}_{t,{\rm box}}
    \, \right|_{\mu=m_e} \: \right\}
    \nl
  \label{tloopW}
\ea
\\
where
\beq
  V^{\alpha\beta}_{t,{\rm aux}} =
    \frac{\,C^{\alpha\beta}_{t,W}\,}{16\pi^2} 
    - \ri \left. \, \mu^{(4-n)} \! \int \! \frac{\d^n p}{(2\pi)^n}
    \left( V^{\alpha\beta}_{t,{\rm vert1}} +
           V^{\alpha\beta}_{t,{\rm vert2}} +
           V^{\alpha\beta}_{t,{\rm self}}\right) \right|_{\mu=m_e} .
\label{virtaux}
\eeq
The virtual and counter term contributions are those
of~(\ref{tloopparts}) to~(\ref{poleIR}) with corresponding $Z$ couplings
replaced by $W$ couplings.  
The electron mass which, in principle, occurs in the \nctwo\
$t$-channel propagators is absent in the charged current case where
a neutrino is exchanged in the $t$-channel instead.
Further, the $u$-channel diagrams are absent.

%
The virtual initial state QED matrix element for the $s$-channel
annihilation represented by the amputated Feynman diagram in
figure~\ref{svirt}
is given by
\beq
  \cm^V_{s,V^0} \: = \:
    \frac{g_{\gamma \gamma'}}{\,D_{V^0}(s)\,}\:
    \frac{g_{\beta \beta'}}{\,D_W(\SONE)\,}\:
    \frac{g_{\alpha \alpha'}}{\,D_W(\STWO)\,} \;
    V^{\gamma'}_{s,V^0}\,M_{12,W}^{\beta'}\,M_{34,W}^{\alpha'} \;
    g_3(V^0)\,T^{\alpha \beta \gamma}
\nl
\eeq
with
\beq
  V^{\gamma'}_{s,V^0} \; = \;
    e^2 \; \left\{ \,\frac{C^{\gamma'}_{s,V^0}}{16\pi^2} 
    \; - \; \ri \left. \, \mu^{(4-n)} \! \int \hspace{-.1cm}
    \frac{\d^n p}{(2\pi)^n} \; V^{\gamma'}_{s,{\rm vert}}\:
    \right|_{\mu=m_e} \right\}
  \label{sloop}
\eeq
and
\\
\ba
  V^{\gamma'}_{s,{\rm vert}} & = &
    {\bar u}(-k_2) \;
    \frac{ \left( -2 k_2^\mu - \gamma^\mu \dagg{p} \right)
           \gamma^{\gamma'}
           \left( 2 k_1^\mu - \dagg{p} \gamma^\mu \right) }
         { \left[ \rule[0cm]{0cm}{.35cm} p^2 + 2 k_2 p - \ieps \right]
           \left[ \rule[0cm]{0cm}{.35cm} p^2 - 2 k_1 p - \ieps \right]
           \left[ \rule[0cm]{0cm}{.35cm} p^2 - \ieps \right] }
 \nl & & \hspace{1.8cm}
\times \left[ \rule[0cm]{0cm}{.35cm} L(e,V^0) \left( 1\!+\!\gamma_5
\right)  + R(e,V^0) \left( 1\!-\!\gamma_5 \right) \right] u(k_1).
\ea
\\
The $s$-channel counterterm
part is a mere vertex counterterm given by
\beq
  C^{\gamma'}_{s,V^0} \; = \;
    B^{\gamma'}_{s,V^0} \: \left( 2 {\rm P} + 4 {\rm P^{IR}} - 4 \right)
\eeq
with $B^{\gamma'}_{s,V^0}$\ from equation~(\ref{scurrent}).
%
%
\subsection{Initial State Bremsstrahlung Amplitudes}
\label{bremamp}
%
As for the case of virtual initial state QED corrections, initial
state bremsstrahlung matrix elements are obtained by altering the
amplitudes in equations~(\ref{sZamp}) to~(\ref{uZamp}).
%
From the Feynman diagrams of figure~\ref{convISR},
one deduces the $t$-channel Bremsstrahlung matrix element for the
\nctwo\ case:
\beq
  \cm_{t,Z}^{R,\lambda} \; = \;
    \frac{g_{\beta \beta'}}{\,D_Z(\SONE)\,}\:
    \frac{g_{\alpha \alpha'}}{\,D_Z(\STWO)\,}\;
    M_{12,Z}^{\beta'}\,M_{34,Z}^{\alpha'} \;
    e \, R^{\:\!\alpha \beta \mu}_{t,Z} \, \varepsilon^\lambda_\mu(p)
  \label{brtmat}
\eeq
with
\\
\ba
  R^{\:\!\alpha \beta \mu}_{\:\!t,Z} & = & {\bar u}(-k_2) \;
    \left\{ \rule[0cm]{0cm}{.6cm} \: \gamma^\alpha
      \frac{\,\dagg{v}_2 - \dagg{k}_2\,}{t_2} \, \gamma^\beta
      \frac{\,2k_1^\mu - \dagg{p} \gamma^\mu\,}{z_1}
    \right.
  \nl & & \hspace{2cm} + \;
      \gamma^\alpha \frac{\,\dagg{v}_2 - \dagg{k}_2\,}{t_2} \,
      \gamma^\mu \frac{\,\dagg{k}_1 - \dagg{v}_1}{t_1} \, \gamma^\beta
  \nl & & \hspace{2cm} + \;
    \left. \rule[0cm]{0cm}{.6cm}
      \frac{\,\gamma^\mu \dagg{p} - 2k_2^\mu\,}{z_2} \, \gamma^\alpha
      \frac{\,\dagg{k}_1 - \dagg{v}_1}{t_1} \, \gamma^\beta
    \; \right\}
  \nl & & \hspace{1cm} \times
    2 \left[ L^2(e,Z) \left( 1+\gamma_5 \right)
             + R^2(e,Z)  \left( 1-\gamma_5 \right)
        \right] u(k_1)~.
\label{brtcurrent}
\ea
\\

%
An analogous result is obtained for the \nctwo\ $u$-channel initial
state bremsstrahlung matrix element,
\\
\ba
  \cm_{u,Z}^{R,\lambda} & = &
    \frac{g_{\alpha \alpha'}}{\,D_Z(\SONE)\,}\:
    \frac{g_{\beta \beta'}}{\,D_Z(\STWO)\,}\;
    M_{12,Z}^{\alpha'} \, M_{34,Z}^{\beta'} \;
    e \, R^{\:\!\alpha \beta \mu}_{\:\!u,Z} \,
    \varepsilon^\lambda_\mu(p) \nl \nl
  R^{\:\!\alpha \beta \mu}_{\:\!u,Z} & = & {\bar u}(-k_2) \;
    \left\{ \rule[0cm]{0cm}{.6cm} \: \gamma^\alpha
      \frac{\,\dagg{v}_1 - \dagg{k}_2\,}{u_2} \, \gamma^\beta
      \frac{\,2k_1^\mu - \dagg{p} \gamma^\mu\,}{z_1}
    \right.
  \nl & & \hspace{2cm} + \;
      \gamma^\alpha \frac{\,\dagg{v}_1 - \dagg{k}_2\,}{u_2} \,
      \gamma^\mu \frac{\,\dagg{k}_1 - \dagg{v}_2\,}{u_1} \,
      \gamma^\beta
  \nl & & \hspace{2cm} + \; \left. \rule[0cm]{0cm}{.6cm}
      \frac{\,\gamma^\mu \dagg{p} - 2k_2^\mu\,}{z_2} \, \gamma^\alpha
      \frac{\,\dagg{k}_1 - \dagg{v}_2\,}{u_1} \, \gamma^\beta \right\}
  \nl & & \hspace{1cm} \times
      2 \left[  L^2(e,Z) \left( 1+\gamma_5 \right)
              + R^2(e,Z) \left( 1-\gamma_5 \right)
       \right] u(k_1).
  \label{brumat}
\ea
\\

%
For the $s$-channel bremsstrahlung matrix element of the \ccthree\
process one has the Feynman diagrams of figure~\ref{annISR}
and obtains
\\
\ba
  \cm^{R,\lambda}_{s,V^0} & = &
    \frac{g_{\gamma \gamma'}}{\,D_{V^0}(s)\,}
    \frac{g_{\beta \beta'}}{\,D_W(\SONE)\,}
    \frac{g_{\alpha \alpha'}}{\,D_W(\STWO)\,} 
    M_{12,W}^{\beta'} M_{34,W}^{\alpha'} \,
    g_3(V^0)\,T^{\alpha \beta \gamma} \,
    e R^{\gamma' \mu}_{s,V^0} \,\varepsilon^\lambda_\mu(p)
  \nl
\ea
\\
with
\ba
  R^{\gamma' \mu}_{s,V^0} & = & {\bar u}(-k_2) 
    \left\{ \rule[0cm]{0cm}{.6cm} \gamma^{\gamma'}
            \frac{\,2k_1^\mu - \dagg{p} \gamma^\mu\,}{z_1} \; + \;
            \frac{\,\gamma^\mu \dagg{p} - 2k_2^\mu\,}{z_2}
            \gamma^{\gamma'}
    \right\} 
  \nl & & \hspace{1.7cm}\times
    \left[  \rule[0cm]{0cm}{.35cm}
            L(e,V^0) \left( 1+\gamma_5 \right)
          + R(e,V^0) \left( 1-\gamma_5 \right)
    \right]  u(k_1).
\ea

All the initial state currents introduced so far fulfill the condition
of U(1) current conservation,
\ba
\nl
  p_{\mu} R^{\gamma \mu}_{s,V^0} =
  p_{\mu} R^{\alpha \beta \mu}_{t} =
  p_{\mu} R^{\alpha \beta \mu}_{u} = 0.
\label{u1}
\ea

There is no \ccthree\ $u$-channel diagram.
For the \ccthree\ $t$-channel diagram, the situation is slightly more
complicated.
Naively, only the two left diagrams in figure~\ref{convISR} contribute
to the \ccthree\ $t$-channel ISR corrections.
For these two, current conservation is violated and may be established
by adding as an auxiliary current the additional Feynman diagram with
radiation from the $t$-channel propagator~\cite{WWnuni} (cf. the
discussion in section~\ref{cccase}).
Then, the charged current $t$-channel matrix element is just given
by~(\ref{brtmat}) with \zz\ couplings replaced by $W$\ couplings.

%
In the above matrix elements the radiatively changed Mandelstam
variables were introduced. In terms of the five-particle phase space
parametrization from appendix~\ref{ps2to5} they are given by
\\
\ba
  t_1 & = & \left( \rule[0cm]{0cm}{.35cm} k_1 - v_1 \right)^2 \;+\;m^2_v
        \; = \; a_1^t \; + \; b_1 \, \mycos\theta_R \; + \;
            c \, \mysin\theta_R \mycos\phi_R , \nl
  t_2 & = & \left( \rule[0cm]{0cm}{.35cm} v_2 - k_2 \right)^2 \;+\;m^2_v
        \; = \; a_2^t \; + \; b_2 \, \mycos\theta_R \; + \;
                  c \, \mysin\theta_R \mycos\phi_R , \nl
  u_1 & = & \left( \rule[0cm]{0cm}{.35cm} k_1 - v_2 \right)^2 \;+\;m^2_v
        \; = \; a_1^u \; - \; b_1 \, \mycos\theta_R \; - \;
                  c \, \mysin\theta_R \mycos\phi_R , \nl
  u_2 & = & \left( \rule[0cm]{0cm}{.35cm} v_1 - k_2 \right)^2 \;+\;m^2_v
        \; = \; a_2^u \; - \; b_2 \, \mycos\theta_R \; - \;
                  c \, \mysin\theta_R \mycos\phi_R .
\label{tuz1}
\ea
\\
Here $m_v=m_e$ for \nctwo~and $m_v=m_{\nu}=0$ for \ccthree.
In (\ref{tuz1}), the kinematical parameters
\\
\ba
  a_1^t & = &m^2_v \;-\;m^2_e
  -\,\SONE \; + \; \frac{s'+\delta}{\,4\,s'\,} \, \left(\sprp
              \; - \; \sprm \, \beta \, \mycos\theta \right) , \nl
  a_2^t & = &m^2_v \;-\;m^2_e
  -\,\STWO \; + \; \frac{s'-\delta}{\,4\,s'\,} \, \left(\sprp
              \; + \; \sprm \, \beta \, \mycos\theta \right) , \nl
  a_1^u & = &m^2_v \;-\;m^2_e
  -\,\STWO \; + \; \frac{s'-\delta}{\,4\,s'\,} \, \left(\sprp
              \; - \; \sprm \, \beta \, \mycos\theta \right) , \nl
  a_2^u & = &m^2_v \;-\;m^2_e
  -\,\SONE \; + \; \frac{s'+\delta}{\,4\,s'\,} \, \left(\sprp
              \; + \; \sprm \, \beta \, \mycos\theta \right) , \nl \nl
  b_1 & = & \frac{\SLAMP}{\,4\,s'\,}
            \left( \sprp \, \beta \, \mycos\theta  - \sprm \right) , \nl
  b_2 & = & \frac{\SLAMP}{\,4\,s'\,}
            \left( \sprp \, \beta \, \mycos\theta  + \sprm \right) ,
            \nl \nl
  c & = & - \, \frac{\SLAMP}{2} \, \sqrt{\frac{\,s\,}{s'}} \, \beta \,
          \mysin\theta 
\label{abctu}
\ea
\\
are derived from the representations~(\ref{kvectrad}) and (\ref{v5vect}) in
appendix~\ref{ps2to5}.
The variables $\delta, \sprm, \sprp, \SLAMP$ are introduced in
sections~\ref{xsISR1} and~\ref{xsISR2}.

The two invariants related to radiation from external legs are
\ba
  z_1 & = & -2\,k_1 p \; = \;
            \frac{\sprm}{\,2\,} \left( \rule[0cm]{0cm}{.35cm}
            1+\beta\mycos\theta \right) \;\;\; \equiv \;\;\;
            \frac{\sprm}{\,2\,} \; {\bar z_1} , \nl
  z_2 & = & -2\,k_2 p \; = \;
            \frac{\sprm}{\,2\,} \left( \rule[0cm]{0cm}{.35cm}
            1-\beta\mycos\theta \right) \;\;\; \equiv \;\;\;
            \frac{\sprm}{\,2\,} \; {\bar z_2} .
\label{tuz2}
\ea
\\
The equations (\ref{abctu}) and (\ref{tuz2}) are exact. 
In (\ref{tuz2}), one has to keep the electron mass, since it gives rise
to mass singularities. The electron mass in (\ref{abctu}) may be
neglected, but it was retained in some integrals for the sake of
numerical stability.
%
%
\section{Phase Space Integrals}
\label{appints}
\ezero
In this appendix, we collect the analytical solutions of the integrals
needed for the semi-analytical \xsec\ calculation of
processes~(\ref{4fproc}).
These integrals are necessary to obtain the twofold differential
cross-section $\d^2\sigma/ \d\SONE \d\STWO$ in the case of virtual ISR
and the threefold differential \xsec\ $\d^3\sigma/ds' \d\SONE \d\STWO$
for initial state bremsstrahlung.
Integrals are presented in the ultrarelativistic approximation,
i.e. the electron mass is neglected wherever possible.
The integrals have been numerically checked.
The azimuth around the beam yields a factor $2\pi$.
We do not elaborate on the pure $s$-channel in this appendix, because
it is well-known that its initial state radiation is fully covered by
the universal ISR corrections.
%
%
\subsection{Fermion Decay Angle Tensor Integrals}
\label{tenint}
Integrations over final state fermion decay angles are carried out
with the help of invariant tensor integration.
Using the decay matrix element $M_{12,V}^\mu$\ from
equation~(\ref{fcurrent}) one obtains
\\
\ba
  \lefteqn{ \sum_{Spins} \, \int \d \Gamma_{12} \:
     M_{12,V}^\mu {M_{12,V}^\nu}^{\ds \!\!*} }
  \nl & & \equiv \;
    \frac{\sqrt{\lambda(\SONE,m_1^2,m_2^2)}}{8 \, \SONE} \:
    \sum_{Spins} \: \int_{-1}^{+1} \d \mycos\theta_1 \:
    \int_0^{2\pi} \d \phi_1 \: M_{12,V}^\mu {M_{12,V}^\nu}^{\ds \!\!*}
  \nl & & = \;
    \frac{\sqrt{\lambda(\SONE,m_1^2,m_2^2)}}{8 \, \SONE}
    \; \frac{4\pi}{3} \left[ L^2(f_1,V) + R^2(f_1,V) \right]
  \nl & & \hspace{3.5cm} \times\;
    \left[ \rule[0cm]{0cm}{.35cm} \SONE \, g^{\mu\nu} +
      \left(p_1^\mu + p_2^\mu\right) \left(p_1^\nu + p_2^\nu\right)
    \right]
  \label{dectenint}
\ea
\\
with $L(f_1,V)$\ and $R(f_1,V)$\ from~(\ref{LRcoup})
and~(\ref{Wcoup}).
An analogous result is valid for the integration of
$M_{34,V}^\mu {M_{34,V}^\nu}^{\ds \!\!*}$\ over $\d \Gamma_{34}$.
The result of the above decay phase space integral factorizes in the
cross-section.
Therefore one can use formula~(\ref{dectenint}) to integrate
the final state fermion decay angles for both the $2\!\to\!4$\ and the
initial state bremsstrahlung $2\!\to\!5$\ particle phase space.
%
%
%
\subsection{Loop Integrals}
\label{loopint1}
For the computation of the virtual initial state QED corrections to
the studied processes, apart from the integration over
final state fermion decay angles, two sets of integrals are needed.
The first set is presented in this subsection and consists of the
integration over the loop momentum $p$:
\beq
  \left[A\right]_n \; = \;
    \left. \, \mu^{(4-n)} \! \int \frac{\d^n p}{(2\pi)^n} \; A
    ~\right|_{\mu=m_e}
    \hspace{2.1cm} {\rm with} \hspace{.5cm}
    \left| n-4 \right| \; \ll \; 1 .
\eeq

The second set, the integration over $\mycos \vartheta$ (in $2 \!\to\! 4$\
kinematics), is given in section~\ref{loopint2}.
It should be noticed that we present only the loop integrals for the
virtual $s$- and $t$-channel graphs, because, in the twofold
differential \xsec, the interferences of the virtual $u$-channel
graphs with the tree level $t$- and $u$-channel graphs equal the
interferences of the virtual $t$-channel graphs with the tree level
$u$-channel and $t$-channel graphs.
This is easily verified by symmetry arguments.
We use the definitions
\\
\ba
  \hspace*{.75cm} \nl
  \frac{1}{\Pi_0} & = & \frac{1}{p^2-\ieps} \nl
  \frac{1}{\Pi_1} & = & \frac{1}{(p-k_1)^2+\ME^2-\ieps} \; = \;
                        \frac{1}{p^2-2pk_1-\ieps} \nl
  \frac{1}{\Pi_2} & = & \frac{1}{(p+k_2)^2+\ME^2-\ieps} \; = \;
                        \frac{1}{p^2+2pk_2-\ieps} \nl
  \frac{1}{\Pi_q} & = & \frac{1}{(p-q_t)^2-\ieps}
\ea
\\
and $q_t^2\!=\!t$.
For later use we define $q_u^2\!=\!u$.
Dimensional regularization is used to consistently treat divergent
loop integrals.
The ultraviolet and infrared poles are taken from
equations~(\ref{poleUV}) and~(\ref{poleIR}).
For the sake of compact presentation of the subsequent integrals we
introduce two symmetry operations, namely
\\
\ba
  \ct\left[\rule[0cm]{0cm}{.35cm} f(\SONE,\STWO)\right] & \equiv &
   f(\STWO,\SONE) \\
  \cu\left[\rule[0cm]{0cm}{.35cm} g(k_1,k_2)\right] & \equiv &
   g(-k_2,-k_1).
\ea
\\
Further we introduce the notations
\\
\ba
  \lz & \equiv & \myln \left( -\frac{s}{\ME^2-\ieps} \right)
      \;\; = \;\;
      \myln \frac{s}{\ME^2} - \ri\pi \; \equiv \; l_\beta - \ri\pi~,
  \nl
  l_1 & \equiv & \myln \left( -\frac{\SONE}{\ME^2-\ieps} \right)
      \;\; = \;\;
      \myln \frac{\SONE}{\ME^2} - \ri\pi~,
  \nl
  l_2 & \equiv & \myln \left( -\frac{\STWO}{\ME^2-\ieps} \right)
      \;\; = \;\;
      \myln \frac{\STWO}{\ME^2} - \ri\pi~,
  \nl \rule[-.1cm]{0cm}{.8cm}
  l_t & \equiv & \myln\frac{q^2}{\ME^2}~.
  \label{logdefs}
\ea
\\

One obtains the following table of canonical integrals:
\bigskip
$\begin{array}{llll}
 1) & {\ds \left[ \frac{1}{\Pi_0\Pi_1} \right]_n } & = &
      {\ds \left[ \frac{1}{\Pi_0\Pi_2} \right]_n } \; = \;
      {\ds \iospi \left( \rule[0cm]{0cm}{.4cm}
        -2{\rm P} + 2 \right) } \hugeskip
 2) & {\ds \left[ \frac{p^{\,\mu}}{\Pi_0\Pi_1} \right]_n } & = & \;
      {\ds \cu \left( \rule[0cm]{0cm}{.4cm}
        \left[ \frac{p^{\,\mu}}{\Pi_0\Pi_2} \right]_n \right) } \;=\;
      {\ds \frac{\,\ri \; k_1^{\,\mu}}{16\,\pi^2}
           \left( \rule[-.1cm]{0cm}{.7cm} \!\!
                 -{\rm P} + \frac{1}{2}\right) } \hugeskip
 3) & {\ds \left[ \frac{1}{\Pi_0\Pi_q} \right]_n } & = &
       {\ds \iospi \left( \rule[0cm]{0cm}{.4cm}
         - 2{\rm P} - l_t + 2 \right) }
    \hugeskip
 4) & {\ds \left[ \frac{p^{\,\mu}}{\Pi_0\Pi_q} \right]_n } & = & \;
      {\ds \frac{q^\mu}{2} \;
        \left[ \frac{1}{\Pi_0\Pi_q} \right]_n } \;=\;
      {\ds \frac{\,\ri \; q^{\,\mu}}{16\,\pi^2}
        \left( \rule[-.1cm]{0cm}{.7cm}
              - {\rm P} - \frac{l_t}{2} + 1 \right) }
\end{array}$
\newline
$\begin{array}{llll}
 5) & {\ds \left[ \frac{1}{\Pi_1\Pi_2} \right]_n } & = &
       {\ds \iospi \left( \rule[0cm]{0cm}{.4cm}
         - 2{\rm P} - \lz + 2 \right) }
       \bigskip
 6) & {\ds \left[ \frac{p^{\,\mu}}{\Pi_1\Pi_2} \right]_n } & = &
      {\ds \frac{\left(k_1 - k_2 \right)^\mu}{2} \;
        \left[ \frac{1}{\Pi_1\Pi_2} \right]_n } \;=\;
      {\ds \iospi \left( k_1 - k_2 \right)^\mu
        \left( \rule[-.1cm]{0cm}{.7cm}
              - {\rm P} - \frac{\lz}{2} + 1 \right) }
       \bigskip
 7) & {\ds \left[ \frac{1}{\Pi_1\Pi_q} \right]_n } & = & \;
      {\ds \ct \left( \rule[0cm]{0cm}{.7cm}
        \left[ \frac{1}{\Pi_2\Pi_q} \right]_n \right) } \; = \;
      {\ds \iospi \left( \rule[0cm]{0cm}{.4cm}
         - 2{\rm P} - l_1 + 2 \right) }
       \bigskip
 8) & {\ds \left[ \frac{p^{\,\mu}}{\Pi_1\Pi_q} \right]_n } & = & \;
      {\ds \ct\cu \left( \rule[0cm]{0cm}{.7cm}
        \left[ \frac{p^{\,\mu}}{\Pi_2\Pi_q} \right]_n \right) } \;=\;
      {\ds \iospi \left( q^{\,\mu} + k_1^{\,\mu} \right)
        \left( \rule[0cm]{0cm}{.4cm}
              -{\rm P} - \frac{l_1}{2} +1 \right) }
       \bigskip
 9) & {\ds \left[ \frac{1}{\Pi_0\Pi_1\Pi_2} \right]_n } & = &
       {\ds \iospi \left[ \rule[-.1cm]{0cm}{.8cm}
         - \frac{2\,\lz}{s} \; {\rm P^{IR}} \; + \;
         \frac{1}{s} \left( \frac{\,\pi^2}{6} - \frac{\lz^2}{2}
       \right)
         \right] }
       \bigskip
 10) & {\ds \left[ \frac{p^{\,\mu}}{\Pi_0\Pi_1\Pi_2} \right]_n } & =
 &
       {\ds - \iospi \left( k_1 - k_2 \right)^\mu \; \frac{\lz}{s}
         }
       \bigskip
 11) & {\ds \left[ \frac{p^{\,\mu} p^{\,\nu}}
                         {\Pi_0\Pi_1\Pi_2} \right]_n } & = &
        {\ds \iospi \left[ \rule[-.1cm]{0cm}{.8cm} \;
          g^{\mu\nu} \; \left( -\frac{1}{2} {\rm P}
          - \frac{\lz}{4} + \frac{3}{4} \right) \right. } \\
      & & & \hspace{.8cm} {\ds \left. \rule[-.1cm]{0cm}{.8cm} + \;
         \left( k_1^{\,\mu} k_1^{\,\nu} + k_2^{\,\mu} k_2^{\,\nu}
       \right)
         \; \frac{1-\lz}{2s} \; + \;
         \left( k_1^{\,\mu} k_2^{\,\nu} + k_2^{\,\mu} k_1^{\,\nu}
       \right)
         \; \frac{1}{2s}
         \; \right] }
       \bigskip
 12) & {\ds \left[ \frac{1}{\Pi_0\Pi_1\Pi_q} \right]_n } & = & \;
       {\ds \ct \left( \rule[0cm]{0cm}{.7cm}
         \left[ \frac{1}{\Pi_0\Pi_2\Pi_q} \right]_n \right) }
     \smskip & & = &
       {\ds \iospi \; \frac{1}{t+\SONE} \;
             \left[ \, \rule[-.1cm]{0cm}{.8cm}
                 \frac{1}{2}
                 \left( \rule[0cm]{0cm}{.4cm} l_t \!-\! l_1 \right) \!
                 \left( \rule[0cm]{0cm}{.4cm} 3l_1 \!-\! l_t \right)
          - 2 \, \mysp \left( \rule[-.1cm]{0cm}{.7cm}
          \frac{\SONE \!+\! t \!-\! \ieps}{\SONE} \right) \,
        \right] }
       \bigskip
 13) & {\ds \left[ \frac{p^{\,\mu}}{\Pi_0\Pi_1\Pi_q} \right]_n } & = & \;
       {\ds \ct\cu \left( \rule[0cm]{0cm}{.7cm}
         \left[ \frac{p^{\,\mu}}{\Pi_0\Pi_2\Pi_q} \right]_n \right) }
       \smskip & & = &
       {\ds \iospi \; \frac{1}{t+\SONE} \; \left\{
          \rule[-.3cm]{0cm}{1.2cm} \;
          q^{\,\mu} \left( \rule[0cm]{0cm}{.4cm} l_t - l_1 \right)
          \right. } \\
      & & & {\ds \hspace{1cm} \left. + \;
          k_1^{\,\mu} \left( \rule[-.1cm]{0cm}{.8cm} - l_1 \; + \; t
          \!
          \; \!\!
          \left[ \frac{1}{\Pi_0\Pi_1\Pi_q} \right]_n
          - \; \frac{2t}{t+\SONE}
          \left( \rule[0cm]{0cm}{.4cm} l_t - l_1 \right) \right)
          \rule[-.3cm]{0cm}{1.2cm} \right\} }
\end{array}$
$\begin{array}{llll}
 14) & {\ds \left[ \frac{p^{\,\mu}p^{\,\nu}}{\Pi_0\Pi_1\Pi_q}
         \right]_n } & = & \;
       {\ds \ct\cu \left( \rule[0cm]{0cm}{.7cm}
         \left[ \frac{p^{\,\mu}p^{\,\nu}}{\Pi_0\Pi_2\Pi_q}
         \right]_n \right) }
       \smskip & & = &
       {\ds \iospi \; \left\{ \rule[-.3cm]{0cm}{1.2cm}
          \; g^{\mu\nu} \;
          \left( \!\! \rule[-.2cm]{0cm}{1cm} -
          \frac{1}{2} {\rm P} + \frac{3}{4} \; - \;
          \frac{1}{4\left(t+\SONE\right)} \left( \rule[0cm]{0cm}{.4cm}
          t \; l_t + \SONE \; l_1 \right) \right) \right. } \\
      & & & {\ds + \;
          \frac{k_1^{\,\mu} k_1^{\,\nu}}{t+\SONE} \;
          \left\{ \rule[-.2cm]{0cm}{1cm} \;
          \left( \frac{1}{2} + \frac{t}{t+\SONE} \right)
          \left( \rule[0cm]{0cm}{.4cm} 1 - l_1 \right) \right.
                                 - \;
          \frac{3\, t^2}{\left( t+\SONE \right)^2}
          \left( \rule[0cm]{0cm}{.4cm} l_t - l_1 \right) } \\
      & & & {\ds \left. \hspace{5.7cm} + \;
          \frac{t^2}{t+\SONE} \;
          \left[ \frac{1}{\Pi_0\Pi_1\Pi_q} \right]_n
          \rule[-.2cm]{0cm}{1cm} \; \right\} } \\
      & & & {\ds + \;
          \frac{\,k_1^{\,\mu} q^{\,\nu} + k_1^{\,\nu} q^{\,\mu}}
               {2 \left(t+\SONE\right)} \;
          \left\{ \rule[-.2cm]{0cm}{1cm} \frac{t}{t+\SONE}
          \left(\rule[0cm]{0cm}{.4cm} l_t - l_1 \right) \; - \; 1
          \right\} } \\
      & & & {\ds \left.\;+ \;
          \frac{\,q^{\,\mu} q^{\,\nu}}{2 \left(t+\SONE\right)} \;
          \left(\rule[0cm]{0cm}{.4cm} l_t - l_1 \right)
          \rule[-.3cm]{0cm}{1.2cm} \;\right\} }
     \bigskip
 15) & {\ds \left[ \frac{1}{\Pi_1\Pi_2\Pi_q} \right]_n } & = &
       {\ds\rm \iospi \; I_{12q} }
     \bigskip
 16) & {\ds \left[ \frac{p^{\,\mu}}{\Pi_1\Pi_2\Pi_q} \right]_n }
     & = &
        {\ds \frac{\rm i}{16\,\pi^2 \; \lambda} }
        \\
     & & & \times {\ds \left\{ \rule[-.3cm]{0cm}{1.2cm} \;
          q^{\,\mu} \left[ \rule[-.2cm]{0cm}{1cm}
          s \left(\rule[-.05cm]{0cm}{.4cm} s-\sigma\right) {\rm
            I_{12q}}
          \; - \; \left( s+\delta \right) \myln \frac{\SONE}{s}
          \; - \; \left( s-\delta \right) \myln \frac{\STWO}{s}
          \right] \right. } \\
     & & & {\ds \;\;\; + \; k_1^{\,\mu} \left[ \rule[-.2cm]{0cm}{1cm}
          -\STWO\left(\rule[-.05cm]{0cm}{.4cm} s+\delta\right)
          {\rm I_{12q}} \; + \; \left( s-\sigma \right) \myln
          \frac{\SONE}{s} \; + \; 2\,\STWO \myln \frac{\STWO}{s}
          \right] } \\
     & & & {\ds \left. \hspace{.31cm} + \; k_2^{\,\mu}
          \left[ \rule[-.2cm]{0cm}{1cm}
          \SONE\left(\rule[-.05cm]{0cm}{.4cm} s-\delta\right)
          {\rm I_{12q}} \; - \; 2\,\SONE \myln \frac{\SONE}{s} \; - \;
          \left( s-\sigma \right) \myln \frac{\STWO}{s} \right]
          \rule[-.3cm]{0cm}{1.2cm} \; \right\} }
     \bigskip
 17) & {\ds \left[ \frac{p^{\,\mu} p^{\,\nu}}
                         {\Pi_1\Pi_2\Pi_q} \right]_n } & = &
        {\ds \iospi \;\left\{ \rule[-.1cm]{0cm}{.6cm}
          \;\: g^{\mu\nu} \; F_{21} \; + \;
          q^{\,\mu} q^{\,\nu} \; F_{22} \; + \;
          k_1^{\,\mu} k_1^{\,\nu} \; F_{23} \; + \;
          k_2^{\,\mu} k_2^{\,\nu} \; F_{24} \right. } \\
      & & & \hspace{1.35cm} {\ds \; + \;
          \left(\rule[-.05cm]{0cm}{.4cm}
          k_1^{\,\mu} k_2^{\,\nu} + k_1^{\,\nu} k_2^{\,\mu} \right)
          \; F_{25} \; + \;
          \left(\rule[-.05cm]{0cm}{.4cm}
          k_1^{\,\mu} q^{\,\nu} + k_1^{\,\nu} q^{\,\mu} \right)
          \; F_{26} } \\
      & & & \hspace{1.31cm} {\ds \left. \; + \;
          \left(\rule[-.05cm]{0cm}{.4cm}
          k_2^{\,\mu} q^{\,\nu} + k_2^{\,\nu} q^{\,\mu} \right)
          \; F_{27} \hspace{.3cm} \rule[-.1cm]{0cm}{.6cm}
          \right\} }
        \bigskip
      & \hspace{1.5cm} F_{21} & = &
          {\ds \frac{1}{4}
            \left\{ \rule[-.3cm]{0cm}{1.2cm} \; \frac{1}{\lambda}
            \left[ \rule[-.2cm]{0cm}{1cm}
            \SONE\!\left(\rule[-.05cm]{0cm}{.4cm} s \!-\! \delta
          \right)
            \myln \frac{\SONE}{s} \; + \;
            \STWO\!\left(\rule[-.05cm]{0cm}{.4cm} s \!+\! \delta
          \right)
            \myln \frac{\STWO}{s} \; - \;
            2 s\SONE\STWO \, {\rm I_{12q}} \right]
            \right. } \\
      & & & \hspace{.7cm} {\ds \left. \rule[-.3cm]{0cm}{1.2cm}
          - 2 {\rm P} + 3 - l_0 \rule[-.3cm]{0cm}{1.2cm} \;
          \right\} }
\end{array}$
\clearpage
$\begin{array}{llll}
      & \hspace{1.6cm} F_{22} & = &
          {\ds \frac{1}{\lambda}
            \left\{ \rule[-.3cm]{0cm}{1.2cm} -s \; - \:
            \frac{1}{2} \left[ \rule[-.2cm]{0cm}{1cm} 3s \, + \,
            \delta  \, + \, \frac{6s\SONE}{\lambda}
            \left(\rule[-.05cm]{0cm}{.4cm} s \!-\! \delta \right)
            \right] \; \myln \frac{\SONE}{s} \right. } \\
      & & & \hspace{1.52cm} {\ds \; - \;\,
            \frac{1}{2} \left[ \rule[-.2cm]{0cm}{1cm} 3s \, - \,
            \delta  \, + \, \frac{6s\STWO}{\lambda}
            \left(\rule[-.05cm]{0cm}{.4cm} s \!+\! \delta \right)
            \right] \; \myln \frac{\STWO}{s} } \\
      & & & \hspace{1.6cm} {\ds \left. + \;
            s^2 \left[ \rule[-.2cm]{0cm}{1cm}
            1 \, + \, \frac{6\,\SONE\STWO}{\lambda} \right] \;
            {\rm I_{12q}} \;
            \rule[-.3cm]{0cm}{1.2cm} \right\} }
        \medskip
      & \hspace{1.6cm} F_{23} & = &
          {\ds \frac{1}{\lambda}
            \left\{ \rule[-.3cm]{0cm}{1.2cm} -\STWO \; + \:
            \frac{1}{2} \left[ \rule[-.2cm]{0cm}{1cm}
            s \, - \,  \SONE \, - \, 3\STWO \, - \,
            \frac{6\,\SONE\STWO}{\lambda}
            \left(\rule[-.05cm]{0cm}{.4cm} s \!-\! \delta \right)
            \right] \; \myln \frac{\SONE}{s} \right. } \\
      & & & \hspace{.76cm} {\ds \left. - \;
            \frac{3\,\STWO^2}{\lambda}
            \left(\rule[-.05cm]{0cm}{.4cm} s \!+\! \delta \right)
             \myln \frac{\STWO}{s} \;\; + \;\;
            \STWO^2 \left[ \rule[-.2cm]{0cm}{1cm}
            1 \, + \, \frac{6\,s\SONE}{\lambda} \right] \;
            {\rm I_{12q}} \;
            \rule[-.3cm]{0cm}{1.2cm} \right\} }
        \medskip
      & \hspace{1.6cm} F_{24} & = &
          {\ds \frac{1}{\lambda}
            \left\{ \rule[-.3cm]{0cm}{1.2cm} -\SONE \; + \:
            \frac{1}{2} \left[ \rule[-.2cm]{0cm}{1cm}
            s \, - \,  \STWO \, - \, 3\SONE \, - \,
            \frac{6\,\STWO\SONE}{\lambda}
            \left(\rule[-.05cm]{0cm}{.4cm} s \!+\! \delta \right)
            \right] \; \myln \frac{\STWO}{s} \right. } \\
      & & & \hspace{.76cm} {\ds \left. - \;
            \frac{3\,\SONE^2}{\lambda}
            \left(\rule[-.05cm]{0cm}{.4cm} s \!-\! \delta \right)
             \myln \frac{\SONE}{s} \;\; + \;\;
            \SONE^2 \left[ \rule[-.2cm]{0cm}{1cm}
            1 \, + \, \frac{6\,s\STWO}{\lambda} \right] \;
            {\rm I_{12q}} \;
            \rule[-.3cm]{0cm}{1.2cm} \right\} }
        \medskip
      & \hspace{1.6cm} F_{25} & = &
          {\ds \frac{1}{2\lambda}
            \left\{ \rule[-.3cm]{0cm}{1.2cm} \; s - \sigma \; + \;
            \SONE \left[ \rule[-.2cm]{0cm}{1cm}
            1 \, + \, \frac{6\,\STWO}{\lambda}
            \left(\rule[-.05cm]{0cm}{.4cm} s \!+\! \delta \right)
            \right] \; \myln \frac{\SONE}{s} \right. } \\
      & & & \hspace{2.1cm} {\ds \; + \hspace{.195cm}
            \STWO \left[ \rule[-.2cm]{0cm}{1cm}
            1 \, + \, \frac{6\,\SONE}{\lambda}
            \left(\rule[-.05cm]{0cm}{.4cm} s \!-\! \delta \right)
            \right] \; \myln \frac{\STWO}{s} } \\
      & & & \hspace{2.165cm} {\ds \left. + \; 2\,\SONE\STWO
            \left[ \rule[-.2cm]{0cm}{1cm}
            1 \, - \, \frac{3\,s}{\lambda}
            \left(\rule[-.05cm]{0cm}{.4cm} s \!-\! \sigma \right)
            \right] \; {\rm I_{12q}} \rule[-.3cm]{0cm}{1.2cm}
            \rule[-.3cm]{0cm}{1.2cm} \; \right\} }
      \bigskip
      & \hspace{1.6cm} F_{26} & = &
          {\ds \frac{1}{\lambda}
            \left\{ \rule[-.3cm]{0cm}{1.2cm} \;\; \frac{s-\delta}{2}
            \; + \; \left[ \rule[-.2cm]{0cm}{1cm}
            \frac{s+\STWO}{2} \, + \, \frac{6\,s\SONE\STWO}{\lambda}
            \right] \; \myln \frac{\SONE}{s} \right. } \\
      & & & \hspace{2.13cm} {\ds  - \hspace{.145cm}
            \left[ \rule[-.2cm]{0cm}{1cm}
            \frac{\STWO}{2} \, - \, \frac{3\,s\STWO}{\lambda}
            \left(\rule[-.05cm]{0cm}{.4cm} s \!-\! \sigma \right)
            \right] \; \myln \frac{\STWO}{s} } \\
      & & & \hspace{2.115cm} {\ds \left. - \; s\STWO
            \left[ \rule[-.2cm]{0cm}{1cm}
            1 \, + \, \frac{3\,\SONE}{\lambda}
            \left(\rule[-.05cm]{0cm}{.4cm} s \!-\! \delta \right)
            \right] \; {\rm I_{12q}}
            \rule[-.3cm]{0cm}{1.2cm} \; \right\} }
\end{array}$
\clearpage
$\begin{array}{llll}
      & \hspace{1.8cm} F_{27} & = &
          {\ds - \frac{1}{\lambda}
            \left\{ \rule[-.3cm]{0cm}{1.2cm} \;\; \frac{s+\delta}{2}
            \; + \; \left[ \rule[-.2cm]{0cm}{1cm}
            \frac{s+\SONE}{2} \, + \, \frac{6\,s\SONE\STWO}{\lambda}
            \right] \; \myln \frac{\STWO}{s} \right. } \\
      & & & \hspace{2.45cm} {\ds  - \hspace{.145cm}
            \left[ \rule[-.2cm]{0cm}{1cm}
            \frac{\SONE}{2} \, - \, \frac{3\,s\SONE}{\lambda}
            \left(\rule[-.05cm]{0cm}{.4cm} s \!-\! \sigma \right)
            \right] \; \myln \frac{\SONE}{s} } \\
      & & & \hspace{2.43cm} {\ds \left. - \; s\SONE
            \left[ \rule[-.2cm]{0cm}{1cm}
            1 \, + \, \frac{3\,\STWO}{\lambda}
            \left(\rule[-.05cm]{0cm}{.4cm} s \!+\! \delta \right)
            \right] \; {\rm I_{12q}}
            \rule[-.3cm]{0cm}{1.2cm} \; \right\} }
        \bigskip
 18) & {\ds \left[ \frac{2\,p\;q}
                         {\Pi_0\Pi_1\Pi_2\Pi_q} \right]_n } & = &
        {\ds \iospi \; \left\{ \rule[-.3cm]{0cm}{1.2cm}
          {\rm I_{12q}} \; + \;
          \frac{1}{s} \left[ \rule[-.2cm]{0cm}{.9cm}
          l_0 \left(\rule[-.05cm]{0cm}{.4cm} l_1 + l_2 - 2\,l_t
        \right)
          \; + \; \frac{1}{2} \, l_-^2 \right]
          \right\} }
        \bigskip
\end{array}$
%
%
\subsection{Virtual Corrections' Phase Space Integrals}
\label{loopint2}
After the integration over the loop momentum, there is one non-trivial
integration left when the virtual initial state corrections to the
\nctwo\ or \ccthree\ processes are evaluated.
This is the integration over the boson scattering angle $\vartheta$ in
the center of mass system.
We recall that the below integrals do not cover the interferences of
virtual $u$-channel graphs with tree level $t$- and $u$-channel
graphs, because these interferences are deduced from the interferences
of the virtual $t$-channel graphs with the tree level $u$-channel and
$t$-channel graphs by symmetry arguments.
We introduce the notation
\beq
  \left[A\right]_V \;\;\; \equiv \;\;\; \frac{\SLAM}{2}
    \int\limits_{-1}^{+1} \d \! \mycos\vartheta~A \;\;\; = \;\;
    \int\limits_{t_{min}}^{t_{max}} \d t~A
\eeq
with $t_{min}$\ and $t_{max}$\ as defined in
equation~(\ref{nunivnota1}).

Below, we use
\\
\ba
 \TONE & \equiv & \SONE +t \nl
 \TTWO & \equiv & \STWO +t
\ea
\\
and adopt the notations of equations~(\ref{nunivnota1})
and~(\ref{nunivnota2}).
The following list of integrals is obtained:
\bigskip
$\begin{array}{ll}
 1) &  \left[\, 1 \, \right]_V \; = \; \SLAM
    \medskip
 2) & \left[\, t \, \right]_V \: = \:
       {\ds \frac{\SLAM}{2}\; \left(s-\sigma\right) }
    \medskip
 3) & \,\! \left[\, t^2 \,\right]_V \: = \: {\ds \frac{\SLAM}{3}
       \;
       \left\{ \left(s \! - \!\sigma\right)^2 - \SONE\STWO \right\} }
    \medskip
 4) & {\ds \left[ \frac{1}{\,t\,} \right]_V \: = \:
           \left[ \frac{1}{\,u\,} \right]_V \: = \: }
      {\ds \myln
         \frac{s\!-\!\sigma\!+\!\SLAM}{s\!-\!\sigma\!-\!\SLAM}
       \; = \; \cl_0 }
    \medskip
 5) & {\ds \left[ \frac{1}{t^2} \right]_V } \: = \:
 {\ds \left[ \frac{1}{u^2} \right]_V }  = 
      {\ds \frac{\SLAM}{\SONE\STWO} }
     \medskip
 6) & {\ds \left[ \frac{1}{t \; u} \right]_V \: = \:
        \frac{1}{s-\sigma} \left( \rule[-.1cm]{0cm}{.8cm}
                 \left[ \frac{1}{\,t\,} \right]_V \, + \,
                 \left[ \frac{1}{\,u\,} \right]_V \right) \: = \:
       \frac{2}{s-\sigma} \; \cl_0 }
\end{array}$

The above integrals 1) to 6) are familiar from the calculation of the
tree level \xsec, while the integrals 7) to 29) genuinely originate
from the virtual corrections.

$\begin{array}{llll}
 7) & {\ds \left[ \frac{1}{t^2_{12}} \right]_V \: = \:
         \ct \left( \rule[0cm]{0cm}{.7cm}
                    \left[ \frac{1}{t^2_{34}} \right]_V \right) \: = \:
         \frac{\SLAM}{s\SONE} }
     \medskip
 8) & {\ds \left[ \frac{1}{\TONE} \right]_V \: = \:
        \ct \left( \rule[0cm]{0cm}{.7cm}
                   \left[ \frac{1}{\TTWO} \right]_V \right) \: = \:
        \myln \frac{s\!+\!\delta\!+\!\SLAM}{s\!+\!\delta\!-\!\SLAM}
       \; = \; \cl_{12} }
    \medskip
 9) & {\ds \; \left[\, l_t \,\right]_V } \: = \:
        {\ds \frac{s-\sigma}{2} \; \cl_0 \; + \;
          \frac{\SLAM}{2}\; \cl_S \; - \; \SLAM }
        \medskip
    & \hspace{.5cm} \cl_S \: = \:
        {\ds \left( \rule[-.15cm]{0cm}{.8cm} 2 \,l_\beta \, + \,
          \myln \frac{\SONE}{s} \, + \, \myln \frac{\STWO}{s}
         \right) }
\medskip
 10)& {\ds \left[\, t \; l_t \,\right]_V } =
        {\ds \frac{\left(s \!-\!\sigma\right)^2 - 2 \SONE \STWO}{4}
             \; \cl_0 \; + \;
             \frac{\left( s \!-\!\sigma \right)\SLAM}{4}
             \; \left( \cl_S - 1 \right) }
        \bigskip
 11) & {\ds \left[ \frac{\,l_t\,}{t} \right]_V } =
       {\ds \frac{1}{2} \, \cl_0 \; \cl_S }
        \bigskip
 12) & {\ds \left[ \frac{\,l_t\,}{u} \right]_V } =
        {\ds \myln \frac{s-\sigma}{\ME^2} \; \cl_0 \; - \;
          \mysp \left( \frac{\,t_{max}\,}{s-\sigma} \right) \; + \;
          \mysp \left( \frac{\,t_{min}\,}{s-\sigma} \right) }
\end{array}$
\clearpage
$\begin{array}{llll}
 13) & {\ds \left[ \frac{l_t}{\TONE} \right]_V } & = & \;
       {\ds \ct \left( \rule[0cm]{0cm}{.7cm}
         \left[ \frac{l_t}{\TTWO} \right]_V \right) }
     \smskip & & = & \;
       {\ds \frac{\cl_{12} \; \cl_S}{2} \; - \;
             \frac{\cl_0}{2} \; \myln \frac{\SONE}{s}
          \; + \; \mysp \left( -\,\frac{\,t_{max}\,}{\SONE} \right)
          \; - \; \mysp \left( -\,\frac{\,t_{min}\,}{\SONE} \right) }
      \bigskip
 14) & {\ds \left[ \frac{\,l_t\,}{t^2} \right]_V } & = &
        {\ds \frac{1}{2\,\SONE\STWO} \left( \rule[-.05cm]{0cm}{.6cm}
          \SLAM \; \left( \cl_S + 2 \right) \; - \;
          \left( s-\sigma \right) \; \cl_0 \right) }
      \bigskip
 15) & {\ds \left[ \frac{l_t}{t^2_{12}} \right]_V } & = & \;
       {\ds \ct \left( \rule[0cm]{0cm}{.7cm}
          \left[ \frac{l_t}{t^2_{34}} \right]_V \right) \: = \:
          \frac{1}{\SONE} \left( \rule[-.25cm]{0cm}{1cm}
          \frac{s-\delta}{2s} \; \cl_0 \; - \;
          \cl_{12} \; + \; \frac{\SLAM}{2s} \; \cl_S
          \right) }
      \bigskip
 16) & {\ds \hspace{.015cm} \left[\, {l_t}^2 \, \right]_V } & = &
        {\ds \frac{s-\sigma}{2}\;
          \left( \rule[-.05cm]{0cm}{.4cm}\cl_S - 2 \right)
          \; \cl_0 \; + \; \SLAM \;
          \left( \rule[-.15cm]{0cm}{.8cm} \frac{\,{\cl_0}^2}{4}
          +\frac{\,{\cl_S}^2}{4} - \cl_S +2 \right) }
        \bigskip
 17) & {\ds \left[\, t \; l_t^2 \,\right]_V } & = &
        {\ds \frac{\left( s-\sigma \right)^2 - 2 \SONE \STWO}{4}
             \; \cl_0 \; \left( \cl_S - 1 \right) }
        \medskip & & & \hspace{1cm} {\ds \; + \;
             \frac{\left( s-\sigma \right) \, \SLAM}{4} \;
             \left( \rule[-.15cm]{0cm}{.8cm}
                    1 - \cl_S + \frac{{\cl_0}^2 + {\cl_S}^2}{2}
             \right) }
        \bigskip
 18) & {\ds \left[ \frac{{l_t}^2}{t} \right]_V } & = &
        {\ds \cl_0 \; \left( \rule[-.25cm]{0cm}{1cm}
          \frac{\,{\cl_0}^2}{12} \: + \:
          \frac{\,{\cl_S }^2}{4} \right) }
      \bigskip
 19) & {\ds \left[ \frac{{l_t}^2}{u} \right]_V } & = &
       {\ds  \left( \ln^2 \frac{s\!-\!\sigma}{\ME^2} - \frac{\pi^2}{6}
             \right) \cl_0}
       {\ds \; - \;
         \cl_S
         \left[ \rule[0cm]{0cm}{.6cm}
                \mysp \left( \frac{t_{max}}{s\!-\!\sigma} \right) -
                \mysp \left( \frac{t_{min}}{s\!-\!\sigma} \right)
         \right] }
       \smskip & & &
       {\ds + \; 2 \,
         \left[ \rule[0cm]{0cm}{.6cm}
                \mytri \left( \frac{t_{max}}{s-\sigma} \right) -
                \mytri \left( \frac{t_{min}}{s-\sigma} \right)
         \right] }
\bigskip
 20) & {\ds \left[ \frac{{l_t}^2}{\TONE} \right]_V } & = & \;
       {\ds \ct \left(\rule[0cm]{0cm}{.7cm}
         \left[ \frac{{l_t}^2}{\TTWO} \right]_V \right) \;\; = \;\;
         \frac{{\cl_0}^2 + {\cl_S}^2}{4} \; \cl_{12} \; - \;
         \frac{\cl_0 \, \cl_S}{2} \; \myln \frac{\SONE}{s} }
      \smskip & & &
        {\ds \; + \; \left( \cl_0 + \cl_S \right) \;
             \mysp \left( -\frac{\,t_{max}\,}{\SONE} \right) \; + \;
             \left( \cl_0 - \cl_S \right) \;
             \mysp \left( -\frac{\,t_{min}\,}{\SONE} \right)}
      \medskip & & & {\ds \; - \; 2 \,
          \mytri \left( -\frac{\,t_{max}\,}{\SONE} \right)
          \;+\; 2 \, \mytri \left( -\frac{\,t_{min}\,}{\SONE} \right)}
      \bigskip
 21) & {\ds \left[ \frac{{l_t}^2}{t^2} \right]_V } & = & \;
        {\ds \frac{1}{\SONE\STWO}
          \left[ \rule[-.25cm]{0cm}{1cm}
                 \SLAM \;\! \left( \frac{{\cl_0}^2}{4}
                 + \frac{{\cl_S}^2}{4} \: + \: \cl_S \: + \: 2 \right)
                 \; - \; \frac{s \!-\! \sigma}{2} \;
                 \left( \rule[-.05cm]{0cm}{.4cm}
                 \cl_S + 2 \right) \; \cl_0
          \right] }
      \bigskip
\end{array}$
\clearpage
$\begin{array}{llll}
 22) & {\ds \left[ \frac{{l_t}^2}{\TONE^2} \right]_V } & = & \;
       {\ds \ct \left(\rule[0cm]{0cm}{.7cm}
         \left[ \frac{{l_t}^2}{\TTWO^2} \right]_V \right)
                \; = \;
            \frac{1}{\SONE}
         \left[ \rule[-.25cm]{0cm}{1cm} \;
                 \frac{s \!-\! \delta}{2\,s} \; \cl_0 \;
                 \cl_S \; + \;
                 \myln \frac{\SONE}{s} \; \cl_0 \; - \;
                 \cl_S \; \cl_{12}   \right. }
      \medskip & & & {\ds \; + \;
                 \frac{\SLAM}{4s} \, \left( {\cl_0}^2 + {\cl_S}^2 \right)
             \left. \rule[-.25cm]{0cm}{1cm} - \;
          2 \, \mysp \left( -\frac{\,t_{max}\,}{\SONE} \right) \; + \; 
          2 \, \mysp \left( -\frac{\,t_{min}\,}{\SONE} \right) \right] }
      \bigskip
\end{array}$
\newline
$\begin{array}{llll}
 23) & {\ds \left[ \rule[-.15cm]{0cm}{.8cm}
          \mysp \left( \frac{\TONE-\ieps}{\SONE} \right) \right]_V }
        & = & \;
        {\ds \ct \left(\rule[0cm]{0cm}{.9cm}
          \left[ \rule[-.15cm]{0cm}{.8cm}
          \mysp \left( \frac{\TTWO-\ieps}{\STWO} \right) \right]_V
          \right) }
      \medskip & & = & \;
        {\ds \frac{s-\sigma}{2} \;\! \cl_0 \; + \;
          \frac{\SLAM}{2} \;\! \left( \rule[-.05cm]{0cm}{.4cm}
          \cl_S - 2\,l_1 -2 \right) }
        \medskip & & & {\ds + \;
          \left( \rule[-.05cm]{0cm}{.4cm} \SONE + t_{max} \right)
          \; \mysp \left(
          1 + \frac{t_{max}-\ieps}{\SONE} \right) }
        \medskip & & & {\ds - \;
          \left( \rule[-.05cm]{0cm}{.4cm} \SONE + t_{min} \right)
          \; \mysp \left(
          1 + \frac{t_{min}-\ieps}{\SONE} \right) }
      \bigskip
\end{array}$
\newline
$\begin{array}{ll}
  24) & {\ds \left[ \rule[-.15cm]{0cm}{.8cm} t \;
          \mysp \left( \frac{\TONE-\ieps}{\SONE} \right) \right]_V
          \; = \;
          \ct \left(\rule[0cm]{0cm}{.9cm}
          \left[ \rule[-.15cm]{0cm}{.8cm} t \;
          \mysp \left( \frac{\TTWO-\ieps}{\STWO} \right) \right]_V
          \right) \; = \; - \frac{\SLAM}{8}\,(s \!-\! 5\SONE \!-\!
          \STWO) }
       \medskip
       & \hspace{1cm}
         {\ds +\; \frac{(s\!-\!\sigma)(s \!-\! 3\SONE \!-\! \STWO) - 2
                        \SONE \STWO}{8} \; \cl_0 \; - \;
              \frac{\SLAM}{8}\,(s \!-\! 3\SONE \!-\! \STWO)
              \;
              \left( \rule[0cm]{0cm}{.35cm} l_- - 2\pi\ri \right) }
       \medskip
       & \hspace{1cm}
         {\ds + \; \frac{{t_{max}}^2 - \SONE^2}{2} \;
           \mysp \left( 1 + \frac{t_{max} \!-\! \ieps}{\SONE} \right)
           \; - \; \frac{{t_{min}}^2 - \SONE^2}{2} \;
           \mysp \left( 1 + \frac{t_{min} \!-\! \ieps}{\SONE} \right) }
\bigskip
\end{array}$
\\
$\begin{array}{llll}
 25) & {\ds \left[ \rule[-.15cm]{0cm}{.8cm} \frac{1}{\,t\,} \;\!
          \mysp \left( \frac{\TONE-\ieps}{\SONE} \right) \right]_V }
        & = & \;
       {\ds \ct \left(\rule[0cm]{0cm}{.9cm}
         \left[ \rule[-.15cm]{0cm}{.8cm} \frac{1}{\,t\,} \;\!
          \mysp \left( \frac{\TTWO-\ieps}{\STWO} \right) \right]_V
          \right) }
     \medskip & & = & \;
       {\ds \frac{\pi^2}{6}\;\! \cl_0 \; - \;
          2\,\mytri \left( -\frac{\,t_{max}\,}{\SONE}  \right) \; + \;
          2\,\mytri \left( -\frac{\,t_{min}\,}{\SONE}  \right) }
     \medskip
& & & {\ds + \;
          \left( \rule[-.15cm]{0cm}{.8cm}
          \myln \frac{\,t_{max}\,}{\ME^2} - l_1 \right) \;
          \mysp \left( -\frac{\,t_{max}\,}{\SONE} \right) }
     \medskip
& & & {\ds - \;
          \left( \rule[-.15cm]{0cm}{.8cm}
          \myln \frac{\,t_{min}\,}{\ME^2} - l_1 \right) \;
          \mysp \left( -\frac{\,t_{min}\,}{\SONE} \right) }
\end{array}$
\clearpage 
$\begin{array}{ll}
 26) & {\ds \re \left(\rule[-.25cm]{0cm}{1cm}
          \left[ \rule[-.15cm]{0cm}{.8cm} \frac{1}{\,u\,} \;\!
          \mysp \left( \frac{\TONE-\ieps}{\SONE} \right) \right]_V
          \right) \; = \;
          \ct \left\{ \rule[0cm]{0cm}{1cm}
          \re \left( \rule[0cm]{0cm}{.9cm}
          \left[ \rule[-.15cm]{0cm}{.8cm} \frac{1}{\,u\,} \;\!
          \mysp \left( \frac{\TTWO-\ieps}{\STWO} \right) \right]_V
          \right) \right\} }
        \medskip & \hspace{.1cm} {\ds = \;
          \cl_0 \; \left[ \rule[-.15cm]{0cm}{.8cm}
          \hspace{.4cm} \frac{{\cl_0}^2}{6} \; - \;
          \frac{1}{2} \; \myln \left(\DONE\right) \;
          \myln \frac{\SONE}{\STWO} \; - \;
          \myln^2 \left(\DONE\right) \; - \;
          2\, \mysp \left( \frac{1}{\DONE} \right) \; + \;
          \frac{\,2\pi^2}{3} \rule[-.15cm]{0cm}{.8cm} \;\; \right] }
        \medskip & \hspace{.7cm} {\ds - \;\;\!
          \myln \frac{t_{max}}{\SONE} \;
          \mysp \left( \frac{t_{min}}{\,s-\STWO\,} \right)
          \;\; + \;\; \myln \frac{t_{min}}{\SONE} \;
          \mysp \left( \frac{t_{max}}{\,s-\STWO\,} \right) }
        \medskip & \hspace{.7cm} {\ds + \;\:
          \mytri \left( -\frac{\,t_{max}\,}{\SONE} \right) \; - \;
          \mytri \left( -\frac{\,t_{min}\,}{\SONE} \right) \; - \;
          \mytri \left( \frac{t_{max}}{\,s\!-\!\STWO\,} \right)
          \; + \;
          \mytri \left( \frac{t_{min}}{\,s\!-\!\STWO\,} \right) }
        \medskip & \hspace{.7cm} {\ds + \;\:
          \mytri \left( -\frac{t_{max}}{\,\DONE\,t_{min}}\, \right)
          \;\: - \;\:
          \mytri \left( -\frac{t_{min}}{\,\DONE\,t_{max}}\, \right) }
      \smskip \smskip
\end{array}$
\newline
$\begin{array}{llll}
 27) & {\ds \left[ \rule[-.15cm]{0cm}{.8cm} \frac{1}{\TONE} \;\!
          \mysp \left( \frac{\TONE-\ieps}{\SONE} \right) \right]_V }
       & = & \;
       {\ds \ct \left(\rule[0cm]{0cm}{.9cm}
          \left[ \rule[-.15cm]{0cm}{.8cm} \frac{1}{\TTWO} \;\!
          \mysp \left( \frac{\TTWO-\ieps}{\STWO} \right) \right]_V
          \right) }
        \smskip & & = & \;
        {\ds \mytri \left( 1 + \frac{t_{max}-\ieps}{\SONE} \right)
          \;-\;
          \mytri \left( 1 + \frac{t_{min}-\ieps}{\SONE} \right) }
      \bigskip
\end{array}$
\newline
$\begin{array}{ll}
 28) & {\ds \left[ \rule[-.15cm]{0cm}{.8cm} \frac{1}{\,t^2} \;\!
         \mysp \left( \frac{\TONE-\ieps}{\SONE} \right) \right]_V
         \; = \;
         \ct \left(\rule[0cm]{0cm}{.9cm}
         \left[ \rule[-.15cm]{0cm}{.8cm} \frac{1}{\,t^2} \;\!
          \mysp \left( \frac{\TTWO-\ieps}{\STWO} \right) \right]_V
          \right) }
        \medskip & \hspace{1.25cm} {\ds \; = \;
          - \frac{1}{\,t_{max}\,} \;
          \mysp \left( 1 + \frac{t_{max}-\ieps}{\SONE} \right)
          \;\; + \;\; \frac{1}{\,t_{min}\,} \;
          \mysp \left( 1 + \frac{t_{min}-\ieps}{\SONE} \right) }
        \medskip & \hspace{2cm} {\ds + \;
          \frac{1}{\SONE} \; \left[ \rule[-.15cm]{0cm}{.8cm} \;
          \mysp \left( 1 + \frac{\SONE}{t_{max}-\ieps} \right) \;\; -
          \;\;
          \mysp \left( 1 + \frac{\SONE}{t_{min}-\ieps} \right) \;
          \right] }
      \smskip \smskip
\end{array}$
\newline
$\begin{array}{ll}
 29) & {\ds \left[ \rule[-.15cm]{0cm}{.8cm} \frac{1}{\TONE^2} \;\!
        \mysp \left( \frac{\TONE-\ieps}{\SONE} \right) \right]_V
        \; = \;
        \ct \left(\rule[0cm]{0cm}{.9cm}
        \left[ \rule[-.15cm]{0cm}{.8cm} \frac{1}{\TTWO^2} \;\!
        \mysp \left( \frac{\TTWO-\ieps}{\STWO} \right) \right]_V
        \right) }
      \medskip & \; = \;
      {\ds \frac{1}{\SONE} \left[ \rule[-.25cm]{0cm}{1cm} \,
        \cl_{12} \; - \;
        \frac{1-\SONE+\STWO}{2\,s} \; \cl_0 \; + \;
        \frac{\SLAM}{2\,s} \; \left( \l_- - 2\pi\ri \right) \right.}
      \medskip & \hspace{1.5cm}
      {\ds \left. - \: \frac{t_{min}\!+\!\SONE}{s} \;
        \mysp \left( 1 + \frac{t_{max}\!-\!\ieps}{\SONE} \right) \;+\;
        \frac{t_{max}\!+\!\SONE}{s} \;
        \mysp \left( 1 + \frac{t_{min}\!-\!\ieps}{\SONE} \right)
        \rule[-.25cm]{0cm}{1cm} \right] }
\end{array}$
\clearpage
%
\subsection{Bremsstrahlung Integrals -- First Series}
\label{bremint1}
%
For the calculation of bremsstrahlung to processes~(\ref{4fproc}),
the phase space was parametrized as exposed in
equation~(\ref{par25b}) and appendix~\ref{ps2to5}. After integration
over the final state fermion decay angles, the first series of
bremsstrahlung integrals is over the scattering azimuth and polar
angles $\phi_R$ and $\theta_R$ of the boson three-vector $v_1$ in the
two-boson rest frame:
\begin{equation}
  \left[A\right]_R \equiv \frac{1}{4\pi}\int\limits_{0}^{2\pi} d\phi_R
                   \int\limits_{-1}^{+1} d\mycos\theta_R~A~.
\end{equation}

We introduce a third symmetry operation,
\\
\ba
  \cs\left[\rule[0cm]{0cm}{.35cm}
    f(\mycos\theta)\right] \;\; \equiv \;\; f(-\mycos\theta)
\ea
\\
Taking into account that the contribution of the bremsstrahlung
$u$-channel matrix element squared to the threefold differential
\xsec\ equals the contribution of the bremsstrahlung $t$-channel
matrix element, the following bremsstrahlung integrals are needed for
the first series:
\vspace{.3cm} \\
$\begin{array}{llll}
 1) & \left[1\right]_R & = & 1 \hugeskip
 3) & \left[\mycos\theta_R\right]_R & = & 0 \hugeskip
\end{array}$
$\begin{array}{llll}
\hspace{3.5cm} 2) & \left[\mycos\phi_R \; f(\mycos\theta_R)\right]_R
                    & = & 0 \hugeskip
\hspace{3.5cm} 4) & \left[\mycos^2\phi_R\right]_R & = & 1/2 \hugeskip
\end{array}$
\newline
$\begin{array}{llll}
 5) & \left[\mycos^2\theta_R\right]_R & = & 1/3 \bigskip
 7) & {\ds \left[\frac{1}{t_1}\right]_R } & = &
       {\ds \frac{2s'}{\SLAMP (\sprp - \sprm\mycos\theta)} \: \; }
         \myln \left(
            \frac{\sprp(s'+\SLAMP) - 2s'\sigma + \sprm\delta -
                (s'+\delta+\SLAMP)\sprm \mycos\theta}
                {\sprp(s'-\SLAMP) - 2s'\sigma + \sprm\delta -
                (s'+\delta-\SLAMP)\sprm \mycos\theta}
            \right)\smskip 
         & & \equiv &
         {\ds \frac{2s'}{\SLAMP \, S_{d1}} \; l_{t1} }
    \bigskip
 8) & {\ds \left[\frac{1}{t_2}\right]_R } & = & \;
      {\ds \cs\ct \!\left(\left[\frac{1}{t_1}\right]_R\right)}
      \; \equiv \; {\ds \frac{2s'}{\SLAMP \, S_{d2}} \; l_{t2} }
    \bigskip
 9) & {\ds \left[\frac{1}{u_1}\right]_R } & = & \;
      {\ds \ct \! \left(\left[\frac{1}{t_1}\right]_R\right)
         \; \equiv \; \frac{2s'}{\SLAMP \, S_{d1}} \; l_{u1}  }
    \bigskip
 10) & {\ds \left[\frac{1}{u_2}\right]_R } & = & \;
       {\ds \cs \!\left(\left[\frac{1}{t_1}\right]_R\right)
         \; \equiv \; \frac{2s'}{\SLAMP \, S_{d2}} \; l_{u2}  }
     \bigskip
\end{array}$
\newline
$\begin{array}{llll}
 11) & {\ds \left[ \frac{\cos\theta_R}{t_1}\right]_R } & = &
        {\ds \frac{16\;s'^2 b_1}{\lambda^{'} \; S_{d1}^2} \;
            \left( 1 - a^t_1 \left[ \frac{1}{t_1}\right]_R \right) }
      \bigskip
 12) & {\ds \left[ \frac{\cos\theta_R}{t_2}\right]_R } & = &
        {\ds \frac{16\;s'^2 b_2}{\lambda^{'} \; S_{d2}^2} \;
            \left( 1 - a^t_2 \left[ \frac{1}{t_2}\right]_R \right) }
        \bigskip
\end{array}$
\newline
$\begin{array}{llll}
 13) & {\ds \left[ \frac{1}{t_1^2}\right]_R } & = &
       {\ds \frac{4s'}{\SONE
         \left(s - \delta + \SLAMP - \sprm \;\mycos\theta \right) 
         \left(s - \delta - \SLAMP - \sprm \;\mycos\theta \right) }
         \;\equiv \;\frac{4s'}{\SONE  d_1^+  d_1^- } }
     \bigskip 
 14) & {\ds \left[ \frac{1}{t_2^2}\right]_R } & = & \;
       {\ds \cs\ct \!\left( \left[\frac{1}{t_1^2}\right]_R \right)
         \; \equiv \; \frac{4s'}{\STWO \; d_2^+ \, d_2^- } }
     \bigskip 
\end{array}$
\newline
%
$\begin{array}{llll}
 15) & {\ds \left[ \frac{1}{t_1t_2}\right]_R } & = &
        {\ds \frac{1}{2 S_{t_1t_2}} \;
         \myln \left( \frac{A^t_{1-} \; A^t_{2+}}
                        {A^t_{1+} \; A^t_{2-}} \right)
         \;\; \equiv \;\; \frac{1}{2 S_{t_1t_2}} \; l_{t12} }
       \bigskip
     & \hspace{.5cm} S_{t_1t_2} & = & {\ds \frac{\SLAMP}{8s'} \: \;
         \left( \sqrt{\bar{\lambda}} + \delta + \sprm\mycos\theta
         \right)
         \left( \sqrt{\bar{\lambda}} - \delta - \sprm\mycos\theta
         \right) } \miniskip
     & & \equiv & {\ds \frac{\SLAMP}{8s'} \: \;
                       S^t_{12+} \, S^t_{12-} }
        \smskip
     & \hspace{.5cm} A^t_{1-} & = & a_{1-} \; + \; b_{1-} \; \mycos\theta
       \hspace{2cm}
               A^t_{1+} \; = \; a_{1+} \; + \; b_{1+} \; \mycos\theta
       \smskip
     & \hspace{.5cm} A^t_{2-} & = & a_{2-} \; + \; b_{2-} \; \mycos\theta
       \hspace{2cm}
               A^t_{2+} \; = \; a_{2+} \; + \; b_{2+} \; \mycos\theta
       \bigskip
     & \hspace{.5cm}\; a_{1-} & = & {\ds s^2\sigma - s\delta^2 +
         \frac{s}{2} \sprm \left( s' + \delta - 2\sigma\right ) -
         \SLAMP \left( s\delta - \frac{s}{2}\sprm \right) }
       \smskip
     & \hspace{.5cm}\; b_{1-} & = & 
         {\ds - \left( \SLAMP + s' + \delta \right) \frac{s}{2} \sprm }
       \smskip
     & \hspace{.5cm}\; a_{1+} & = & {\ds s^2\sigma - s\delta^2 +
         \frac{s}{2} \sprm \left( s' + \delta - 2\sigma\right ) +
         \SLAMP \left( s\delta - \frac{s}{2}\sprm \right) }
       \smskip
     & \hspace{.5cm}\; b_{1+} & = &
         {\ds + \left( \SLAMP - s' - \delta \right) \frac{s}{2} \sprm }
       \smskip
     & \hspace{.5cm}\; a_{2-} & = & {\ds s^2\sigma - s\delta^2 +
           \frac{s}{2} \sprm \left( s' - \delta - 2\sigma\right ) -
           \SLAMP \left( s\delta + \frac{s}{2}\sprm \right) }
       \smskip
     & \hspace{.5cm}\; b_{2-} & = &
         {\ds - \left( \SLAMP - s' + \delta \right) \frac{s}{2} \sprm }
       \smskip
     & \hspace{.5cm}\; a_{2+} & = & {\ds s^2\sigma - s\delta^2 +
           \frac{s}{2} \sprm \left( s' - \delta - 2\sigma\right ) +
           \SLAMP \left( s\delta + \frac{s}{2}\sprm \right) }
       \smskip
     & \hspace{.5cm}\; b_{2+} & = &
         {\ds + \left( \SLAMP + s' - \delta \right) \frac{s}{2} \sprm }
\end{array}$
\newpage
$\begin{array}{llll}
 16) & {\ds \left[ \frac{1}{u_1u_2}\right]_R } & = &
        {\ds \cs \! \left(\left[\frac{1}{t_1t_2}\right]_R\right)
         \; \equiv \; \frac{1}{2 S_{u_1u_2}} \; l_{u12} }
     \bigskip
 17) & {\ds \left[ \frac{1}{t_1u_1}\right]_R } & = &
        {\ds \frac{1}{a^{ut}_1} \left(
            \left[\frac{1}{t_1}\right]_R + \left[\frac{1}{u_1}\right]_R
                             \right) }
        \medskip
      & \hspace{.8cm} a^{ut}_1 & = &
          {\ds \frac{1}{2} \left( \sprp - \sprm\mycos\theta \right)
            - \sigma
          \; \equiv \; \frac{a}{2}
            \left( 1 - b \! \; \! \mycos\theta \right) } \\
      & \hspace{1cm} a & = & {\ds \sprp - 2 \sigma ,
        \hspace{1cm} b \;= \; \frac{\sprm}{\sprp - 2 \sigma} ,
        \hspace{1cm} 0 \;\leq\; b \;<\; 1 }
      \bigskip
\end{array}$
\newline
$\begin{array}{llll}
 18) & {\ds \left[ \frac{1}{t_2u_2}\right]_R } & = &
        {\ds \frac{1}{a^{ut}_2} \left(
            \left[\frac{1}{t_2}\right]_R + \left[\frac{1}{u_2}\right]_R
                             \right) }
        \medskip
      & \hspace{.8cm} a^{ut}_2 & = &
          {\ds \frac{1}{2} \left( \sprp + \sprm\mycos\theta \right)
           - \sigma 
          \;\; = \;\; \frac{a}{2}
            \left( 1 + b \! \; \! \mycos\theta \right) }
     \bigskip
 19) & {\ds \left[ \frac{1}{t_1u_2}\right]_R } & = &
        {\ds \frac{1}{2 \sqrt{C_{12}}} \;
          \left[ \rule[-.2cm]{0cm}{1cm} \;
          \myln \left( \frac{A^-_1 \; B^-_2}{A^+_1 \; B^+_2}
                \right) 
          + 2 \myln \left( \frac{a_{s1} + b_d}{a_{s1} - b_d} \right)
                                       \; \right] }
          \equiv
      {\ds \frac{1}{2 \sqrt{C_{12}}} \; l_{t_1u_2} }
      \bigskip
      & \hspace{.8cm} A_1^- & = &
          \sqrt{C_{12}} \; b_d \; \left( a^t_1 + b_1 \right) +
          C_{12} + B^t_1 \; \left( a_{s1} - b_d \right)/2
        \smskip
      & \hspace{.8cm} A_1^+ & = &
          \sqrt{C_{12}} \; b_d \; \left( a^t_1 - b_1 \right) +
          C_{12} + B^t_1 \; \left( a_{s1} + b_d \right)/2
        \smskip
      & \hspace{.8cm} B_2^- & = &
          \sqrt{C_{12}} \; b_d \; \left( a^u_2 - b_2 \right) +
          C_{12} + B^u_2 \; \left( a_{s1} - b_d \right)/2
        \smskip
      & \hspace{.8cm} B_2^+ & = &
          \sqrt{C_{12}} \; b_d \; \left( a^u_2 + b_2 \right) +
          C_{12} + B^u_2 \; \left( a_{s1} + b_d \right)/2
        \bigskip
      & \hspace{.8cm} C_{12} & = &
          {\ds \left( \frac{\SLAMP}{2s'} \right)^2 \;
          \left( A_{12} \; \mycos^2\theta + B_{12} \right) }
        \smskip
      & \hspace{.8cm} A_{12} & = &
          - s'^{\;2}_- \; \SONE \; \left( s - \SONE \right)
        \smskip
      & \hspace{.8cm} B_{12} & = &
          s'^{\;2}_- \; s \; \left( s - \SONE \right) -
          2 s \sprm \left( s - \sigma \right) \left( s - \SONE \right)
          + s^2 \left( s - \sigma \right)^2
        \smskip
      & \hspace{.8cm} B^t_1 & = &
          -2 \left[ \rule[0cm]{0cm}{.4cm}
               b_1 \left( a^t_1 b_2 + a^u_2 b_1 \right) + a_{s1}c^2 
             \right]
        \smskip
      & \hspace{.8cm} B^u_2 & = &
          -2 \left[ \rule[0cm]{0cm}{.4cm}
               b_2 \left( a^t_1 b_2 + a^u_2 b_1 \right) + a_{s1}c^2 
             \right]
        \smskip
      & \hspace{.8cm} a_{s1} & = & {\ds a^t_1 + a^u_2 \; = \;
          \frac{\sprp}{2s'} \left( s' + \delta \right) - 2 \SONE }
        \smskip
      & \hspace{.8cm} b_d & = & b_2 - b_1 \; = \; 
          {\ds \frac{\SLAMP}{2s'} \sprm }
\end{array}$
\newpage
$\begin{array}{llll}
 20) & {\ds \left[ \frac{1}{t_2u_1}\right]_R } & = &
       {\ds \ct \! \left( \left[\frac{1}{t_1u_2}\right]_R \right)
         \; \equiv \;
         \frac{1}{2 \sqrt{C_{21}}} \; l_{t_2u_1} }
        \smskip
        \\
 21) & {\ds \left[ \frac{\cos\theta_R}{t_1^2}\right]_R } & = &
        {\ds \frac{\,16\:s'^2 \, b_1\,}{\lambda^{'} S_{d1}^2} \;
             \left( \left[ \frac{1}{t_1}\right]_R
                   - a^t_1 \left[ \frac{1}{t_1^2}\right]_R \right) }
        \smskip
        \\
 22) & {\ds \left[ \frac{\cos\theta_R}{t_2^2}\right]_R } & = &
        {\ds \frac{\,16\:s'^2\,b_2\,}{\lambda^{'} {S^2_{d2}}} \;
             \left( \left[ \frac{1}{t_2}\right]_R
                   - a^t_2 \left[ \frac{1}{t_2^2}\right]_R \right) }
        \smskip
        \\
 23) & {\ds \left[ \frac{\cos^2\theta_R}{t_1}\right]_R } & = &
        {\ds \frac{\,16\:s'^2\,}{\lambda^{'} {S^2_{d1}}} \;
          \left\{ -a_1^t + \frac{1}{2} \left( 2{a_1^t}^2 + c^2 \right)
                           \left[ \frac{1}{t_1} \right]_R
          \right\} } \smskip
      & & & + \;\;
        {\ds \frac{\,384\:s'^4\,}{{\lambda^{'}}^2 {S_{d1}}^4} \;
          a_1^t c^2 \;
          \left\{ 1 - a_1^t \left[ \frac{1}{t_1} \right]_R
          \right\} }
        \smskip
        \\
 24) & {\ds \left[ \frac{\cos^2\theta_R}{t_2}\right]_R } & = &
       {\ds \cs\ct \! \left( \rule[0cm]{0cm}{.7cm}
         \left[ \frac{\cos^2\theta_R}{t_1}\right]_R \right)}
        \smskip
        \\
 25) & {\ds \left[ \frac{1}{t_1^2t_2}\right]_R } & = &
        {\ds \frac{1}{S^2_{t_1t_2}}
          \left\{ -a^t_d \; + \;  a^t_1
                 \left( a^t_1\,a^t_d - b_1\,b_d \right)
                 \left[ \frac{1}{t_1^2}\right]_R \right. }
        \smskip & & & \hspace{1.17cm} \;\: {\ds \left. +
                 \left( b_2 \, (a^t_1 b_2 - a^t_2 b_1) - a^t_d \, c^2
                 \right)
                 \left[ \frac{1}{t_1t_2}\right]_R 
             \right\} }
        \medskip
      & \hspace{1.3cm} a^t_d & = & a^t_2 -a^t_1 \; = \;
        {\ds - \frac{\sprm}{2\,s'}
             \left(\delta - s' \, \mycos\theta \right) }
        \smskip
        \\
 26)  & {\ds \left[ \frac{1}{t_1t_2^2}\right]_R } & = &
        {\ds \cs\ct \! \left( \rule[0cm]{0cm}{.7cm}
         \left[ \frac{1}{t_1^2t_2}\right]_R \right) }
        \smskip
        \\
 27)  & {\ds \left[ \frac{1}{t_1t_2u_1}\right]_R } & = &
         {\ds \frac{1}{a^{ut}_1} \left(
           \left[ \frac{1}{t_1t_2} \right]_R +
           \left[ \frac{1}{t_2u_1} \right]_R
                                 \right) }
        \smskip
        \\
 28)  & {\ds \left[ \frac{1}{t_1t_2u_2}\right]_R } & = &
         {\ds \frac{1}{a^{ut}_2} \left(
           \left[ \frac{1}{t_1t_2} \right]_R +
           \left[ \frac{1}{t_1u_2} \right]_R
                                 \right) }
\end{array}$
\newpage
$\begin{array}{llll}
 29)  & {\ds \left[ \frac{1}{t_1u_1u_2}\right]_R } & = &
         {\ds \frac{1}{a^{ut}_1} \left(
           \left[ \frac{1}{u_1u_2} \right]_R +
           \left[ \frac{1}{t_1u_2} \right]_R
                                 \right) }
        \bigskip \\
 30)  & {\ds \left[ \frac{1}{t_2u_1u_2}\right]_R } & = &
         {\ds \frac{1}{a^{ut}_2} \left(
           \left[ \frac{1}{u_1u_2} \right]_R +
           \left[ \frac{1}{t_2u_1} \right]_R
                                 \right) }
	        \bigskip \\
 31)  & {\ds \left[ \frac{1}{t_1t_2u_1u_2}\right]_R } & = &
         {\ds \frac{1}{a^{ut}_1} \; \frac{1}{a^{ut}_2}
         \left(
           \left[ \frac{1}{t_1t_2} \right]_R +
           \left[ \frac{1}{u_1u_2} \right]_R +
           \left[ \frac{1}{t_1u_2} \right]_R +
           \left[ \frac{1}{t_2u_1} \right]_R
         \right) }
        \bigskip \\
\end{array}$
%
%
%
\subsection{Bremsstrahlung Integrals -- Second Series}
\label{bremint2}
Below, the integrals needed for the last analytical integration in the
bremsstrahlung case are given. This integration is over the 
photon scattering angle $\theta$. Using the ultrarelativistic
approximation and the notation
\begin{equation}
  \left[A\right]_\theta \equiv
    \frac{1}{2} \int\limits_{-1}^{+1} d\mycos\theta~A,
\end{equation}
the following integrals are required for the ``second series'' of
bremsstrahlung integrals:
\vspace{.3cm} \\
$\begin{array}{llrl}
 1) & {\ds \left[ 1 \right]_\theta} \; = \; 1
    \smskip \\
 2) & {\ds \left[ \frac{1}{\bar z_1} \right]_\theta \; = \;
           \left[ \frac{1}{\bar z_2} \right]_\theta \; = \;
        \frac{1}{2} \; l_\beta } &
 3) & {\ds \left[ \frac{m_e^{\;2}}{\bar z_1^{\;2}} \right]_\theta
        \; = \; 
           \left[ \frac{m_e^{\;2}}{\bar z_2^{\;2}} \right]_\theta
        \; = \; \frac{s}{4} }
     \smskip \\
 4) & {\ds \left[ \frac{1}{S_{d1}} \right]_\theta \: = \:
           \left[ \frac{1}{S_{d2}} \right]_\theta \: = \:
        \frac{1}{2 \sprm} \,\myln\left( \frac{s}{s'} \right) } &
       \hspace{1.5cm}
 5) & {\ds \left[ \frac{1}{S^2_{d1}}  \right]_\theta \: = \:
           \left[ \frac{1}{S^2_{d2}}  \right]_\theta \: = \:
        \frac{1}{4\,ss'} }
    \smskip \\
 6) & {\ds \left[ \frac{1}{S^3_{d1}}  \right]_\theta \; = \;
           \left[ \frac{1}{S^3_{d2}}  \right]_\theta \; = \;
        \frac{\sprp}{16\,s^2 {s'}^2} }
\end{array}$
\clearpage
$\begin{array}{llll}
 7) & {\ds \left[ l_{t1} \right]_\theta} & = &
        {\ds - \;\left(1 + \frac{s'-\delta}{2 \sprm} \right) \; L_{c1}
          \; + \; \frac{\SLAMP}{2 \sprm} \; L_{c3}
          \; + \; L_{c5} }
      \smskip
      & \hspace{.3cm} L_{c1} & = &
        {\ds \myln \left( \frac{\sprm(s'+\SLAMP) + s'\sigma -
            s\delta}
                               {\sprm(s'-\SLAMP) + s'\sigma -
                                 s\delta}
                   \right) }
        \smskip
      & \hspace{.3cm} L_{c3} & = &
        {\ds \myln \left( 1 + \frac{\sprm (s-\delta)}{s'\STWO} \right)
          }
        \smskip
      & \hspace{.3cm} L_{c5} & = &
        {\ds \myln \left(
                     \frac{s' - \sigma + \SLAMP}{s' - \sigma - \SLAMP}
                   \right) }
        \bigskip \\
 8) & {\ds \left[ l_{t2} \right]_\theta} & = &
      {\ds \ct \! \left( \rule[0cm]{0cm}{.35cm}
        \left[ l_{t1} \right]_\theta \right) \;=\;
           - \;\left( 1 + \frac{s'+\delta}{2 \sprm} \right) \; L_{c2}
        \; + \; \frac{\SLAMP}{2 \sprm} \; L_{c4}
        \; + \; L_{c5} }
      \smskip
      & \hspace{.3cm} L_{c2} & = &
        {\ds \myln \left( \frac{\sprm(s'+\SLAMP) + s'\sigma + s\delta}
                               {\sprm(s'-\SLAMP) + s'\sigma + s\delta}
                   \right) }
      \smskip
      & \hspace{.3cm} L_{c4} & = &
        {\ds \myln \left( 1 + \frac{\sprm (s+\delta)}{s'\SONE} \right)
          }
        \bigskip \\
 9) & {\ds \left[ l_{t12} \right]_\theta} & = &
        {\ds \left( 1 + \frac{s'-\delta}{2 \sprm} \right)
             \; L_{c1}
          \; + \; \left( 1 + \frac{s'+\delta}{2 \sprm} \right)
             \; L_{c2}
          \; - \; \frac{\SLAMP}{2 \sprm} \;
             \left( L_{c3} + L_{c4} \right) }
      \bigskip \\
 10) & {\ds \left[ \frac{1}{\bar z_1}  \, l_{t1} \right]_\theta} & = &
       {\ds \left[ \frac{1}{\bar z_2}  \, l_{u2} \right]_\theta \;=\;
         \frac{1}{2} \left( \frac{}{} l_\beta \; L_{t1}
         \; - \; D_{z1t1} \right) }
        \smskip
     & \hspace{.4cm} L_{t1} & = & L_{c5} - L_{c1} \smskip
     & \hspace{.4cm} D_{z1t1} & = & {\ds
            \mysp \left( \frac{\sprm(s + \sprm - \delta - \SLAMP)}
                              {2s(\sprm - \delta) + s'(\sigma+\delta)}
                    \right) 
          - \mysp \left( \frac{\sprm(s + \sprm - \delta + \SLAMP)}
                              {2s(\sprm - \delta) + s'(\sigma+\delta)}
                  \right) }
      \bigskip \\
 11) & {\ds \left[ \frac{1}{\bar z_1}  \, l_{t2} \right]_\theta} & = &
       {\ds \left[ \frac{1}{\bar z_2}  \, l_{u1} \right]_\theta \;=\;
            \frac{1}{2} \left( \frac{}{} l_\beta \; L_{c5}
            \; - \; D_{z1t2} \right) }
        \bigskip
      & \hspace{.4cm}  D_{z1t2} & = & {\ds
            \mysp \left( \frac{-\sprm(s' + \delta - \SLAMP)}
                                {s'(\sigma+\delta)}
                    \right)
          - \mysp \left( \frac{-\sprm(s' + \delta + \SLAMP)}
                                {s'(\sigma+\delta)}
                    \right) }
\end{array}$
\clearpage
$\begin{array}{llll}
 12) & {\ds \left[ \frac{1}{\bar z_2}  \, l_{t1} \right]_\theta} & = &
       {\ds \left[ \frac{1}{\bar z_1}  \, l_{u2} \right]_\theta \;=\;
            \frac{1}{2} \left( \frac{}{} l_\beta \; L_{c5}
            \; - \; D_{z2t1} \right) }
        \smskip
      & \hspace{.4cm}  D_{z2t1} & = & {\ds
           \mysp \left( \frac{-\sprm(s' - \delta - \SLAMP)}
                               {s'(\sigma-\delta)}
                    \right)
         - \mysp \left( \frac{-\sprm(s' - \delta + \SLAMP)}
                               {s'(\sigma-\delta)}
                    \right) }
      \hugeskip
 13) & {\ds \left[ \frac{1}{\bar z_2}  \, l_{t2} \right]_\theta} & = &
       {\ds \left[ \frac{1}{\bar z_1}  \, l_{u1} \right]_\theta \;=\;
            \frac{1}{2} \left( \frac{}{} l_\beta \; L_{t2}
             \; - \; D_{z2t2} \right) }
        \smskip
     & \hspace{.4cm} L_{t2} & = & L_{c5} - L_{c2} \smskip
     & \hspace{.4cm}  D_{z2t2} & = & {\ds
           \mysp \left( \frac{\sprm(s + \sprm + \delta - \SLAMP)}
                        {2s(\sprm \!+\! \delta) + s'(\sigma\!-\!\delta)}
                    \right)
         - \mysp \left( \frac{\sprm(s + \sprm + \delta + \SLAMP)}
                        {2s(\sprm \!+\! \delta) + s'(\sigma\!-\!\delta)}
                    \right) }
      \hugeskip
\end{array}$
\footnotetext[7]{
A comment on the treatment of arguments of logarithms and
dilogarithms is in order here. The logarithm is undefined for
negative real values, the dilogarithm has a cut for real
numbers larger than 1.
This means that, in principle, infinitesimal imaginary parts from the
propagator denominators have to be carried along in the whole
calculation. This, however, is an unnecessary nuisance, because all
integrands are real and regular. If infinitesimal imaginary parts were
needed, because logarithm or dilogarithm arguments lay on cuts in final
results, they were attributed to the invariant masses
$\SONE$~and $\STWO$. As integrands are real and regular, this is a
correct treatment, because it is then irrelevant for the
integrand how an infinitesimal imaginary part is entered. This
technique will be used in subsequent integrals without further notice.}
\newline
$\begin{array}{llll}
 14) & {\ds \left[\frac{1}{S_{d1}} \, l_{t1}\right]_\theta} & = &
       {\ds \left[\frac{1}{S_{d2}} \, l_{u2}\right]_\theta}
       \smskip & & = &
       {\ds \frac{1}{2 \sprm} \; \re \! \left[ \rule[-.2cm]{0cm}{1cm} \;
             - \; \mysp \left( \frac{s\,(s'+\delta+\SLAMP)}
                                   {s'(\sigma + \delta)} \right)
             \;+\; \mysp \left( \frac{s'+\delta+\SLAMP}
                                   {\sigma + \delta} \right) 
                           \right. }
          \smskip & & & \hspace{1.71cm} {\ds \left.
             + \; \mysp \left( \frac{s\,(s'\!+\!\delta\!-\!\SLAMP)}
                                   {s'(\sigma + \delta)} \right)
             \;-\; \mysp \left( \frac{s'\!+\!\delta\!-\!\SLAMP}
                                   {\sigma + \delta} \right)
             \rule[-.2cm]{0cm}{1cm} \right]^{\,\footnotemark[7]} }
      \miniskip
        & & \equiv & {\ds \frac{1}{2\, \sprm} \; D^t_1 }
      \hugeskip
 15) & {\ds \left[\frac{1}{S_{d2}} \, l_{t2}\right]_\theta} & = &
       {\ds \left[\frac{1}{S_{d1}} \, l_{u1}\right]_\theta
         \; = \;
         \ct \! \left( \left[\frac{1}{S_{d1}} \,
                             l_{t1}\right]_\theta \right)
          \; \equiv \; \frac{1}{2\, \sprm} \; D^t_2 }
\end{array}$
\clearpage
$\begin{array}{llll}
 16) & {\ds \left[\frac{1}{\bar z_1} \, l_{t12} \right]_\theta} & = &
       {\ds \left[\frac{1}{\bar z_2} \, l_{u12} \right]_\theta \;=\;
            \frac{1}{2} \left( \frac{}{} l_\beta \; L_{c1}
                               + D_{z1t1} + D_{z1t2} \right) }
      \hugeskip
 17) & {\ds \left[\frac{1}{\bar z_2} \, l_{t12} \right]_\theta} & = &
       {\ds \left[\frac{1}{\bar z_1} \, l_{u12} \right]_\theta \;=\;
            \frac{1}{2} \left( \frac{}{} l_\beta \; L_{c2}
                               + D_{z2t1} + D_{z2t2} \right) }
      \hugeskip
 18) & {\ds \left[\frac{1}{a^{ut}_1} l_{t12} \right]_\theta} & = &
       {\ds \left[\frac{1}{a^{ut}_2} l_{u12} \right]_\theta \;=\;
            \frac{1}{\sprm} \; \re \!
            \left[ \rule[-.2cm]{0cm}{.35cm} \,
                   L_{c8} \; L_{c6} \; + \; D^{tu}_{a1} \, \right] }
       \medskip & \hspace{1cm}
         L_{c8} & = & {\ds \myln \left( \frac{s-\sigma}{s'-\sigma}
                                 \right) }
       \medskip & \hspace{1cm}
         L_{c6} & = & {\ds \myln
            \frac{(a_{1-}\,b + b_{1-})(a_{2+}\,b + b_{2+})}
                 {(a_{1+}\,b + b_{1+})(a_{2-}\,b + b_{2-})} }
       \medskip & \hspace{1cm}
         D^{tu}_{a1} & = & {\ds
          - \; \mysp \left( \frac{b_{1-}(1+b)}{a_{1-}\,b + b_{1-}} \right)
       \; + \; \mysp \left( \frac{b_{1-}(1-b)}{a_{1-}\,b + b_{1-}} \right) }
          \smskip & & & {\ds
          - \; \mysp \left( \frac{b_{2+}(1+b)}{a_{2+}\,b + b_{2+}} \right)
       \; + \; \mysp \left( \frac{b_{2+}(1-b)}{a_{2+}\,b + b_{2+}} \right) }
          \smskip & & & {\ds
          + \; \mysp \left( \frac{b_{1+}(1+b)}{a_{1+}\,b + b_{1+}} \right)
       \; - \; \mysp \left( \frac{b_{1+}(1-b)}{a_{1+}\,b + b_{1+}} \right) }
          \smskip & & & {\ds
          + \; \mysp \left( \frac{b_{2-}(1+b)}{a_{2-}\,b + b_{2-}} \right)
       \; - \; \mysp \left( \frac{b_{2-}(1-b)}{a_{2-}\,b + b_{2-}} \right) }
       \hugeskip
 19) & {\ds \left[\frac{1}{a^{ut}_2} l_{t12} \right]_\theta } & = &
       {\ds \left[\frac{1}{a^{ut}_1} l_{u12} \right]_\theta \;=\;
            \frac{1}{\sprm} \; \re \!
            \left[ \rule[-.2cm]{0cm}{.35cm} \,
                   L_{c8} \; L_{c7} \; + \; D^{tu}_{a2} \, \right] }
       \smskip & \hspace{1cm}
         L_{c7} & = & \ct \left( L_{c6} \right)
       \smskip & \hspace{1cm}
         D^{tu}_{a2} & = & \ct \left( D^{tu}_{a1} \right)
       \hugeskip
 20) & {\ds \left[\frac{1}{S^2_{d1}} \, l_{t1}\right]_\theta} & = &
       {\ds \ct \! \left( \rule[0cm]{0cm}{.7cm}
         \left[\frac{1}{S^2_{d2}} \, l_{t2}\right]_\theta \right)}
       \smskip & & = &
       {\ds \frac{1}{4\,\sprm s' \SONE} \left(
          \SLAMP \, \myln \frac{s'}{s} \; + \;
          \frac{\SLAMP}{2} \, L_{c3} \; - \;
          \frac{s' - \sigma}{2} \, L_{c5} \right. }
          \smskip & & & \hspace{2cm} {\ds \left. + \;
          \frac{s(s'+\delta) - s'(\sigma+\delta)}{2s} \, L_{t1}
          \right) }
\end{array}$
\clearpage
$\begin{array}{llll}
 21) & {\ds \left[\frac{1}{S^3_{d1}} \, l_{t1}\right]_\theta} & = &
       {\ds \ct \! \left( \rule[0cm]{0cm}{.7cm}
         \left[\frac{1}{S^3_{d2}} \, l_{t2}\right]_\theta \right)}
       \smskip & & = &
       {\ds \frac{1}{16 \, \sprm} \left[ \rule[-.2cm]{0cm}{1cm} \:
             - \; \frac{\sprm \SLAMP}{s {s'}^2 \SONE}
          \; + \; \frac{(s' + \delta) \SLAMP}{2\,{s'}^2 \SONE^2} \,
          L_{c3} \right. }
          \smskip & & & \hspace{1.25cm} {\ds + \;
          \frac{\sigma^2 - \delta^2 + 2\sigma\delta +
                2s'(\sigma-\delta) - 2{s'}^2}
               {4\,{s'}^2 \SONE^2} \, L_{c5} }
          \smskip & & & \hspace{1.2cm} {\ds \left. + \;
          \frac{(s' + \delta) \SLAMP}{{s'}^2 \SONE^2}
          \myln\frac{s'}{s} \; +
          \left( \frac{{s'}^2 - s'(\sigma \!-\! \delta) + \delta^2}
                      {2\,{s'}^2 \SONE^2} - \frac{1}{s^2} \right) \! L_{t1}
          \, \rule[-.2cm]{0cm}{1cm} \right] } \hspace{-.5cm}
      \bigskip
 22) & {\ds \left[\frac{1}{S^4_{d1}} \, l_{t1}\right]_\theta} & = &
       {\ds \ct \! \left( \rule[0cm]{0cm}{.7cm}
         \left[\frac{1}{S^4_{d2}} \, l_{t2}\right]_\theta \right)}
       \smskip & & = &
       {\ds \frac{1}{48 \, \sprm} \left[ \rule[-.2cm]{0cm}{1cm} \:
          - \; \frac{\sprm \SLAMP}{s {s'}^3 \SONE} \;
            \left( \frac{\sprp}{2s} + \frac{s'\!+\!\delta}{\SONE}
            \right) 
          \; - \; \frac{1}{s^3} \; L_{t1}
          \; + \; \frac{1}{{s'}^3} \; L_{c5} \right. }
      \smskip
      & & &  \hspace{1.46cm}
        {\ds + \; \frac{2{s'}^2 + 2\delta^2 + 3s'\delta - s'\sigma}
                       {4 {s'}^3 \SONE^3}
          \; \SLAMP \;
          \left( 2 \myln\frac{s'}{s} + L_{c3} \right) }
      \smskip
      & & &  \hspace{1.46cm}
        {\ds - \: \left. \rule[-.2cm]{0cm}{1cm}
          \frac{(s'\!+\!\delta)
                (2{s'}^2 + 2\delta^2 + s'\delta - 3s'\sigma)}
               {4 {s'}^3 \SONE^3} \; L_{c1}
          \; \right] }
      \bigskip
\end{array}$
\newline
$\begin{array}{llll}
 23) & {\ds \left[\frac{m_e^{\;2}}{\bar z_1^{\;2}} \, l_{t2}
            \right]_\theta } & = &
       {\ds \left[\frac{m_e^{\;2}}{\bar z_2^{\;2}} \, l_{t1}
            \right]_\theta \; = \;
            \left[\frac{m_e^{\;2}}{\bar z_1^{\;2}} \, l_{u2}
            \right]_\theta \; = \;
            \left[\frac{m_e^{\;2}}{\bar z_2^{\;2}} \, l_{u1}
            \right]_\theta \; = \;
            \frac{s}{4} \; L_{c5} }
      \bigskip
 24) & {\ds \left[\frac{1}{d_1^+ \, d_1^-} \right]_\theta} & = &
       {\ds \ct \! \left( \rule[0cm]{0cm}{.7cm}
         \left[\frac{1}{d_2^+ \, d_2^-} \right]_\theta \right) \;=\;
         \frac{1}{4\,\sprm\SLAMP} \; L_{c1} }
      \bigskip
 25) & {\ds \left[\frac{\mycos\theta}{d_1^+ \, d_1^-} \right]_\theta} & = &
       {\ds \frac{1}{4\,{\sprm}^2} \left( \,
         \frac{s-\delta}{\SLAMP} \, L_{c1} \; - \; L_{c3} \, \right) }
      \bigskip
 26) & {\ds \left[\frac{\mycos\theta}{d_2^+ \, d_2^-} \right]_\theta} & = &
       {\ds - \, \frac{1}{4\,{\sprm}^2} \left( \,
         \frac{s+\delta}{\SLAMP} \, L_{c2} \; - \; L_{c4} \, \right) }
      \bigskip
\end{array}$
\newline
$\begin{array}{llll}
 27) & {\ds \left[\frac{1}{S^t_{12+} \, S^t_{12-} } \right]_\theta} & = &
        {\ds \frac{1}{4\,\sprm\SLAMB}  \; L_{\bar \lambda} }
        \bigskip
      & \hspace{1.5cm} L_{\bar \lambda} & = & {\ds
          \myln \left(
            \frac{\sprm + \sigma + \SLAMB}{\sprm + \sigma - \SLAMB}
          \right) }
\end{array}$
\newpage
$\begin{array}{llll}
 28) & {\ds \left[\frac{\mycos\theta}{S^t_{12+} \, S^t_{12-} }
             \right]_\theta} & = &
        {\ds \frac{1}{4\,{\sprm}^2} \left( \,
          \myln \frac{\SONE}{\STWO}
          \; - \; \frac{\delta}{\SLAMB}\,L_{\bar \lambda} \, \right) }
      \hugeskip
 29) & {\ds \left[\frac{1}{(S^t_{12+})^2\,(S^t_{12-})^2}
             \right]_\theta} & = &
        {\ds \frac{1}{8\,\sprm {\bar \lambda}} \left( \,
          \frac{1}{\SLAMB}\,L_{\bar \lambda} \; + \;
          \frac{\sprm \sigma + \delta^2}{2\,\sprm \SONE \STWO} \, \right) }
      \hugeskip
 30) & {\ds \left[\frac{\mycos\theta}{(S^t_{12+})^2\,(S^t_{12-})^2}
             \right]_\theta} & = &
        {\ds \frac{\delta}{8\,{\sprm}^2\LAMB} \left(
          \: - \; \frac{1}{\SLAMB}\,L_{\bar \lambda} \; + \;
          \frac{\sprm + \sigma}{2\,\SONE \STWO} \, \right) }
      \hugeskip
\end{array}$
\newline
$\begin{array}{llll}
 31) & {\ds \left[\frac{l_{t12}}{S^t_{12+} \, S^t_{12-}}
             \right]_\theta } & = &
        {\ds \frac{1}{4\, \sprm \SLAMB} \; \re \!
        \left[ \rule[-.2cm]{0cm}{1cm}
           - \, \mysp \left( + \frac{c_{++} a_{--} e_+}{d} \right)
        \; + \; \mysp \left( + \frac{c_{-+} a_{--} e_+}{d} \right)
        \right. \! }
        \smskip & & & \hspace{2.2cm} {\ds
        \; + \; \mysp \left( - \frac{c_{+-} a_{--} e_+}{d} \right)
        \; - \; \mysp \left( - \frac{c_{--} a_{--} e_+}{d} \right) }
        \smskip & & & \hspace{2.2cm} {\ds
        \; - \; \mysp \left( - \frac{c_{++} a_{-+} e_+}{d} \right)
        \; + \; \mysp \left( - \frac{c_{-+} a_{-+} e_+}{d} \right) }
        \smskip & & & \hspace{2.21cm} {\ds
        \; + \; \mysp \left( + \frac{c_{+-} a_{-+} e_+}{d} \right)
        \; - \; \mysp \left( + \frac{c_{--} a_{-+} e_+}{d} \right) }
        \smskip & & & \hspace{2.2cm} {\ds
        \; - \; \mysp \left( + \frac{c_{+-} a_{++} e_-}{d} \right)
        \; + \; \mysp \left( + \frac{c_{--} a_{++} e_-}{d} \right) }
        \smskip & & & \hspace{2.21cm} {\ds
        \; + \; \mysp \left( - \frac{c_{++} a_{++} e_-}{d} \right)
        \; - \; \mysp \left( - \frac{c_{-+} a_{++} e_-}{d} \right) }
        \smskip & & & \hspace{2.2cm} {\ds
        \; - \; \mysp \left( - \frac{c_{+-} a_{+-} e_-}{d} \right)
        \; + \; \mysp \left( - \frac{c_{--} a_{+-} e_-}{d} \right) }
        \smskip & & & \hspace{2.165cm} {\ds \left.
        \; + \; \mysp \left( + \frac{c_{++} a_{+-} e_-}{d} \right)
        \; - \; \mysp \left( + \frac{c_{-+} a_{+-} e_-}{d} \right)
        \rule[-.2cm]{0cm}{1cm} \right] }
        \bigskip
          & & \equiv & {\ds \frac{1}{4\, \sprm \SLAMB} \; D^t_{12} }
        \bigskip
      & \hspace{1.5cm}
          a_{\pm\pm} & = & s \pm \SLAMP \pm \SLAMB  \smskip
      & \hspace{1.5cm}
          c_{\pm\pm} & = & \delta \pm \sprm \pm \SLAMB  \smskip
      & \hspace{1.6cm}
          e_{\pm} & = & s' - \sigma \pm \SLAMP   \smskip
      & \hspace{1.7cm}
          d & = & 8\,s\,\SONE\,\STWO
\end{array}$
\newpage
$\begin{array}{llll}
 32) & {\ds \left[\frac{l_{t12}}{(S^t_{12+})^2 \, (S^t_{12-})^2}
             \right]_\theta } & = &
        {\ds \frac{1}{2\,{\bar \lambda}}
          \left[\frac{l_{t12}}{S^t_{12+} \, S^t_{12-}} \right]_\theta }
        \smskip & & &
        {\ds + \; \frac{1}{8\,\sprm \LAMB} \left[ \;
          \frac{\SLAMP\SLAMB}{2\,s\SONE\STWO} \, L_{\bar \lambda} \; + \;
          \frac{\sprm - \delta}{2\,\sprm \SONE} \, L_{c1} \; + \;
          \frac{\sprm + \delta}{2\,\sprm \STWO} \, L_{c2} \right. }
        \smskip & & & \hspace{1.8cm} {\ds
          \; + \; \frac{{\bar \lambda} - s(\sprm+\sigma)}{4\,s\SONE\STWO}
          \left( L_{c1} + L_{c2} \right) }
        \smskip & & & \hspace{1.75cm} {\ds \left.
          \; - \; \frac{\SLAMP(\sprm+\sigma)}{4\,s\SONE\STWO}
          \left( L_{c3} + L_{c4} \right)
        \; \right] }
        \hugeskip \\
 33) & {\ds \left[\frac{\mycos\theta \; \; \; l_{t12}}
                        {(S^t_{12+})^2 \, (S^t_{12-})^2}
             \right]_\theta } & = &
        {\ds - \; \frac{\delta}{\sprm}
          \left[\frac{l_{t12}}{(S^t_{12+})^2 \, (S^t_{12-})^2}
          \right]_\theta }
        \smskip & & &
        {\ds + \; \frac{1}{8\,{\sprm}^2} \left[ \;
          \frac{\SLAMP}{2\,s\SONE\STWO} \myln \frac{\SONE}{\STWO}
          \; - \;
          \frac{1}{2\,\sprm \SONE} \, L_{c1} \; + \;
          \frac{1}{2\,\sprm \STWO} \, L_{c2} \right. }
        \smskip & & & \hspace{1.58cm} {\ds \left.
          \; - \; \frac{s'-\sigma}{4\,s\SONE\STWO}
          \left( L_{c1} \! - \! L_{c2} \right)
          \; - \; \frac{\SLAMP}{4\,s\SONE\STWO}
          \left( L_{c3} \! - \! L_{c4} \right) \; \right] }
        \bigskip \\
 34) & {\ds \left[\frac{l_{t12}}{(S^t_{12+})^3 \, (S^t_{12-})^3}
             \right]_\theta } & = &
        {\ds \frac{3}{4 \LAMB}
          \left[\frac{l_{t12}}{(S^t_{12+})^2 \, (S^t_{12-})^2}
          \right]_\theta
          \;\;  + \;\; \frac{1}{32\,\sprm \SLAMB^{\,3}} }
        \smskip & & & {\ds \times \left[ \;
          \frac{\SLAMB\,\SLAMP\,\left( \sprm \sigma + 2\delta^2 \right)}
               {4 \, s \sprm \SONE^2 \STWO^2} \; + \;
          \frac{\SLAMB\,\delta}{4 \,\SONE^2 \STWO^2} \, L_{c12}
          \; + \; \frac{\SLAMB}{4 \,\sprm \SONE^2} \, L_{c1} \right. }
        \smskip & & & \hspace{.32cm} {\ds
             + \; \frac{\SLAMB}{4 \,\sprm \STWO^2} \, L_{c2} \; - \;
             \frac{\SLAMP \left( 2s\SONE\STWO + \LAMB (s'-\sigma)\right)}
                  {4\,s^2 \SONE^2 \STWO^2} \, L_{\LAMB}}
        \smskip & & & \hspace{.32cm} {\ds
             - \; \frac{\SLAMB \left( 6\,s\SONE\STWO -
                        s^2(\sprm\!+\!\sigma) +
                        \LAMB(\sprp\!-\!\sigma) \right) }
                       {8\,s^2 \SONE^2 \STWO^2}
                  \left( L_{c1} + L_{c2} \right) }
        \smskip & & & \hspace{.32cm} {\ds \left.
             + \; \frac{\SLAMB\,\SLAMP\,\left( s(\sprm+\sigma) -
                        2\,\SONE\STWO - \LAMB \right) }
                       {8\,s^2 \SONE^2 \STWO^2}
                  \left( L_{c3} \! + \! L_{c4} \right)
        \;\right] }
        \medskip
      & \hspace{1.5cm} L_{c12} & = & {\ds
          \frac{1}{\sprm} \; \left( \frac{\SONE^2}{\sprm}\,L_{c2}
            - \frac{\STWO^2}{\sprm}\,L_{c1}
            - \frac{\delta\,\SLAMP}{s} \right) }
\end{array}$
\newpage
$\begin{array}{llll}
 35) & {\ds \left[\frac{\mycos\theta \; \; \; l_{t12}}
                        {(S^t_{12+})^3 \, (S^t_{12-})^3}
             \right]_\theta } & = &
        {\ds \frac{1}{\,4\,\LAMB\,}
          \left[\frac{\mycos\theta \; \; \; l_{t12}}
                     {(S^t_{12+})^2 \, (S^t_{12-})^2}
          \right]_\theta \; + \;
          \frac{\delta}{4\,\sprm\LAMB}
          \left[\frac{l_{t12}}{(S^t_{12+})^2 \, (S^t_{12-})^2}
          \right]_\theta }
        \vspace{.25cm} \\ & & & {\ds 
          - \; \frac{\delta}{\sprm}
          \left[\frac{l_{t12}}{(S^t_{12+})^3 \, (S^t_{12-})^3}
          \right]_\theta  \; + \;
          \frac{1}{32\,{\sprm}^2 \LAMB}
          \left[ \;
          \frac{\SLAMP\,\LAMB\,\delta}{2\,s\,\sprm\,\SONE^2\STWO^2} \right.}
        \vspace{.25cm} \\ & & & {\ds \; + \;
          \frac{\LAMB}{4\,\SONE^2 \STWO^2} \, L_{c12} \; + \;
          \frac{1}{2\,\sprm\,\SONE}\,L_{c1}\;-\;
          \frac{1}{2\,\sprm\,\STWO}\,L_{c2}  }
        \vspace{.25cm} \\ & & & {\ds \; - \;
          \frac{\SLAMP \left( (s'\!-\!\sigma)\,\LAMB +
                2\,s\,\SONE\STWO \right)}{8\,s^2 \SONE^2 \STWO^2}
         \; \left( 2\myln \frac{\SONE}{\STWO} - L_{c3} + L_{c4} \right) }
         \smskip & & & {\ds \left. \; + \;
          \frac{\LAMB\LAMP + 2\,\SONE\STWO \left( s\,(s'\!-\!\sigma)
                + \LAMB \right)}
               {8\,s^2 \SONE^2 \STWO^2}
          \left( L_{c1} - L_{c2} \right)
        \hspace{.1cm} \right] } \hspace{-.75cm}
        \bigskip
\end{array}$
\newline
$\begin{array}{llll}
 36) & {\ds \left[ \frac{l_{t_1u_2}}{\,2\,\sqrt{C_{12}}\,} \right]_\theta }
      & = &
        {\ds \frac{\ri \; c_{12}}{x_0} \;
          \left[ \rule[-.2cm]{0cm}{1cm} \;
          \mysp \left( \frac{\beta_{12} + \ri\:\!x_0}{\tau_1} \right)
          \; + \;
          \mysp \left( \frac{\beta_{12} + \ri\:\!x_0}{\tau_2} \right)
          \right. }
        \vspace{.1cm} \\ & & & \hspace{1.63cm} {\ds + \;
          \mysp \left( \frac{\beta_{12} + \ri\:\!x_0}{\tau_1^*} \right) 
          \; + \;
          \mysp \left( \frac{\beta_{12} + \ri\:\!x_0}{\tau_2^*} \right) }
        \vspace{.25cm} \\ & & & \hspace{1.625cm} {\ds - \;
          \mysp \left(-\frac{\beta_{12} + \ri\:\!x_0}{\tau_1} \right)
          \; - \;
          \mysp \left(-\frac{\beta_{12} + \ri\:\!x_0}{\tau_2} \right) }
        \vspace{-.2cm} \\ & & & \hspace{1.58cm} {\ds \left. - \;
          \mysp \left(-\frac{\beta_{12} + \ri\:\!x_0}{\tau_1^*}
        \right)
          \; - \;
          \mysp \left(-\frac{\beta_{12} + \ri\:\!x_0}{\tau_2^*}
        \right)
        \hspace{.2cm} \rule[-.2cm]{0cm}{1cm} \right]^{\;\footnotemark[8]} }
        \\
      & & \equiv & c_{12} \; D_{t1u2}
        \vspace{.25cm} \\
      & \hspace{1.5cm} c_{12} & = &
        {\ds \frac{s'}{\sprm\sqrt{\LAMP\,s\,\SONE}} } \vspace{.18cm} \\
      & \hspace{1.5cm} x_0 & = &
        {\ds \sqrt{\frac{s-\SONE}{s}} } \vspace{.18cm} \\
      & \hspace{1.5cm} \beta_{12} & = &
        {\ds \sqrt{\frac{\SONE-4\ME^2}{s}} }
\end{array}$
\newline
$\begin{array}{llll}
      \hspace{1.5cm} & \tau_1 & = &
        {\ds \frac{+ q_{12} - \ri\:\!p_{12}}{b_{12} + \ri\:\!r_{12}} }
        \smskip 
      \hspace{1.5cm} & \tau_2 & = &
        {\ds \frac{- q_{12} - \ri\:\!p_{12}}{b_{12} + \ri\:\!r_{12}} }
        \bigskip
\end{array}$
$\begin{array}{llll}
      & \hspace{1.5cm} q_{12} & = &
        {\ds \left(ss' - s\STWO -s'\SONE\right)
             \sqrt{\frac{\SONE}{s}} } \vspace{.1cm} \vspace{-.05cm} \\
      & \hspace{1.5cm} p_{12} & = & \sprm\,\SONE \; x_0
      \vspace{.12cm} \\
      & \hspace{1.5cm} r_{12} & = &
        \sqrt{\LAMP\,s\,\SONE} \; x_0 \vspace{.12cm} \\
      & \hspace{1.5cm} b_{12} & = & \SONE \left(s-\delta\right)
\end{array}$
\footnotetext[8]{Integrals 36) to 41) were not directly calculated
  from the result of the $\Omega_R$ integration. Instead, a Feynman
  parametrization was used to ``linearize'' $1/t_1u_2$
  and $1/t_2u_1$: $1/t_1u_2 = \int_0^1 d\alpha/\left[t_1\alpha +
  u_2(1-\alpha)\right]^2$. Then, $\Omega_R$, $\theta$, and finally the
  Feynman parameter $\alpha$~were integrated.}
\newpage
$\begin{array}{llll}
 37) & {\ds \left[ \frac{l_{t_2u_1}}{\,2\,\sqrt{C_{21}}\,}
         \right]_\theta } \;\; & = &
        {\ds \ct \! \left( \left[
          \frac{l_{t_1u_2}}{\,2\,\sqrt{C_{12}}\,}
        \right]_\theta \right) } \;\; \equiv \;\;
        c_{34} \; D_{t2u1}
\bigskip
\end{array}$
\newline
$\begin{array}{llll}
 38) & {\ds \left[ \frac{l_{t_1u_2}}{\,{\bar z_1} \; 2\,\sqrt{C_{12}}\,}
             \right]_\theta } & = &
       {\ds \left[ \frac{l_{t_1u_2}}{\,{\bar z_2} \;
           2\,\sqrt{C_{12}}\,}
             \right]_\theta } \; = \;
\end{array}$
\\
\ba
  \lefteqn{ \hspace*{-.3cm}
  \frac{s'}{4\,\SLAMP\, \left(ss' - s\STWO -s'\SONE\right)}
  \; \left\{ \rule[-.3cm]{0cm}{1.4cm} \;\;
  2\, l_{\beta} \; \left( 2 L_{c5} - L_{c1} \right) \right. } \nl \nl
  \hspace*{.33cm}
  &  + &  \myln^2 \left(\frac{2\,x_1}{x_1 - 1} \right)
  \; - \; \myln^2 \left(\frac{2\,x_1}{x_1 + 1} \right)
  \; + \; \mysp \left(-\frac{x_1 + 1}{x_1 - 1} \right)
  \; - \; \mysp \left(-\frac{x_1 - 1}{x_1 + 1} \right) \nl \nl
  &  - &  \myln^2 \left(\frac{2\,x_2}{x_2 - 1} \right)
  \; + \; \myln^2 \left(\frac{2\,x_2}{x_2 + 1} \right)
  \; - \; \mysp \left(-\frac{x_2 + 1}{x_2 - 1} \right)
  \; + \; \mysp \left(-\frac{x_2 - 1}{x_2 + 1} \right) \nl \nl
  &  - &  \myln \frac{x_1-x_2}{x_1+x_2} \; \left(
          \myln \frac{x_2 - 1}{x_2 + 1} \; - \; 
          \myln \frac{x_1 - 1}{x_1 + 1} \right) \nl \nl
  &  + &  \myln \frac{x_1 - 1}{x_1-x_2} \;
          \myln \left| \frac{x_2 - 1}{x_1-x_2} \right|
  \; - \; \myln \frac{x_1 + 1}{x_1-x_2} \;
          \myln \left| \frac{x_2 + 1}{x_1-x_2} \right| \nl \nl
  &  + &  \myln \frac{x_1 + 1}{x_1+x_2} \;
          \myln \left| \frac{x_2 - 1}{x_1+x_2} \right|
  \; - \; \myln \frac{x_1 - 1}{x_1+x_2} \;
          \myln \left| \frac{x_2 + 1}{x_1+x_2} \right| \nl \nl
  &  + &  2 \: \mysp \left( - \frac{x_2 - 1}{x_1-x_2} \right)
  \; - \; 2 \: \mysp \left( - \frac{x_2 + 1}{x_1-x_2} \right) \nl \nl
  &  + & \frac{\sprm\,\sqrt{\SONE\rule[0cm]{0cm}{.3cm}}}{\sqrt{s\LAMP}}
         \; \re \left[ \rule[-.3cm]{0cm}{1.3cm} \;
         \sum_{i=1}^2 \;
         \frac{(-1)^{i+1}\;(x_i-a_{12})}
              {\sqrt{\beta^2\,x_i^2-x_0^2}}
         \; \times \right. \nl \nl
  & &    \sum_{k=1}^8 (-1)^{k+1}
         \!\left( \rule[-.3cm]{0cm}{1.2cm} \;
         \myln \frac{\,t^{(i)}_{k+}\,}{t^0_k} \! \; \!\!
         \left( \rule[-.1cm]{0cm}{.6cm}
         \myln \left[u^{(i)}_+(+\beta)\right] -
         \myln \left[u^{(i)}_+(-\beta)\right] -
         2 \pi \ri \; \Theta(x_i) \right) \right. \nl
  & & \hspace{2.37cm} - \myln \frac{\,t^{(i)}_{k-}\,}{t^0_k} \! \; \!\!
         \left( \rule[-.1cm]{0cm}{.6cm}
         \myln \left[u^{(i)}_-(+\beta)\right] -
         \myln \left[u^{(i)}_-(-\beta)\right] -
         2 \pi \ri \; \Theta(-x_i) \right) \nl
  & & \hspace{2.37cm} - \,
         \mysp \left[ -\frac{u^{(i)}_+(+\beta)}{t^{(i)}_{k+}} \right] \; + \;
         \mysp \left[ -\frac{u^{(i)}_+(-\beta)}{t^{(i)}_{k+}} \right] \nl
  & & \hspace{2.25cm} \left. \left. \left. + \,
         \mysp \left[ -\frac{u^{(i)}_-(+\beta)}{t^{(i)}_{k-}} \right] \; - \;
         \mysp \left[ -\frac{u^{(i)}_-(-\beta)}{t^{(i)}_{k-}} \right]
         \; \rule[-.3cm]{0cm}{1.2cm} \right)
         \; \rule[-.3cm]{0cm}{1.3cm} \right]
         \indent\; \rule[-.3cm]{0cm}{1.4cm} \right\} \nl \nl \nl
  & \equiv & \frac{s'}{4\,\SLAMP\, \left(ss' - s\STWO -s'\SONE\right)}
         \;
         \left\{ \rule[-.3cm]{0cm}{.35cm}
                 \, 2 \, l_{\beta} \; \left( 2 L_{c5} - L_{c1} \right)
                 \; + \; D^z_{t1u2} \right\}  \nonumber
\ea
\newpage
$\begin{array}{lcclccl}
  & \hspace{1cm} a_{12}  & = & {\ds \frac{s-\delta}{\sprm} } \miniskip
  & \hspace{1cm} t^0_1 & = & \tau_1   \hspace{4.5cm} &
    \hspace{1cm} t^0_2 & = & - \tau_1 \\
  & \hspace{1cm} t^0_3 & = & \tau_2   \hspace{4.5cm} &
    \hspace{1cm} t^0_4 & = & - \tau_2 \\
  & \hspace{1cm} t^0_5 & = & \tau_2^* \hspace{4.5cm} &
    \hspace{1cm} t^0_6 & = & - \tau_2^* \\
  & \hspace{1cm} t^0_7 & = & \tau_1^* \hspace{4.5cm} &
    \hspace{1cm} t^0_8 & = & - \tau_1^*
  \medskip
  & \hspace{1cm}t^{(i)}_{k\pm}  & = & t^0_k \; + \; t^{(i)}_{\pm}
  & \hspace{1cm}u^{(i)}_{\pm}(x)  & = & t(x) \; - \; t^{(i)}_{\pm}
    \smskip
  & \hspace{1cm}t^{(i)}_{\pm} & = & {\ds
      \frac{\ri\,\!x_0 \pm \sqrt{\beta^2 x_i^2 - {x_0}^2}}{x_i} }
  & \hspace{1.1cm}    t(x) & = & {\ds
      \frac{\ri\,\!x_0 + \sqrt{x^2-{x_0}^2}}{x} } \medskip
\end{array}$
\newline
$\begin{array}{lccl}
  & \hspace{1.04cm} x_{1/2} \hspace{1pt} & = & {\ds
      \frac{\sprm\,\SONE(s-\delta) \pm \SLAMP\,(ss' - s\STWO - s'\SONE)}
           {{\sprm}^2\SONE - s \LAMP} }
    \smskip
  & \hspace{1cm} \beta  & = &
    {\ds \sqrt{1 - \frac{4\,\ME^2}{s}} }
\end{array}$
\hugeskip
\hugeskip
$\begin{array}{llll}
 39) & {\ds \left[ \frac{l_{t_2u_1}}
         {\,{\bar z_1} \; 2\,\sqrt{C_{21}}\,} \right]_\theta } & = &
       {\ds \left[ \frac{l_{t_2u_1}}
                        {\,{\bar z_2} \; 2\,\sqrt{C_{21}}\,}
            \right]_\theta \; = \;
         \ct \! \left( \rule[-.2cm]{0cm}{1cm}
                  \left[ \frac{l_{t_1u_2}}
                              {\,{\bar z_1} \; 2\,\sqrt{C_{12}}\,}
                  \right]_\theta
                \right) }
     \medskip & & \equiv &
     {\ds \frac{s'}{4\,\SLAMP\, \left(ss' - s\SONE - s'\STWO\right)}
          \;
          \left\{ \rule[-.3cm]{0cm}{.35cm}
                  \, 2 \, l_{\beta} \; \left( 2 L_{c5} \!-\! L_{c2} \right)
                  \; + \; D^z_{t2u1} \right\} }
\hugeskip
\end{array}$
\newline
$\begin{array}{llll}
 40) & {\ds \left[ \frac{l_{t_1u_2}}{\,a^{ut}_1 \; 2\,\sqrt{C_{12}}\,}
             \right]_\theta } & = &
       {\ds \left[ \frac{l_{t_1u_2}}{\,a^{ut}_2 \;
            2\,\sqrt{C_{12}}\,} \right]_\theta  \; = \;
         \frac{s'}{\sprm} \; \re
         \left( \rule[-.1cm]{0cm}{.5cm} X_0 + X_{1} + X_{2} \right)
         \: \equiv \: \frac{s'}{\sprm} \, D^a_{t1u2} }
\end{array}$
\begin{list}{}{\leftmargin=.95cm}
  \item In the calculation of the integrals $X_i$ the following
    quantities appear: \vspace{-.5cm}
\end{list}
\ba
  \hspace*{2cm}
  A_3 & = & \SONE\,(\sprp -2\sigma)^2 \; - \; \beta^2 s \LAMP
   \hspace{5.4cm} \nl
  B_3 & = & \SONE\,(\sprp -2\sigma)\,(s-\delta) \nl
  \Delta_3 & = & \LAMP \; \left[ \rule[-.1cm]{0cm}{.7cm}
    \left\{ \rule[-.1cm]{0cm}{.5cm}
      ss' - s\STWO -s'\SONE -2\SONE\,(s-\sigma) \right\}^2 \right. \nl
    & & \left. \hspace{1cm} \; - \;
    4\ME^2 \left\{ \SONE\,(s-\delta)^2 + \LAMP\,(s-\SONE) \right\}
    \rule[-.1cm]{0cm}{.7cm} \right]
    \nl \nl
  x_{3/4}  & = & {\ds \beta \; \frac{-B_3 \pm \sqrt{\Delta_3}}{A_3} }
    \nonumber
\ea
\begin{list}{}{\leftmargin=.95cm}
  \item In principle, depending on the values of $x_3$ and $x_4$, four
    cases must be distin\-guished: \vspace{.2cm}
\end{list}
$\begin{array}{lrcccccc}
 \hspace*{1.75cm} & (i) & x_{3/4} \, \in \, \numreal ; & \;\;
                         |x_4| & \geq & |x_3| & > & \beta \\
                 & (ii) & x_{3/4} \, \in \, \numreal ; & \;\;
                         |x_4| & > & \beta & \geq & |x_3| \\
                & (iii) & x_{3/4} \, \in \, \numreal ; & \;\;
                         \beta & \geq & |x_4| & \geq & |x_3| \\
                 & (iv) & \Delta_3 \leq 0 ; & \;\;
                         x_3 & = & x_4^* & \in & \numcomp
 \vspace{.2cm}
\end{array}$
\begin{list}{}{\leftmargin=.95cm}
\item However, since the integrand is regular, it is sufficient to
  present the solution for case $(i)$. The solutions for the other
  three cases are then obtained by analytical continuation. In
  case $(iv)$, which is relevant for only a very small fraction of the
  phase space, the problem of a numerically correct treatment arises.
  This problem was solved by computing the integral's value
  for a very nearby phase space point so that the expression of case
  $(i)$~could be used. Having in mind that the integral is regular, it
  is clear that thus only a negligible error is introduced.
  The expressions $X_0,~X_1$, and $X_2$ for case $(i)$~are presented
  below.
\end{list}
\ba
  \hspace*{.75cm}
  X_0 & = & \frac{2}{\,\sqrt{\Delta_3}\,} \;
            L_{c8} \; \left[ \rule[-.3cm]{0cm}{1.1cm}
            \myln \left( \frac{\beta-x_3}{\beta+x_3} \right) \; - \;
            \myln \left( \frac{\beta-x_4}{\beta+x_4} \right) \right]
            \hspace{1.9cm} \nonumber
\ea
\newline
\ba
  \hspace*{.75cm}
  X_1 & = & \frac{\,\sqrt{\SONE\rule[0cm]{0cm}{.3cm}} \;
                  \left( \sprp-\sigma \right)\,}
                 {2\,\sqrt{s\,\LAMP\,\Delta_3}} \; \; \;
            \sum_{i=3}^4 \frac{(-1)^{i+1}\;(x_i-a_{34})}
                              {\sqrt{x_i^2-x_0^2}}   \times \nl
  & & \sum_{k=1}^8 (-1)^{k+1}
         \!\left( \rule[-.3cm]{0cm}{1.2cm}
         \myln \frac{\,t^{(i)}_{k+}\,}{t^0_k}  \,
         \left( \rule[-.1cm]{0cm}{.6cm}
         \myln \left[u^{(i)}_+(+\beta)\right] -
         \myln \left[u^{(i)}_+(-\beta)\right] -
         2 \pi \ri \: \Theta(x_i) \right) \right. \nl
  & & \hspace{1.5cm} \! - \myln \frac{\,t^{(i)}_{k-}\,}{t^0_k} \!
         \; \!\! \left( \rule[-.1cm]{0cm}{.6cm}
         \myln \left[u^{(i)}_-(+\beta)\right] -
         \myln \left[u^{(i)}_-(-\beta)\right] -
         2 \pi \ri \; \Theta(-x_i) \right) \! \nl
  & & \hspace{1.5cm} \! - \,
         \mysp \left[ -\frac{u^{(i)}_+(+\beta)}{t^{(i)}_{k+}} \right]
         \; + \;
         \mysp \left[ -\frac{u^{(i)}_+(-\beta)}{t^{(i)}_{k+}} \right]
         \nl
  & & \hspace{1.46cm} \! \left. + \,
         \mysp \left[ -\frac{u^{(i)}_-(+\beta)}{t^{(i)}_{k-}} \right]
         \; - \;
         \mysp \left[ -\frac{u^{(i)}_-(-\beta)}{t^{(i)}_{k-}} \right]
         \; \rule[-.3cm]{0cm}{1.2cm} \right) \nl \nonumber
\ea
$\begin{array}{lrclrcl}
  & \hspace{2cm} a_{34} & = & {\ds - \frac{s-\delta}{\sprp - 2\sigma} } 
  \medskip
  & \hspace{.5cm} t^{(i)}_{\pm\,k}  & = & t^0_k \; + \; t^{(i)}_{\pm}
  & \hspace{2cm} u^{(i)}_{\pm}(x)  & = & t(x) \; - \; t^{(i)}_{\pm}
    \smskip
  & \hspace{.5cm} t^{(i)}_{\pm} & = & {\ds
      \frac{\ri\,\!x_0 \pm \sqrt{x_i^2 - {x_0}^2}}{x_i} }
  & \hspace{2cm}    t(x) & = & {\ds
      \frac{\ri\,\!x_0 + \sqrt{x^2-{x_0}^2}}{x} } \bigskip \\
\end{array}$
\ba
  \hspace*{.75cm}
  X_2 & = & -\frac{1}{\,2\,\sqrt{\Delta_3}\,} \; \times \; \nl
      & & \sum_{i=1}^2 \; \sum_{j=1}^2 \; (-1)^{j+1} \,
            \left( \rule[-.3cm]{0cm}{1.2cm} \;
            \myln \frac{x_i-x_j-\ieps}{x_i+x_j-\ieps} \;
            \myln \frac{x_j-\beta}{x_j+\beta} \right. \nl \nl
 & & \hspace{2.36cm} - \; \mysp \left[
            -\frac{x_j-\beta}{x_i-x_j-\ieps} \right] \; + \;
            \mysp \left[
            -\frac{x_j+\beta}{x_i-x_j-\ieps} \right] \nl \nl
 & & \hspace{2.33cm} \left. + \; \mysp \left[
            \frac{x_j-\beta}{x_i+x_j-\ieps} \right] \; - \;
            \mysp \left[
            \frac{x_j+\beta}{x_i+x_j-\ieps} \right]
            \rule[-.3cm]{0cm}{1.2cm} \; \right) \; \nonumber
\ea
\hugeskip \\
$\begin{array}{llll}
 41) & {\ds \left[ \frac{l_{t_2u_1}}{\,a^{ut}_1 \; 2\,\sqrt{C_{21}}\,}
             \right]_\theta } & = &
       {\ds \left[ \frac{l_{t_2u_1}}
                        {\,a^{ut}_2 \; 2\,\sqrt{C_{21}}\,}
             \right]_\theta \; = \;
         \ct \! \left( \rule[-.2cm]{0cm}{1cm}
                   \left[ \frac{l_{t_1u_2}}
                               {{\,a^{ut}_1} \; 2\,\sqrt{C_{12}}\,}
                   \right]_\theta \right)}
         \; \equiv \; {\ds \frac{s'}{\sprm} \; D^a_{t2u1} }
\end{array}$
%
%
\clearpage
%

%
\end{document}